\documentclass[preprint]{elsarticle}

\usepackage[colorlinks,citecolor=blue,linktoc=all,linkcolor=cyan]{hyperref}
\usepackage{graphicx}

\usepackage[T1]{fontenc}
\usepackage{dsfont}               
\usepackage{mathrsfs}             
\usepackage{slashed}              
\usepackage{amsmath}
\usepackage{amssymb}
\usepackage{amsbsy}
\usepackage{amsfonts}

\numberwithin{equation}{section}
\numberwithin{table}{section}
\numberwithin{figure}{section}

\journal{Progress in Particle and Nuclear Physics}

\usepackage{titlesec}
\usepackage{sectsty}
\titleformat{\section}{\normalfont\Large\bfseries}{\thesection}{1em}{}
\titleformat{\subsection}{\normalfont\large\bfseries}{\thesubsection}{1em}{}
\titleformat{\subsubsection}{\normalfont\normalsize\bfseries}{\thesubsubsection}{1em}{}

\def\eg{{\em e.g.}}
\def\ie{{\em i.e.}}

\newcommand{\beq}{\begin{equation}}
\newcommand{\eeq}{\end{equation}}
\newcommand{\bea}{\begin{eqnarray}}
\newcommand{\eea}{\end{eqnarray}}
\newcommand{\ltsim}{\raisebox{-4pt}{$\,\stackrel{\textstyle <}{\sim}\,$}}
\newcommand{\gtsim}{\raisebox{-4pt}{$\,\stackrel{\textstyle >}{\sim}\,$}}
\newcommand{\QQb}{Q\bar{Q}}
\newcommand{\cD}{{\cal D}}
\newcommand{\cQ}{{\cal Q}}
\newcommand{\Tpc}{T_{\rm pc}}
\newcommand{\RAA}{R_{\rm AA}}
\newcommand{\bvec}[1]{\mathbf{#1}}
\newcommand{\dd}{\mathrm{d}}
\newcommand{\erw}[1]{\left \langle #1 \right \rangle}
\newcommand{\MeV}{\mathrm{MeV}}
\newcommand{\GeV}{\mathrm{GeV}}
\newcommand{\TeV}{\mathrm{TeV}}

\begin{document}
	
	\begin{frontmatter}
		
		\title{Heavy-Quark Diffusion in the Quark-Gluon Plasma}

		\author[He]{Min He}
                \author[Hees,Hees2]{Hendrik van Hees}
		\author[Rapp]{Ralf Rapp\corref{mycorrespondingauthor}}
		\cortext[mycorrespondingauthor]{Corresponding author}
		\ead{rapp@comp.tamu.edu}
		
		\address[He]{Department of Applied Physics, Nanjing University of Science and Technology, Nanjing 210094, China}
		\address[Hees]{Institut fuer Theoretische Physik, Goethe-Universit\"at Frankfurt am Main, Max-von-Laue-Strasse 1, D-60438 Frankfurt am Main, Germany}
             \address[Hees2]{Helmholtz Research Academy Hesse for FAIR, Campus Riedberg,  Max-von-Laue-Str. 12, D-60438 Frankfurt am Main, Germany}
              \address[Rapp]{Cyclotron Institute and Department of Physics and Astronomy,
Texas A\&M University, College Station, Texas 77843-3366, U.S.A.}
		
		\begin{abstract}
The diffusion of heavy quarks through the quark-gluon plasma (QGP) as produced in high-energy heavy-ion collisions has long been recognized 
as an excellent probe of its transport properties. In addition, the experimentally observed heavy-flavor hadrons carry valuable information
about the hadronization process of the transported quarks. Here we review recent progress in the theoretical developments of heavy-quark 
interactions in the QGP and how they relate to the nonperturbative hadronization process, and discuss the recent status of the pertinent phenomenology
in heavy-ion collisions at the RHIC and the LHC. The interactions of heavy quarks in the QGP also constitute a central building block in the description
of the heavy quarkonia which controls their transport parameters as well. We will thus focus on theoretical approaches that aim for a unified description 
of open and hidden heavy-flavor particles in medium, and discuss how they can be constrained by lattice QCD ``data'' and utilized to deduce fundamental properties of the microscopic interactions and emerging spectral properties of the strongly coupled QGP.
		\end{abstract}
		
		
	\end{frontmatter}
	
	\newpage
	
	\thispagestyle{empty}
	\tableofcontents
	

	\newpage
	\section{Introduction}
\label{intro}
The theory of the strong nuclear force, Quantum Chromodyanics (QCD), is anchored in a remarkably compact Lagrangian density,
\beq
{\cal L}_{QCD} =   \bar q (i D_\mu \gamma^\mu - m_q) q+ \frac{1}{4} G^{\mu\nu} G_{\mu\nu}    \  ,  
\eeq
formulated in terms of spin-1/2 quark fields, $q$, with bare masses $m_q$, interacting via massless bosonic gluon fields, $A^\mu$, encoded in 
$D^\mu = \partial^\mu -igA^\mu$ with coupling strength $g$. Contrary 
to Quantum Electrodynamics (QED), the quark fields come in three color-charges, and the gluons carry color-charge as well, implying a self-interaction 
which is encoded in the field strength tensor, $G^{\mu\nu}$. This leads to a running coupling constant, $\alpha_s=g^2/4\pi$, that increases with decreasing 
momentum transfer. As a consequence, nonperturbative phenomena emerge at scales of 1\,GeV and below, most notably the confinement of quarks
and gluons into hadrons, and a spontaneous breaking of chiral symmetry causing a mass splitting of about 0.5\,GeV in chiral multiplets of the light 
hadronic spectrum.  
These phenomena are intimately related to the phase structure of QCD matter. Lattice-discretized computations of the finite-temperature partition
function of QCD have demonstrated that the spontaneously broken chiral symmetry is gradually restored around a pseudo-critical temperature of 
$\Tpc\simeq 155$\,MeV~\cite{Aoki:2006br,HotQCD:2014kol}. In the same temperature regime, the thermodynamic bulk properties indicate a 
transition from hadronic to quark-gluon 
degrees of freedom. However, the nature of the quark-gluon plasma (QGP), \ie, its spectral and transport properties, remains under intense debate.
High-energy heavy-ion collisions conducted at the Relativistic Heavy-Ion Collider (RHIC) and the Large Hadron Collider (LHC) have revealed 
that the hot QCD medium formed in these reactions is strongly coupled as deduced from the success of relativistic hydrodynamics. The latter require
a rapid local thermalization of the medium with little friction losses in the subsequent fireball expansion, implying a very small value for the ratio 
of  shear viscosity to entropy density, close to a conjectured quantum lower bound. A key question is how this phenomenon emerges from the
underlying QCD interactions, and which probes in heavy-ion collisions can help to unravel them. 

In the vicinity of the phase transition and at the temperatures typically reached in heavy-ion collisions (up to ca.~500\,MeV at the LHC), the QCD medium
is essentially made of the {\em light} up, down and strange quarks  ($q$=$u$,$d$,$s$), antiquarks and gluons, and/or their composites. In this regime, 
the {\em heavy} charm and bottom quarks ($Q=c,b$) turn out to be excellent probes of the interaction strength in the QGP, and thus its transport 
properties(top quarks, with a lifetime of $\sim$0.15\,fm/$c$, are too short-lived for this purpose). 
When immersed into a QGP, heavy quarks exert a Brownian motion where the pertinent diffusion coefficient gives direct information about their interactions 
with their environment~\cite{Svetitsky:1987gq,Rapp:2009my}. This idea works surprisingly well, in momentum space, in the context of heavy-ion collisions: 
Heavy quarks are produced very early on in the collision, with reasonably well-defined initial momentum spectra (which are substantially harder than the 
asymptotic limit of  thermalized spectra). Most of these quarks are produced at relatively small momenta, and through multiple soft elastic scatterings 
their spectra are driven toward thermalization. Their thermal relaxation rate, $\gamma_Q$, is suppressed relative to the light partonic degrees of 
freedom by a factor of $T/m_Q$, where $m_Q$ is the heavy-quark (HQ) mass. Current estimates of the typical scale of the thermal relaxation time, 
$\tau_Q=1/\gamma_Q$, indicate that it is comparable to the lifetime of the QGP fireball of $\tau_{\rm QGP}\simeq5-10$\,fm/$c$, 
presumably somewhat smaller for charm, but larger for bottom quarks. 
Thus, the finally observed spectra of heavy-flavor hadrons containing a $c$ or $b$ quark, carry a {\em memory} of 
their interaction history, rendering them quantitative gauges of their interaction strength with the 
medium~\cite{Prino:2016cni,Dong:2019unq,Dong:2019byy}. 
Furthermore, the collective flow of the expanding medium (including azimuthal-angular modulations, most notably a so-called ``elliptic flow'') exerts 
an additional drag on the diffusing quarks, which can lead to the development of a characteristic maximum structure in the heavy-quark (HQ) spectrum 
reflecting the radial-flow velocity of the fireball. 
       
Pioneering data of the PHENIX collaboration, which indirectly measured heavy-flavor (HF) spectra in Au-Au collisions through their semi-leptonic decay 
electrons, showed large modifications relative to expectations from proton-proton collisions~\cite{PHENIX:2006iih}. It soon became clear that these 
modifications cannot be reconciled with perturbative-QCD (pQCD) calculations, but require nonperturbative heavy-quark (HQ) interactions in the QGP 
and/or in the hadronization process.
To date these findings have been thoroughly corroborated, quantified and expanded through a wealth of high-quality experimental data, especially for 
charm hadrons ($D$, $D_s$ and $\Lambda_c$ ) from the Relativistic Heavy-Ion Collider (RHIC) and the Large Hadron Collider (LHC). Recent accounts 
of these developments, together with the main theoretical and phenomenological developments, have been given in several recent review 
articles~\cite{Prino:2016cni,Dong:2019unq,Dong:2019byy}. Specifically, the description of $D$-meson observables in Pb-Pb(5\,TeV) and Au-Au(0.2\,TeV) 
collisions has put substantial constraints on the (temperature-scaled) charm-quark diffusion coefficient  in the QGP, 
to values of ${\cal D}_s (2\pi T)$$\simeq$\,2-4 
in the vicinity of the transition temperature. This is about a factor of 10 smaller than leading-order pQCD calculations~\cite{Svetitsky:1987gq} and not 
far from the conjectured quantum lower bound of one~\cite{Son:2007vk}. One of the main objectives we aim to discuss in this review is the 
progress that has been made in 
understanding the microscopic interactions that are responsible for the remarkable coupling strength in the QGP. Also in this context, the large HQ mass 
provides valuable benefits. In a typical elastic scattering process of a heavy quark on light thermal partons, the energy transfer, $q_0\simeq q^2/m_Q$, 
is parametrically suppressed relative to the three-momentum transfer, $q$, thus justifying the notion of potential interactions which facilitate the 
theoretical description in terms of three-dimensional scattering equations. In addition, high quality lattice-QCD (lQCD) results are available for static 
quantities such as the HQ free energy, which is closely related to an in-medium HQ potential.     

\begin{figure}[t]
\centering
\hspace{0.8cm}
 \begin{minipage}{0.4\textwidth}
 \includegraphics[width=0.8\textwidth]{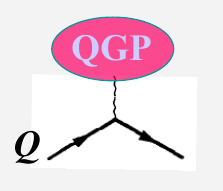}
\end{minipage} 
\hspace{-0.5cm}
\begin{minipage}{0.15\textwidth}
\hbox{ {\Huge $\Rightarrow$} \hspace{0.5cm} }
\end{minipage}
\begin{minipage}{0.4\textwidth}
\includegraphics[width=0.76\textwidth]{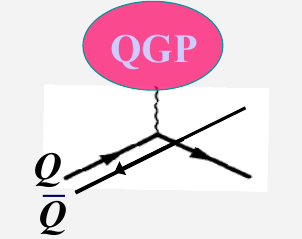}
\end{minipage}
\caption{Schematic representation of how the interactions of an individual heavy quark (left panel) are related to dissociation processes of quarkonia 
(right panel) in the QGP.}
\label{fig_HQscatt}
\end{figure}
Elastic scattering of a heavy quark in the QGP is closely related to the dissociation reactions of heavy quarkonia, ${\cal Q}$ where the same scattering 
occurs except that the incoming heavy quark is off-shell due to its binding in a bound state, cf. Fig.~\ref{fig_HQscatt}. The process has been referred to
as  ``quasi-free" dissociation~\cite{Grandchamp:2001pf} and becomes the dominant process (as compared to, \eg, gluo-absorption,  
$g+{\cal Q}\to Q + \bar{Q}$~\cite{Bhanot:1979vb,Kharzeev:1994pz}, if the binding energy of the quarkonium is of order of the temperature or less. 
Even for larger binding energies, where interference effects between the scattering off the quark and the antiquark become
relevant~\cite{Laine:2006ns,Park:2007zza}, quasi-free dissociation can still be numerically large~\cite{Du:2017qkv}. 
In quantum many-body language, dissociation processes of quarkonia are encoded in the selfenergies of the individual heavy quarks in the bound state.
Therefore, a microscopic description of HQ diffusion directly manifests itself in the in-medium quarkonium properties, and as such can be further
tested against lQCD data including euclidean and spatial correlation functions, as well as experimental observables. Also in this sector, rather extensive 
recent reviews are available~\cite{Mocsy:2013syh,Rothkopf:2019ipj}, and therefore we will focus here on the connections to the open HF sector. 

Our article is organized as follows. In Chap.~\ref{sec_int} we discuss microscopic models for HQ interactions in the QGP, including pQCD calculations
(Sec.~\ref{ssec_pQCD}) and a thermodynamic $T$-matrix approach (Sec.~\ref{ssec_tmat}), constraints from lQCD and pertinent HQ and quarkonium 
spectral functions (Sec.~\ref{ssec_lat}), consequences for and the role of the ambient QCD medium (Sec.~\ref{ssec_med}), and diffusion in hadronic 
matter (Sec.~\ref{ssec_had}).  Chapter~\ref{sec_trans}  is dedicated to transport parameters, specifically the friction and momentum diffusion 
coefficients  of heavy quarks (Sec.~\ref{ssec_Ap}), their implications for quarkonium dissociation  (Sec.~\ref{ssec_Gam}), and the HF spatial diffusion
coefficient  and relations to the shear viscosity  (Sec.~\ref{ssec_Ds}).  In Chap.~\ref{sec_open} we review key features 
of open HF transport approaches in heavy-ion collisions  (Sec.~\ref{ssec_trans-app})
and give a critical discussion of the current phenomenology at RHIC and the LHC (Sec.~\ref{ssec_hf-obs}). 
In Chap.~\ref{sec_onium}, we briefly introduce pertinent transport approaches as well  (Sec.~\ref{ssec_QQ-kin})  and then review recent 
developments in the interpretation of data for charmonia, bottomonia and ``exotic'' states with an emphasis on connections to the open HF
sector (Sec.~\ref{ssec_QQ-obs}). A brief summary and future perspectives are given in Chap.~\ref{sec_sum}.

	\newpage
	\section{Heavy-Flavor Interactions in QCD Matter}
    \label{sec_int}
The basic building block in a quantum many-body description of HQ systems in the QGP is the two-body scattering amplitude of a heavy quark on the 
constituents of the thermal bath, for both on-shell kinematics (\eg, in the semi-classical transport treatment), or off-shell when embedded into bound 
states. We will discuss these amplitudes from the pQCD perspective in Sec.~\ref{ssec_pQCD} and in nonperturbative approaches with focus on the 
thermodynamic $T$-matrix approach in Sec.~\ref{ssec_tmat} . Since we expect very high collision rates in a strongly coupled quark-gluon plasma 
(sQGP), with transport properties close 
to pertinent quantum lower bounds, the spectral functions of the thermal-medium partons will acquire large collisional widths, comparable to their masses.
This suggests that off-shell treatments are also needed in the calculations of in-medium selfenergies and transport coefficients.  
Even for charm quarks, the widths may not be negligible relative to their mass, while for quarkonium properties the relevant scale is the in-medium 
binding energy which may quickly be exceeded by the collisional width as temperature increases, leading to a melting of the bound state. In a 
nonperturbative system, it is therefore essential to constrain the calculated spectral functions with information from lQCD as much as possible. 
This  will be addressed  in Sec.~\ref{ssec_lat}, followed in Sec.~\ref{ssec_med} by a discussion of the QCD medium properties that may emerge 
and have to be fed back into the HQ interactions.  We end this chapter with a brief survey on the investigation of HF diffusion in hadronic matter in 
Sec.~\ref{ssec_had}.

	\subsection{Perturbative QCD}
    \label{ssec_pQCD}


The basic entity for evaluating HQ interactions in the QGP is the two-body amplitude for elastic scattering off the thermal partons in the heat bath. To 
leading order in the strong coupling constant, $\alpha_s $, the pertinent Feynman diagrams encompass $s$-, $t$- and $u$-channels for HQ-gluon 
scattering, as well 
as a $t$-channel for HQ scattering off up, down and strange quarks ($q=u,d$ and $s$, respectively). Since the $s$ and $u$-channel amplitudes are 
suppressed by the HQ mass in the intermediate HQ propagator, the dominant contributions arise from the  two $t$-channel diagrams, specifically 
HQ-gluon scattering due to the larger color-charge factor the three-gluon vertex. In vacuum, both diagrams exhibit an infrared singularity for 
forward scattering, which is, however, naturally regularized through the appearance of a Debye screening mass, $\mu_D \sim gT$, in the $t$-channel 
gluon exchange propagator. 
When utilizing these diagrams to compute the thermal relaxation time of charm quarks in the 
QGP~\cite{Svetitsky:1987gq,Braaten:1991jj,GolamMustafa:1997id,Riek:2010fk}, 
one finds rather long thermalization times, $\tau_c\ge 20$fm/$c$, for temperatures $T\le 300$\,MeV and a strong coupling constant of $\alpha_s$=0.3-0.4. 
The results also depend on the precise value of the charm-quark mass in the medium, since in leading order one expects $\tau_c$ to be proportional to 
$m_c$. Converting the relaxation time into a spatial diffusion coefficient,  ${\cal D}_s = \tau_c (m_Q/T)$, and normalizing it to the thermal wavelength, 
$1/2\pi T$, one finds values of  ${\cal D}_s (2\pi T) \simeq35$.
\begin{figure}[t]
\centering
\begin{minipage}{0.2\textwidth}
 \includegraphics[width=1.2\textwidth]{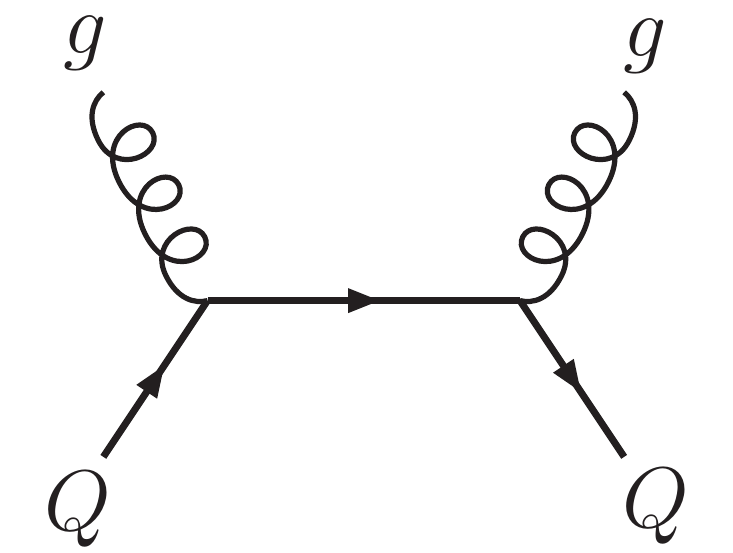}
\end{minipage}
\hspace{1cm}
 \begin{minipage}{0.2\textwidth}
 \includegraphics[width=0.9\textwidth]{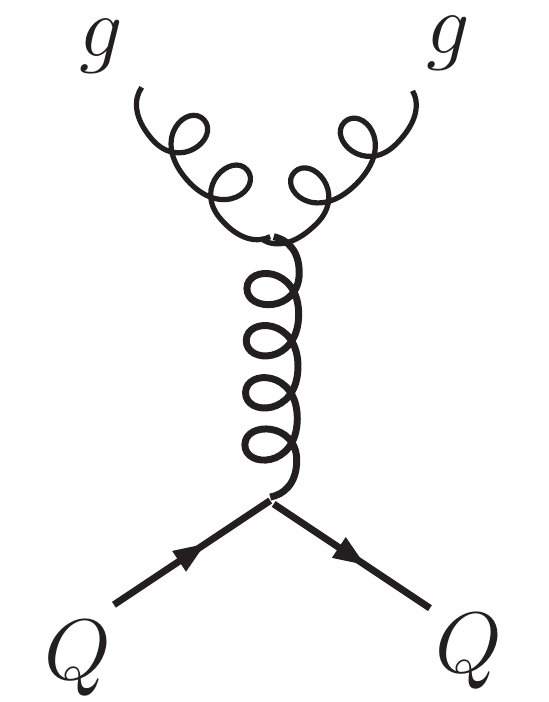}
\end{minipage}
\hspace{0.0cm}
\begin{minipage}{0.2\textwidth}
\includegraphics[width=1.2\textwidth]{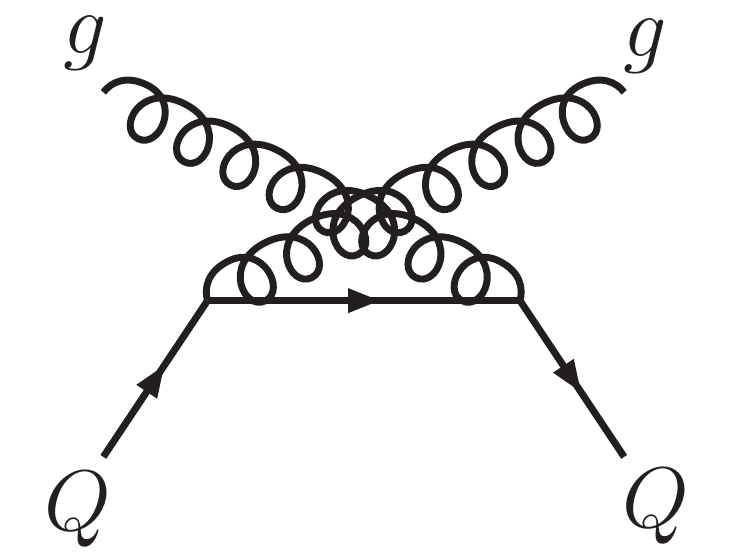}
\end{minipage}
\hspace{1cm}
 \begin{minipage}{0.2\textwidth}
\includegraphics[width=0.9\textwidth]{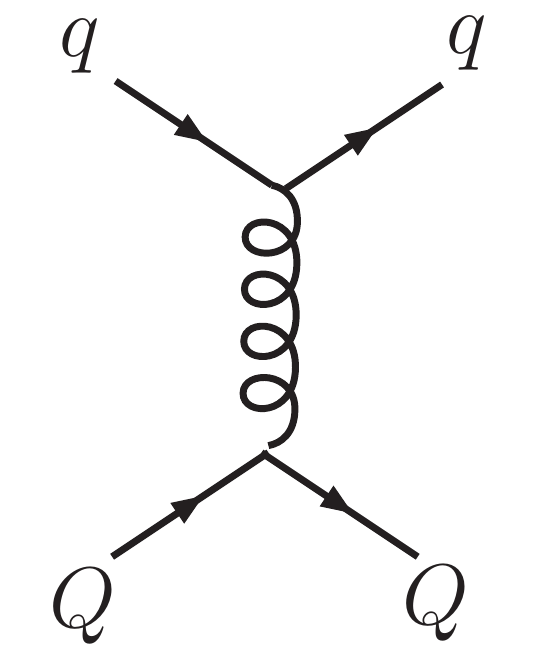}
\end{minipage} 
\caption{Tree-level diagrams for HQ scattering off gluons in the $s$-, $t$- and $u$-channel (from left to right), and off light quarks where only the $t$-channel is available.}
\label{fig_HQpQCD}
\end{figure}

A complete next-to-leading order (NLO) calculation has been carried out in Ref.~\cite{Caron-Huot:2007rwy}, requiring various types of resummations 
including corrections from very soft HQ momenta, $p\simeq \mu_B$, that lead to a non-analytic behavior in the strong coupling constant. The final
result shows that NLO corrections to the LO result are very large, even at relatively small coupling, \eg, more than a factor of 5 for $\alpha_s$=0.2. This
implies that the perturbative series is not under control and cannot be used to assess meaningful corrections to the leading-order result.

Other corrections to the leading-order Born diagrams have been investigated. In Refs.~\cite{Moore:2004tg,Alberico:2011zy} the energy-momentum
dependent polarization tensors have been inserted into the $t$-channel gluon propagators in hard-thermal-loop approximation. A sizable increase in the
interaction strength has been found, which can decrease the HQ diffusion coefficient by a factor of $\sim$2-3 or more , (see, \eg, Fig.\,4 in 
Ref.~\cite{Prino:2016cni}) and also affects the momentum dependence  of the relaxation rate, relative to the LO results. In 
Refs.~\cite{Gossiaux:2008jv,Peshier:2008bg} the running of the QCD coupling constant has been introduced not only as a function of temperature, but also
as a function of the momentum transfer, $Q^2$ of the exchanged gluon, reaching values of close to 1 in the limit of $Q^2\to0$. In addition, it has been 
argued that  a selfconsistent determination of the Debye mass, $\tilde\mu_D^2 \sim \alpha_s(\tilde\mu_D^2) T^2$, leads to much reduced values of 
about a factor  $\sim$2.5 relative to the usual $\mu_D \sim gT$ with a fixed coupling. Put together, these effects lead to a reduction the spatial 
diffusion coefficient down to values of ${\cal D}_s (2\pi T) \simeq$\,3-4. The interested reader may find more detailed discussions on these 
approaches in previous review articles~\cite{Rapp:2009my,Prino:2016cni}.

	\subsection{Thermodynamic $T$-matrix}
    \label{ssec_tmat}


The problems or perturbative calculations to produce an interaction strength of heavy quarks in the QGP needed to account for the experimentally observed 
spectra of heavy-flavor particles in heavy-ion collisions has led to several developments of nonperturbative approaches. In the following we briefly review
some of the older ones and then turn our focus on the thermodynamic $T$-matrix approach that has a number of appealing features and is still actively 
pursued.

The first nonperturbative approach to HQ diffusion in the QGP was based on the conjecture that $D$-meson resonances survive at temperatures above
the critical one~\cite{vanHees:2004gq}. The presence of hadronic states in the QGP is not uncommon in the light-flavor 
sector~\cite{Blaschke:2002ww,Shuryak:2003ty} and has also been put forward early on in the charmonium sector as an interpretation of lQCD data for 
euclidean correlation functions~\cite{Asakawa:2003re}.
In the QGP, $D$-mesons have been utilized as resonant states in HQ scattering off thermal anti/quarks~\cite{vanHees:2004gq}, which was found to
reduce the LO pQCD estimates of the HQ diffusion coefficient by about a factor of five, down to  ${\cal D}_s (2\pi T) \simeq$\,5-6 for charm quarks. 
In addition, this model provides a natural pathway for color neutralization as the system transits from quark to hadronic degrees of freedom.
The HQ diffusion coefficient has also been computed in the strong-coupling limit for a scale-invariant gauge theory utilizing the conjectured equivalence 
between a weakly coupled gravitational and conformal field theory (AdS/CFT 
correspondence)~\cite{Policastro:2002se,Casalderrey-Solana:2006fio,Gubser:2006qh}.  The resulting HQ diffusion coefficient, 
${\cal D}_s = 2/\pi T \sqrt{\lambda}$ still has a weak dependence on the 't Hooft coupling constant, $\lambda=g^2 N_c$, as opposed to pQCD where
${\cal D}_s \propto 1/\alpha_s^2 T$. This renders the HQ diffusion coefficient parametrically smaller compared to the AdS/CFT result for the 
specific shear viscosity, $\eta/s \simeq 1/4\pi$. A mapping of the AdS/CFT result to the QGP depends on the identification of the equivalent temperature 
and coupling constant in the QGP, and produces a range of ${\cal D}_s (2\pi T) \simeq$\,0.9-3~\cite{Gubser:2006qh} (which is implicitly assuming that 
the QCD plasma is driven by the color-Coulomb force). 

\begin{figure}[t]
\centering
\begin{minipage}{0.46\textwidth}
\includegraphics[width=0.99\textwidth]{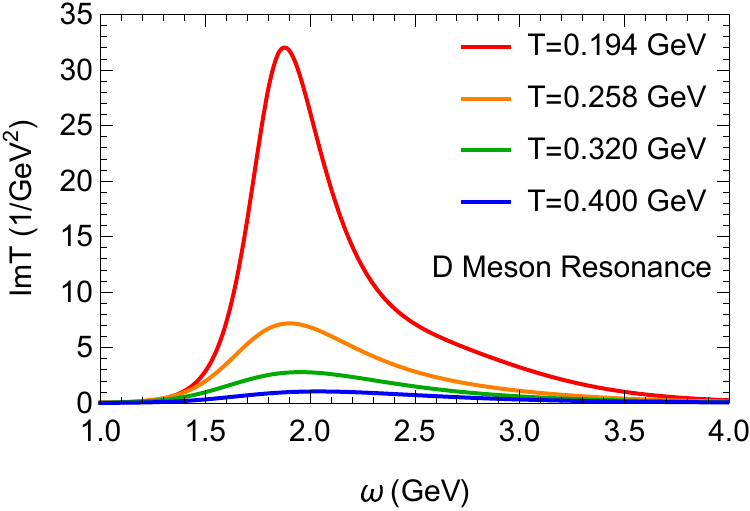} 
\end{minipage}
\hspace{0.3cm}
\begin{minipage}{0.46\textwidth}
\vspace{-0.1cm}
\includegraphics[width=0.95\textwidth]{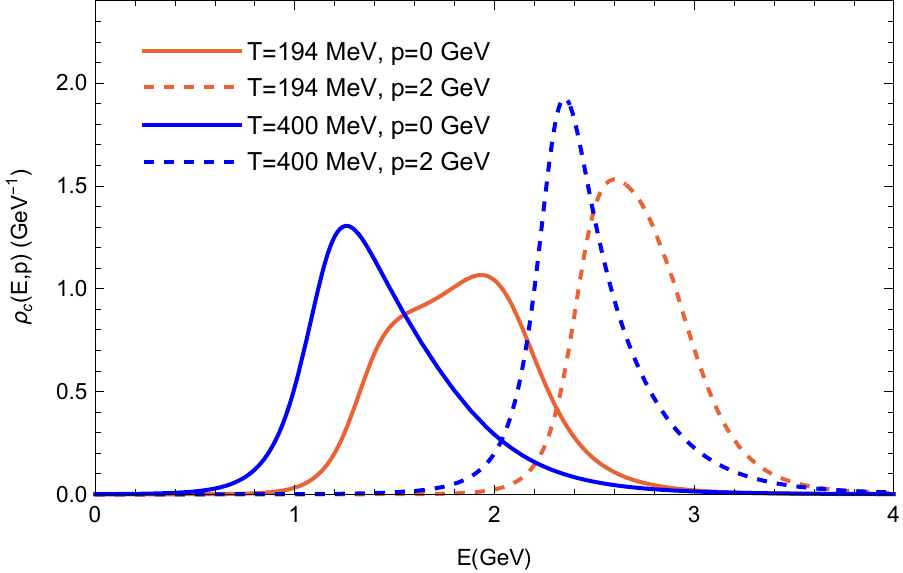}
\end{minipage}
\caption{Left panel: Imaginary part of the scattering amplitude for charm-quark scattering on light antiquarks in the QGP in the color-singlet $S$-wave channel, 
for four different temperatures. Right panel: Charm-quark spectral functions at $T$=194 and 400\,MeV (red and blue lines, respectively) for three-momenta
of 0 and 2\,GeV (solid and dashed lines, respectively).}
\label{fig_c-spec}
\end{figure}

Next, we turn to the thermodynamic $T$-matrix approach~\cite{Mannarelli:2005pz,Riek:2010fk,Liu:2017qah}, which is a rigorous quantum 
many-body framework for a selfconsistent solution of the one- and two-body correlation functions in medium.  Starting point is the scattering equation 
for the interactions of  a heavy quark with thermal partons ($i$=$q, \bar q, g$) in the QGP, 
\beq
T_{Qi}^a(E,{\bf p},{\bf p}')= V_{Qi}^a  + \int \frac{d^3k}{(2\pi)^3} V_{Qi}({\bf p},{\bf k})  G_{Qi}^0(E,{\bf k}) T_{Qi}^a(E,{\bf k},{\bf p}')   \  , 
\eeq 
where  ${\bf p}, {\bf p}'$ and ${\bf k}$ denote the three-momenta of the heavy quark in the incoming, outgoing and intermediate state, respectively 
(and equal opposite for the light parton). The 3D $T$-matrix equation can be obtained from the 4D Bethe-Salpeter equation in the limit of small
energy transfer, $q_0$, which is satisfied if the heavy quark is sufficiently massive, \ie, $q_0 \sim {\bf q}^2/m_Q \ll  q \sim T$. The driving
kernel, $V_{Qi}$, which is the main input to the approach, then becomes amenable to a potential approximation which opens the way for quantitative 
constraints from lQCD (this will be elaborated in Sec.~\ref{ssec_lat}). However, the dynamic (quantum) information of the propagating quarks, encoded 
in the finite-temperature two-body propagator,
\beq
G_{iQ}^0(E,{\bf k}) = \int d\omega_1 d\omega_2  \frac{\rho_i(\omega_1,{\bf k}) \rho_Q(\omega_2,{\bf k})} {E - \omega_1 - \omega_2 +i\epsilon}
[1 \pm f_i(\omega_1) - f_Q(\omega_2)]  \  , 
\label{G2}
\eeq
is retained through the in-medium parton spectral functions, $\rho_{i,Q}(\omega)$ and pertinent thermal distribution functions, $f_{i,Q}$ (with the upper 
and lower sign for bosons or fermions, respectively). The single-particle spectral function is given by the imaginary part of the pertinent propagator, 
\beq
G_Q = 1 / [\omega - \omega_Q(k) - \Sigma_Q(\omega,k)]   \ , \quad \rho_Q = -\frac{1}{\pi} {\rm Im}~G_Q  \  
\eeq 
(where $\omega_Q(k) =\sqrt{m_Q^2 + k^2}$ denotes the on-shell energy). The medium effects are encoded in the complex selfenergy, which, in turn,
is computed from the $T$-matrix summed over all thermal partons ($i$),  
\beq
\Sigma_Q(\omega,k) = \int \frac{d^3p_2}{(2\pi)^3} \int d\omega_2 \sum_i \rho_i(\omega_2,p_2)  \int  \frac{dE}{\pi} 
\frac{{\rm Im} {T}_{Qi}^{al}(E,{\bf k},\bf{p_2})}{E-\omega-\omega_2+i\epsilon} [f_i(\omega_2) \mp f(E)]   \  
\eeq
including all possible color ($a$), flavor and spin channels, as well as a partial-wave expansion in angular momentum ($l$), where 
$l$$\ltsim$5 is safely sufficient for convergence. 
A suitable ansatz for the two-body interaction is an in-medium screened Cornell potential,  
\beq
V_a(r) = C_a \frac{\alpha_s}{r}  {\rm e}^{-\mu_D T} + \frac{\sigma}{\mu_s} (1-{\rm e}^{-\mu_s T})  \ , 
\label{Va}
\eeq
where $C_a$ denotes a color-factor (\eg, $C_a$=4/3 in the color-singlet quark-antiquark channel), $\sigma$ the string tension, and 
$\mu_D$ and $\mu_s$ the screening masses of the color-Coulomb and string interaction. This potential recovers the well-established quarkonium 
spectroscopy in the vacuum (also for heavy-light mesons).
Quantitative determinations of the in-medium screening parameters based on lQCD ``data" will be discussed in the following section.  After subtracting the 
infinite-distance part, $\bar{V}_a(r) \equiv V_a(r) - V_a(\infty)$, this potential can be straightforwardly Fourier-transformed and partial-wave expanded for 
use in a one-dimensional  $T$-matrix equation (which can solved numerically in each $al$ channel through a matrix inversion of the momentum-discretized 
integral). To recover the correct perturbative behavior at short distance (or high parton momenta), one needs to account for relativistic corrections to the 
color-Coulomb interaction~\cite{Brown:2003km,Riek:2010fk}, which essentially correspond to color-magnetic interactions arising from the vector-current 
coupling of one-gluon exchange and ensure that the potential is Ponicar\'e invariant. For simplicity, the string interaction has been assumed to be a 
Lorentz scalar (there are, however, arguments that this characterization might not be complete~\cite{Szczepaniak:1996tk,Brambilla:1997kz,Ebert:2002pp}). 
From the $T$ matrices one computes the in-medium selfenergies and vice versa, which can be rendered selfconsistent through numerical iteration.
In addition to the selfenergies, the HQ masses receive a contribution from the (real) Fock term of the Coulomb and string potentials,  
$\Delta m_Q = \frac{1}{2}(-\frac{4}{3}\alpha_s+\sigma/\mu_s)$.
Examples of the resulting charm-light quark scattering amplitude and charm-quark spectral functions are shown in Fig.~\ref{fig_c-spec}. As the 
temperature is reduced, a broad $D$-meson resonance develops in the color-singlet $c\bar q$ channel  (strictly speaking, they are bound states, as 
their peak mass is below the ``nominal" heavy-light mass threshold). Its mass is close to the vacuum $D$-meson mass (in the absence of hyperfine 
interactions, the spin-0 and spin-1 states are degenerate), a consequence of the vacuum benchmark of the potential, mostly driven by the 
confining part of the interaction. In this way ``hadronization" of the QGP at low temperatures is not a separate phenomenon, but emerges dynamically  
from the quantum many-body theory with an in-medium Cornell potential. 
The pertinent $c$-quark spectral functions carry rather large collisional widths, of about 500-700\,MeV, which slightly {\em increase} with decreasing 
temperature, \ie, the reduction in degrees of freedom in the heat bath (which is about an order of magnitude from $T$=400\,MeV to 194\,MeV), is 
overcompensated by the increasing interaction strength provided by the resonances (which also form in the color antitriplet (diquark) channel, as
a building block of charm baryons).  The $c$-quark  widths
decrease with increasing three-momentum, indicating a gradual transition to a more weakly coupled regime. On the other hand, at $T$=194\,MeV and 
vanishing three-momentum, the $c$-quark spectral function develops a low-energy shoulder, which is a consequence of low-energy (off-mass shell) 
$c$-quark scattering with thermal antiquarks into a $D$-meson resonance.

	\subsection{Constraints from Lattice QCD}
    \label{ssec_lat}


A pivotal ingredient in constructing reliable nonperturbative interactions are constraints from lQCD computations. The latter are difficult to perform
for timelike quantities (such as spectral functions), but they provide precision ``data" on spacelike ones. These, in turn, can be readily calculated from model results for spectral functions. In the context of in-medium HQ physics, the most prominent quantities are the free energy of a static $\QQb$ pair, 
$F_{\QQb}(r;T)$ as a function of the relative distance, $r$ , and euclidean-time correlation functions of quarkonia, as well as 
quark-number susceptibilities including charm.

The HQ free energy computed in lQCD is defined as the difference in the free energies of the QCD heat bath with and without a static $\QQb$ pair. 
In the vacuum, this coincides with the HQ potential, $V(r)$, but at finite temperature an extra entropy term appears, 
$F_{\QQb}(r,T) = U_{\QQb}(r,T) - T S_{\QQb}(r,T)$, representing the competition between minimizing the internal energy and maximizing the 
entropy (caused by interactions with the heat bath). The question then arises what is the appropriate potential to use at finite temperature, see, \eg, 
Refs.~\cite{Shuryak:2003ty,Wong:2006bx}. The answer will likely depend on the context of its definition (as it is not a direct 
observable)~\cite{Rothkopf:2019ipj}.  A pertinent relation can be derived within the $T$-matrix formalism~\cite{Liu:2015ypa,Liu:2017qah}. One starts 
from the definition of the free energy in terms of the $\QQb$ correlation function,  
\beq
F_{\QQb}(r'T) = -T \ln \left( G^>_{\QQb} (-i\beta,r)\right) \ 
\eeq
(where $\beta \equiv 1/T$). Utilizing the (static) uncorrelated two-body propagator from Eq.~(\ref{G2}), 
$[G^0_{\QQb}(E)]^{-1} =  E-2\Delta m_Q-\Sigma^0_{\QQb}(E)$, to define the uncorrelated $\QQb$ selfenergy, $\Sigma^0_{\QQb}(E)$,  
one can obtain the correlated spectral function in energy coordinate representation as
\beq
G_{\QQb}(E,r) = \frac{1}{E-2\Delta m_Q - \bar{V}(r) -\Sigma_{\QQb}(E,r)}  \ , 
\label{GQQ}
\eeq
where the correlated two-body selfenergy additionally contains interference effects. The latter cause a suppression of the imaginary part relative to the 
absorptive effetcs in the uncorrelated part~\cite{Laine:2006ns}. In a diagrammatic description, pertinent processes correspond to three-body diagrams 
which are not easily implemented. Instead, in Ref.~\cite{Liu:2017qah} these effects were approximated in a factorized form with an interference 
function, $\phi(x_er)\leq 1$, motivated by the perturbative calculations in Ref.~\cite{Laine:2006ns} but with an extra scaling factor $x_e$ to allow for 
nonpertubative effects,  $\Sigma_{\QQb}(E,r) = \Sigma^0_{\QQb}(E) \phi(x_e r)$.
The final expression relating the free energy with the potential defined in the $T$-matrix formalism takes the form
\beq
F_{\QQb}(r;T) = -T \ln \left[ \int\limits_{-\infty}^{\infty} \frac{dE}{\pi} {\rm e}^{-\beta E} 
{\rm Im}\left(\frac{-1}{E-\bar{V}(r;T)-2\Delta m_Q -\Sigma_{\QQb}(E,r;T)} \right)\right]   \ .
\eeq 
This expression has several interesting features~\cite{Liu:2015ypa}. For example, in the weakly coupled limit where the selfenergy vanishes, the 
imaginary part approaches
a $\delta$-function, $-\pi\delta(E-\bar{V}-2\Delta m_Q)$. This enables to carry out the energy integration to obtain the result $F_{\QQb}(r;T) = V(r;T)$ 
(including the finite infinite-distance value), \ie, for a weakly coupled system the potential is close to the free energy. On the other hand, for large 
imaginary parts in the $\QQb$ selfenergy, the spectral function is smeared out which generally requires a stronger potential to match the resulting free 
energy. In turn, a stronger potential will increase the scattering rates in the system, requiring a selfconsistent solution of the problem. 
\begin{figure}[t]
\centering
 \includegraphics[width=0.8\textwidth]{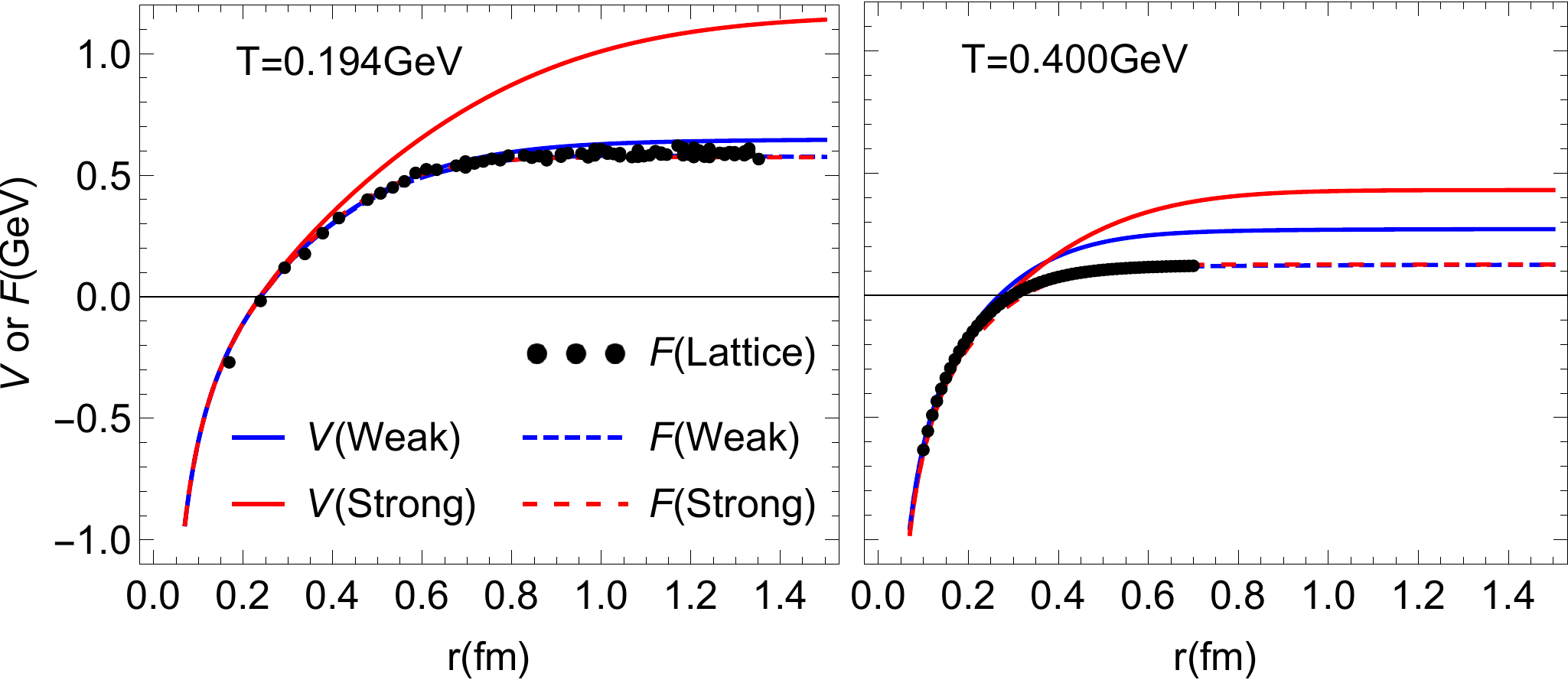} 
\caption{Fits of the color-singlet HQ free energy within the selfconsistent $T$-matrix formalism (dashed lines) to $N_f$=2+1-flavor lQCD 
``data" (black dots)~\cite{Bazavov:2012fk,Mocsy:2013syh}, 
and the underlying two-body potential (solid lines) for a strongly-coupled scenario (SCS, red lines) and a weakly-coupled scenario (WCS, blue lines) 
at $T$=194\,MeV (left panel) and $T$=400\,MeV (right panel). Figure taken from Ref.~\cite{Liu:2016ysz}.}
\label{fig_FvsV}
\end{figure}
This program has been carried out in Ref.~\cite{Liu:2017qah}, cf.~Fig.~\ref{fig_FvsV}. The red lines are the selfconsistent results corresponding the 
HQ spectral functions and $T$-matrices shown in Fig.~\ref{fig_c-spec}, carried to the static limit, \ie, for $m_Q\to \infty$  (in practice, a large value 
of $m_Q$$\simeq$20\,GeV is adopted). The large imaginary part of the HQ selfenergy generates a large broadening of the static quarkonium spectral 
function, and, as a consequence, the underlying potential is driven above the free energy. The effect is most pronounced at low temperatures (left panel 
of Fig.~\ref{fig_FvsV}), where the resulting in-medium potential is rather close to the vacuum Cornell potential, while for higher temperatures, the
Debye screening is much stronger and the gap to the free energy reduced. However, this solution, referred to as a strongly-coupled scenario (SCS), is
not unique. Another selfconsistent solution has been found whose potential is much closer to the free energy and the underlying HQ widths are a factor of 
$\sim$5 smaller at low temperatures, while at higher temperatures the differences are less pronounced but still appreciable. This solution has been
referred to as a weakly-coupled scenario (WCS). 

Another quantity that has been widely used to constrain in-medium properties of quarkonia are their euclidean-time ($\tau$) correlators,  
defined as
\beq
G_\alpha(\tau,P;T)=\int d^3r {\rm e}^{i{\bf P}\cdot {\bf r} } \langle j_\alpha(\tau, {\bf r},j_\alpha(0,0)\rangle 
\label{Galpha}
\eeq
where $j_M= \bar\psi \Gamma_\alpha \psi$ denotes the mesonic current in channel $M$ characterized by its Dirac structure, 
$\Gamma_\alpha$.
Focusing on vanishing total momentum of the $\QQb$ pair, ${\bf P}=0$, the finite-temperature Fourier transform of the euclidean correlator
is related to its spectral function, $\rho_\alpha$, as
\beq
G_\alpha(\tau;T) =\int\limits_0^\infty dE \rho_\alpha(E,T) {\cal K}(\tau,E;T)  \ 
\eeq
with a temperature kernel ${\cal K}(\tau,E;T) = \cosh[E (\tau-1/2T)] / \sinh[E/2T]$.
Since in thermal field theory, $\tau$ is limited to the interval $[0,1/T]$, and due to an extra symmetry about the midpoint, $\tau=1/2T$, this 
expression exhibits the well-known problem of inferring a spectral function (especially with non-trivial structures such as bound states or mass 
thresholds) from a limited $\tau$ range of generally rather smooth correlation functions on a discrete set of numerical points obtained from lQCD 
computations. 
Since the euclidean correlators are typically exponentially falling functions, one defines a ratio,
\beq
R_G^\alpha  = \frac{G_\alpha(\tau;T)}{G^{\rm rec}_\alpha(\tau;T)} \ ,
\eeq
relative to a "reconstructed" correlator, $G^{\rm rec}_\alpha(\tau;T)$,
 which is computed with the same temperature kernel but with a spectral function at a reference temperature 
$T_{\rm ref}$ typically chosen to be (much) smaller than $T$ so that its medium modifications are suppressed. 
The euclidean correlator ratios (ECRs) thus quantify the medium effects as a deviation from one, and augment the sensitivity to the 
large-$\tau$ behavior of the correlation function
associated with the low-energy properties of the spectral functions.     
A broad range of in-medium models for quarkonium  spectral functions (often based on internal- or free-energy potential proxies, and varying degrees of 
absorptive parts) have been explored to investigate the ECRs computed in lQCD, but the conclusions have remained ambiguous on whether quarkonium 
bound states in the QGP can be supported (for charmonia), and if so (for bottomonia), up to which temperatures (see, \eg, the 
reviews~\cite{Rapp:2008tf,Kluberg:2009wc,Mocsy:2013syh,Rothkopf:2019ipj} and references therein). Part of the reason for that is that the ``point-to-point" 
correlation functions involve the creation and annihilation of the $\QQb$ current at a single point, \ie, at vanishing distance of the $\QQb$ pair, which 
somewhat limits the information on the long-distance behavior where the physics of the bound state prevails. Thus, in recent years, the development
of lQCD computations of extended operators is being pursued, which should provide a better sensitivity in this 
regard~\cite{Larsen:2019bwy,Larsen:2020rjk}. 
  
\begin{figure}[t]
\centering
 \includegraphics[width=0.99\textwidth]{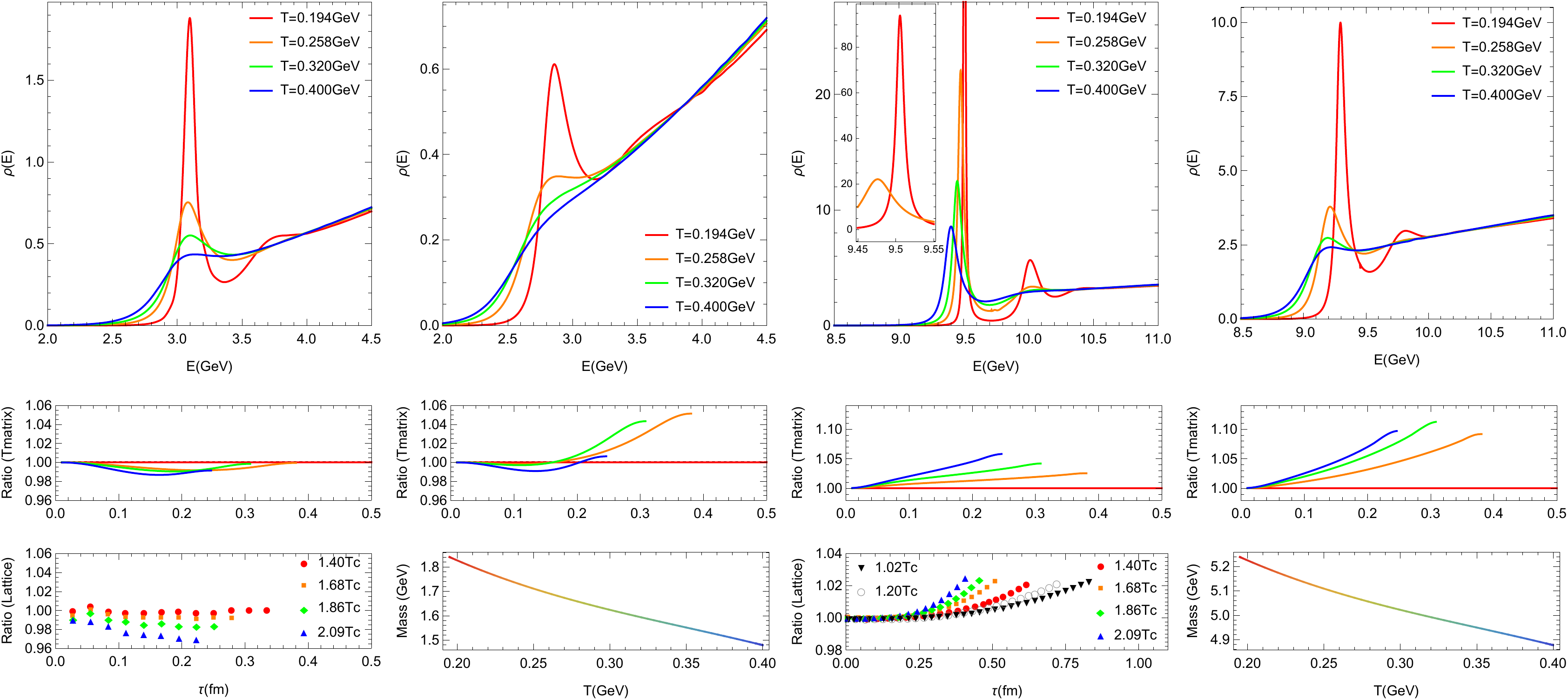} 
\caption{Charmonium and bottomonium spectral functions (top row) and pertinent euclidean-correlator ratios (middle row, with the ``reconstructed" 
one at $T$=194\,MeV) from the thermodynamic $T$-matrix approach using the SCS input potential (as shown by the red lines in Fig.~\ref{fig_FvsV}).
The first (second) and third (fourth) column are calculated with (without) interference effects in the imaginary part of the potential. The bottom row shows 
the ECR results from $N_f$=2-lQCD for charmonia (first panel) and bottomonia (third panel), and the temperature dependence of the charm- (second panel) 
and bottom-quark (fourth panel) masses. Figure taken from Ref.~\cite{Liu:2017qah}.}
\label{fig_corr}
\end{figure}

In Fig.~\ref{fig_corr} we display the selfconsistent $T$-matrix result for charmonium and bottomonium $S$-wave spectral functions and ECRs 
computed from the SCS potential in Fig.~\ref{fig_FvsV} right, in comparison to two-flavor lQCD data~\cite{Aarts:2007pk,Aarts:2011sm} for 
$\eta_c$ and $\eta_b$ mesons (in the pseudoscalar channels, the so-called zero-mode contribution generated by a transport peak in the
low-energy limit of the spectral function is suppressed). 
Both charmonium and bottomonium results are shown with (first and third column) and  without (second and fourth column) 
interference effects in the imaginary part (encoded in the ``interference function", $\phi(x_er)$,  recall the discussion following Eq.~(\ref{GQQ})). 
As to be expected, the bound-state solutions are generally more robust (and narrower) if the imaginary part is suppressed at small distance, and this 
is more pronounced for the more compact $\eta_b$ states. As a consequence, a well-defined $\eta_b(1S)$ ground-state peak persists for temperatures 
well beyond 400\,MeV, while the ground-state charmonium ($\eta_c$ or $J/\psi$) dissolves at around $T$$\sim$300\,MeV, and the $\eta_b(2S)$ around 
$T$$\sim$250\,MeV. It tuns out that the rather small variations of a few percent in the lQCD ECRs are slightly better described with the interference 
effects included, relative to the more rapid melting without interference effects. Also shown are the in-medium charm and bottom-quark masses, 
whose temperature dependence is driven by the infinite-distance limit of the potential and which increase with decreasing temperature, reaching 
values of about $m_c$=1.85\,GeV and $m_b$=5.25\,GeV at $T$=194\,MeV, which essentially agree with the values required by their vacuum 
masses (together with the vacuum potential). Indeed, in the SCS the potential at $T$=194\,MeV is still close to the vacuum one. However, the 
$\eta_c(2S)$ (or $\psi(2S)$) state cannot survive under these conditions since the in-medium quark widths are much larger than the $\eta_c(2S)$ 
binding energy; the same argument applies to the $\eta_b(3S)$ (or $\Upsilon(3S)$), while the $\eta_b(2S)$ (or $\Upsilon(2S)$) can still survive to 
temperatures slightly above 200\,MeV if interference effects are accounted for. 

In WCS (not shown here), the $\eta_c(1S)$ (or $J/\psi$) melts at a lower temperature (ca.~260\,MeV), while the much more compact $\eta_b(1S)$ 
can still persist to temperatures above 400\,MeV, provided the interference effects are included. The smaller interaction strength goes along
with a smaller infinite-distance value of the potential, and consequently the in-medium HQ masses are significantly smaller than in the SCS, by
about 0.25(0.1)\,GeV at $T$=194(400)\,MeV. Nevertheless, due to the reduced bound-state strength, the agreement of the ECRs with lQCD data is 
only marginally worse than in the SCS. Thus, the combination of HQ free energies and point-to-point ECRs does not seem to give a decisive answer on 
the coupling strength in the QGP.

In this context it is interesting to note two recent analysis for potential extractions from lQCD data. In Ref.~\cite{Bala:2021fkm} Wilson-line 
correlation functions in $N_f$=2+1-flavor QCD were analyzed using potential non-relativistic QCD to extract underlying bottomonium spectral 
functions using a variety of fit procedures (hard-thermal-loop (HTL) inspired, Pad{\'e}, Gaussian and Bayesian methods). 
While the HTL method results in a potential that exhibits significant screening and is close to the free energy, the Pad{\'e} and Gaussian methods 
lead to potentials that suggest little screening even at temperatures as large as 400\,MeV. 
In Ref.~\cite{Shi:2021qri} the masses and thermal widths of excited bottomonia extracted from the lQCD results with extended 
operators~\cite{Larsen:2019bwy,Larsen:2020rjk} have been analyzed using a deep-neural-network for an underlying complex-valued trial 
potential within a Schr\"odinger equation. It has been found that the real part of the potential varies rather little with temperature, while its 
imaginary part can reach around 1\,GeV at large distance (recall that the large-distance limit of the potential
essentially corresponds to the sum of the selfenergies of the heavy quark and antiquark).  

\begin{figure}[t]
\centering
 \includegraphics[width=0.45\textwidth]{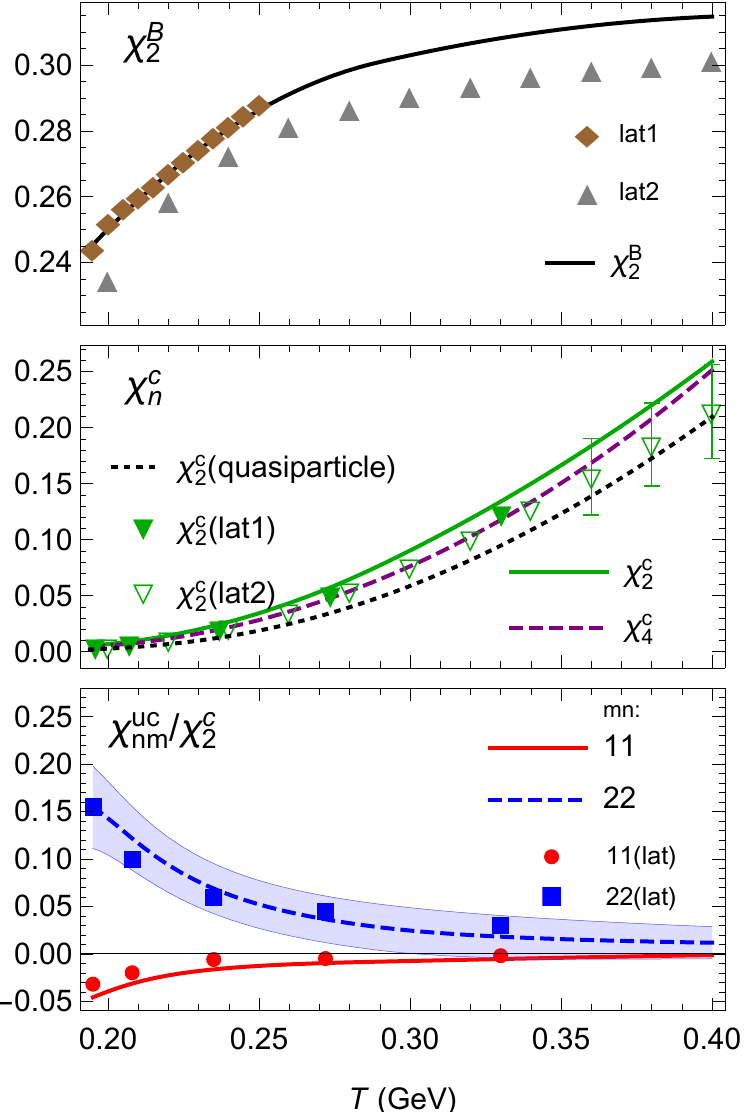} 
\caption{Diagonal baryon- and charm-number (upper and middle panel, respectively), as well as mixed charm-light quark suscpetibilities normalized to the
diagonal one (lower panel) from the $T$-matrix approach, compared to $N_f$=2+1 lQCD 
data~\cite{Borsanyi:2011sw,HotQCD:2012fhj,Bazavov:2014yba,Mukherjee:2015mxc,Bellwied:2015lba}; 
figure taken from Ref.~\cite{Liu:2021rjf}.}
\label{fig_chi}
\end{figure}
Fluctuation quantities of conserved charges, characterizing the medium's response for producing a charge-anticharge pair, are believed to be 
a good probe of the degrees of freedom that carry these charges, specifically hadronic vs. quark carriers. These susceptibilities, which are defined
through derivatives of the pressure of the system with respect to the pertinent chemical potentials,
\beq
\chi^{qc}_{nm}=\frac{\partial^{n+m}\hat{P}}{\partial\hat{\mu}_q^n\partial \hat{\mu}_c^m} \ ,
\eeq
can be computed with good precision in lQCD (here, the light- and charm-quark chemical potentials,  $\hat\mu_{q,c}=\mu_{q,c}/T$ and $\hat{P}/T^4$, 
are scaled dimensionless).
In the charm sector, pertinent $N_f$=2+1 lQCD data for charm-light channels  (including strangeness and baryonic channels) have been analyzed 
in terms of a non-interacting hadron resonance gas (HRG) model~\cite{Bazavov:2014yba}. In particular, by defining judiciously chosen susceptibility 
ratios for charm and charm-light channels, it has been found that the lQCD results start to deviate from the HRG predictions right above the pseudo-critical 
temperature, $\Tpc$, for the chiral transition, interpreted as the onset of charm-hadron melting. On the other hand, utilizing a mixed ansatz of 
non-interacting charm quarks and HRG states, it has been inferred that hadronic correlations persist up to temperatures of 
1.2-1.3\,$\Tpc$~\cite{Mukherjee:2015mxc}. 
In addition, by constructing ratios that are sensitive to the 
charm-baryon relative to the charm-meson content of the system, it has been found that the HRG results near $\Tpc$ underestimate the lQCD by 
a substantial factor when using the charm-baryon spectrum from the particle-data group listings~\cite{ParticleDataGroup:2012pjm}. The discrepancy can 
be resolved when instead using  predictions from the relativistic  quark-diquark model~\cite{Ebert:2011kk}, which includes a large set of additional 
charm-baryon states most of which have not been observed yet. 

In  Ref.~\cite{Riek:2010py} the diagonal charm-quark susceptibility, $\chi_2^c(T) \equiv \chi_{11}^{cc}$, has been calculated from the in-medium 
$T$-matrix using the HQ internal ($U_{\QQb}$) and free energy ($F_{\QQb}$) as potential proxies. Despite the significantly larger charm-quark masses 
implied by the $U$-potential relative to $F$ (not unlike the case of SCS vs. WCS), the larger widths generated by the $U$ potential lead to slightly larger
susceptibilities, in approximate agreement with lQCD data~\cite{Petreczky:2008px}. More systematic calculations have recently been carried
out in Ref.~\cite{Liu:2021rjf}. First, the $\mu_q$-dependence of the input potential has been constrained by the baryon number susceptibility, 
$\chi_2^B$ ($= \chi^q_2/9$), see upper panel of Fig.~\ref{fig_chi}, leading to rather moderate additional screening compared to the 
finite-temperature dependence. The resulting diagonal charm susceptibilities agree fairly well with the lQCD data~\cite{Bazavov:2014yba}; also
here the interaction effects lead to a significant increase over the quasiparticle result. In addition, off-diagonal charm-light susceptibilities have been 
extracted, which turned out to be numerically rather challenging. The negative values for $\chi_{11}^{uc}$, following the trend of the lQCD
data, have not been found in other model calculations to date, and might be an indicator of nonperturbative effects. Overall, the SCS of the 
$T$-matrix approach, with large parton widths and resonant states emerging near $\Tpc$, seem to be largely compatible with the lQCD results.

	\subsection{Role of QCD Medium}
    \label{ssec_med}


The ultimate goal of developing theoretical approaches to describe HF particles in the QCD medium is to learn about the latter's properties, in particular
how they emerge from the underlying interactions. It is therefore natural to ask to what extent the interactions that are being constructed in the HF 
sector are responsible for the transport and spectral properties of the QGP, and also its hadronization. From a practical point of view, the calculation 
of heavy-light scattering amplitudes in the QGP requires to specify the properties (spectral functions) of the thermal constituents in the system. 
Both the HQ selfenergies and transport coefficients involve an integration of the heavy-light amplitudes over the distributions of the thermal constituents. 
When implementing the HQ transport coefficients into a hydrodynamic space-time evolution of heavy-ion collisions, a significant mismatch between the 
medium temperature with the one of the calculated transport coefficients occurs if the latter are computed with thermal degrees of freedom that do
not obey constraints from the equation-of-state (EoS) of the medium, see, \eg, Ref.~\cite{Rapp:2018qla} for an illustration. In the following, we will 
discuss three examples where the evaluation of HQ properties in the QGP has been carried out with thermal-parton degrees of freedom that reproduce 
the EoS computed in lQCD.

\begin{figure}[t]
\centering
 \includegraphics[width=0.45\textwidth]{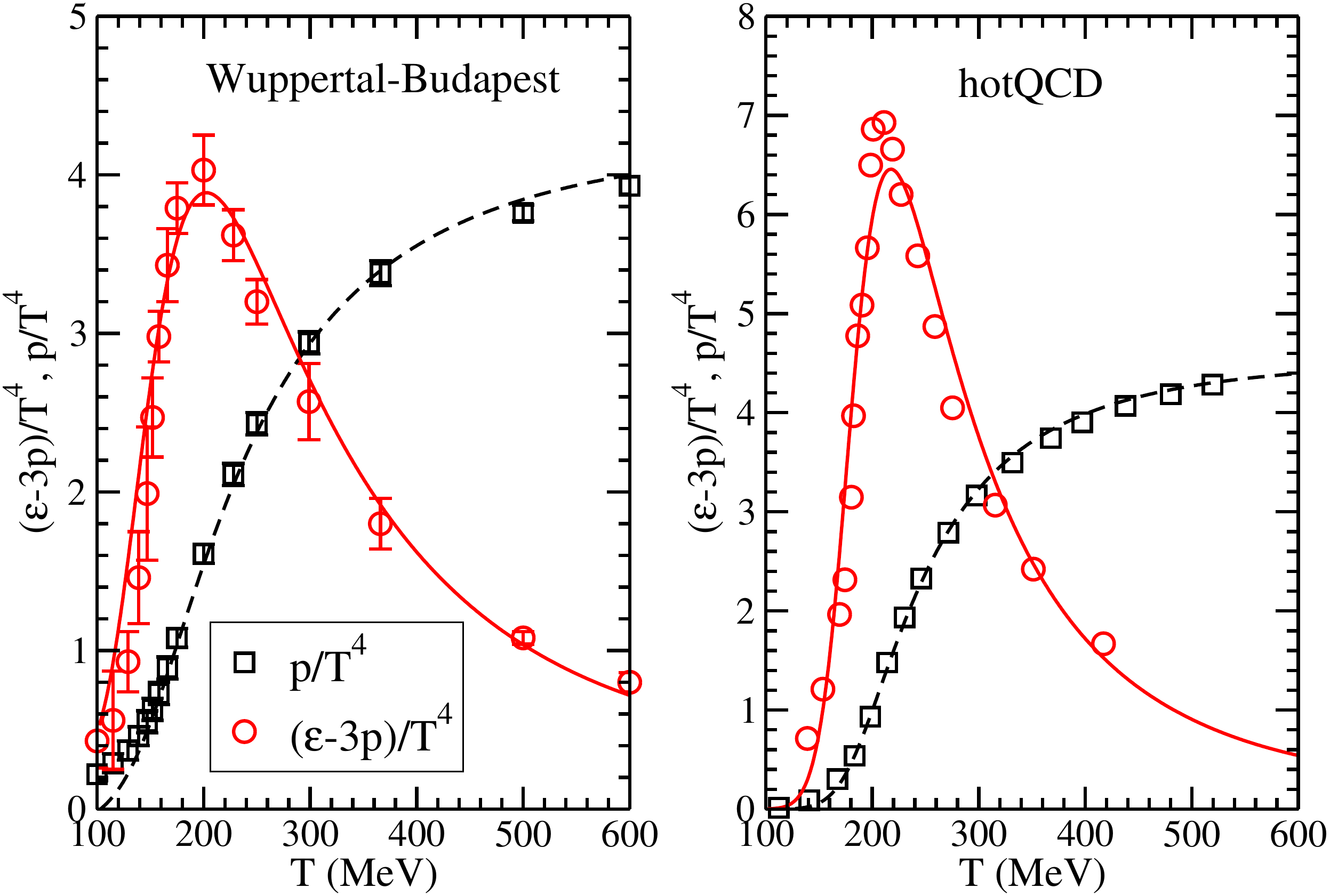} 
 \includegraphics[width=0.54\textwidth]{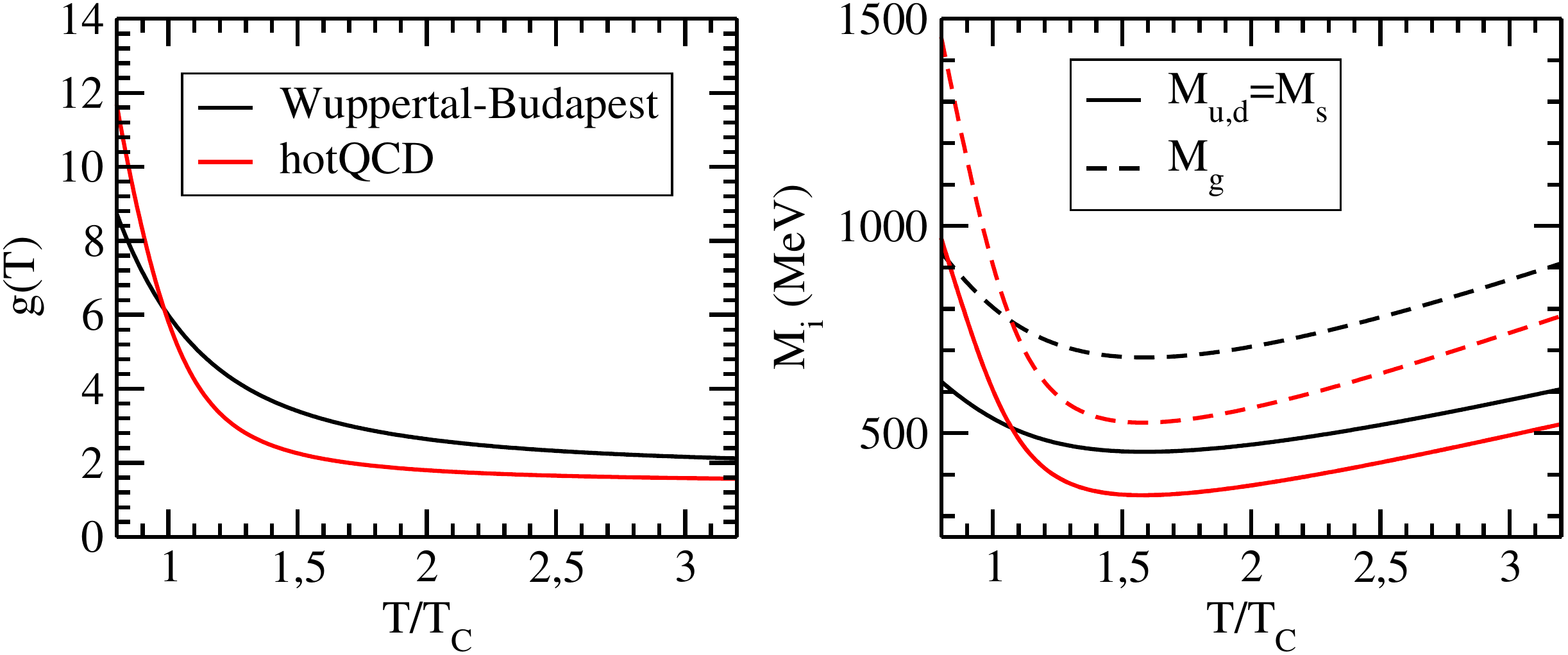} 
\caption{A recent quasiparticle description of the QGP EoS in terms of the pressure and interaction measure (left two panels) and the underlying 
fits for the effective coupling constant (third panel) and the resulting thermal-parton masses (right panel); figures taken from 
Ref.~\cite{Plumari:2011mk}.}
\label{fig_qp}
\end{figure}

In Ref.~\cite{Scardina:2017ipo}, Born diagrams for HQ interactions with thermal quasiparticles have been implemented utilizing thermal light-quark 
and gluon masses with a pQCD motivated ansatz, $m_{q}^2 = g(T)^2T^2/3$ and $m_g^2 = 3g(T)^2T^2/4$, respectively. The key ingredient is the 
effective temperature-dependent coupling constant,  $g(T)$, which is adjusted such that the non-interacting massive parton gas reproduces the lQCD EoS~\cite{Plumari:2011mk} for the pressure and the so-called interaction measure $(\epsilon-3P)/T^4$. The reduction in the pressure found in lQCD 
when going down in temperature toward $\Tpc$ is achieved by quasiparticle masses that start to increase below $T\simeq1.5\Tpc$, which is realized 
through an increase of the effective coupling, rising from perturbative values of about $g\simeq2$ at high $T$ to near 5-6 at $\Tpc$. This construction
also reproduces the characteristic maximum of the interaction measure slightly above the pseudo-critical temperature $\Tpc$, see Fig.~\ref{fig_qp}, and  
furthermore yields a fair description of the diagonal light- and strange-quark number susceptibilities computed in lQCD.

In Ref.~\cite{Song:2015sfa} Born amplitudes for heavy-light scattering have been evaluated with thermal quark and gluon degrees of freedom that follow
from the dynamical quasiparticle model (DQPM)~\cite{Cassing:2008nn,Berrehrah:2016vzw}. In the DQPM, the ansatz for the quasiparticle masses is very 
similar to the one in the zero-width quasiparticle models discussed in the preceding paragraph. In addition, collisional widths of the thermal quarks 
and gluons are introduced through a perturbative expression of the type $\Gamma_{q,g}\sim g(T)^2T \ln(1.1+c/g^2)$  where the effective coupling 
constant again serves as the main fit parameter.  
The effective gluon masses in the QGP tend to be somewhat  higher than in the zero-width quasiparticle model, while the quark masses turn out be be 
rather similar. The collisional widths are typically around 0.2\,GeV with a relatively weak temperature dependence. Also in this approach not only the 
QGP EoS but also the quark-number susceptibilities computed in lQCD can be described. For the computations of HQ transport coefficients, off-shell 
integrations over the thermal parton propagators are carried out.

\begin{figure}[!t]
\centering
 \includegraphics[width=0.33\textwidth]{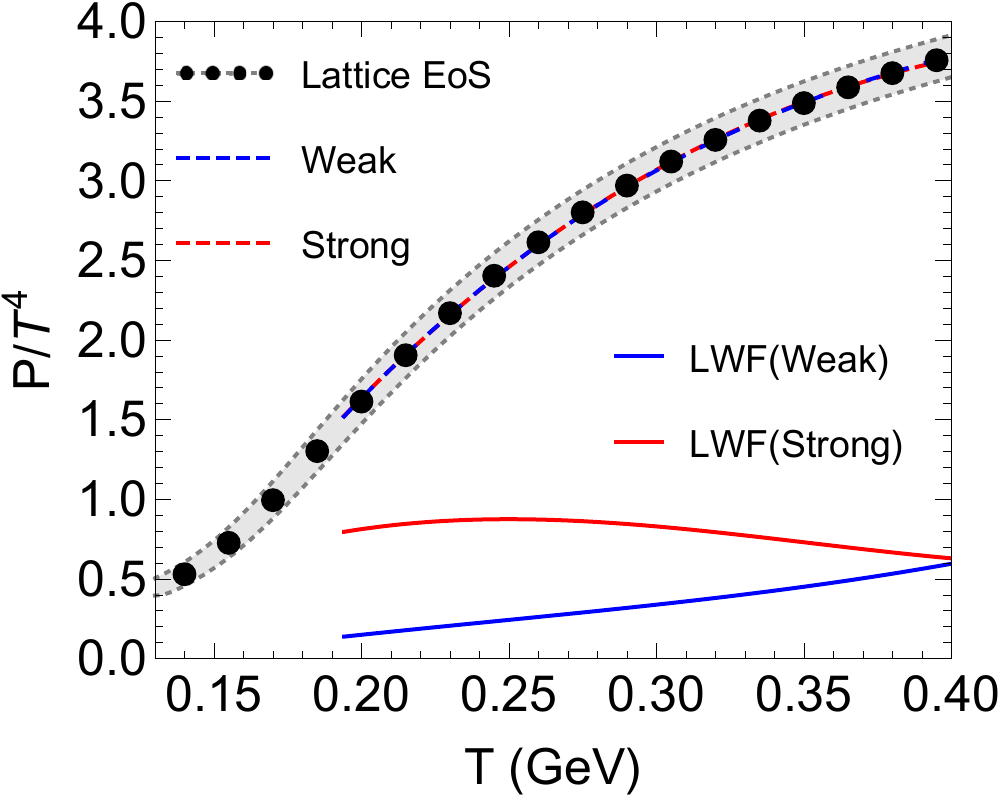} 
 \includegraphics[width=0.3\textwidth]{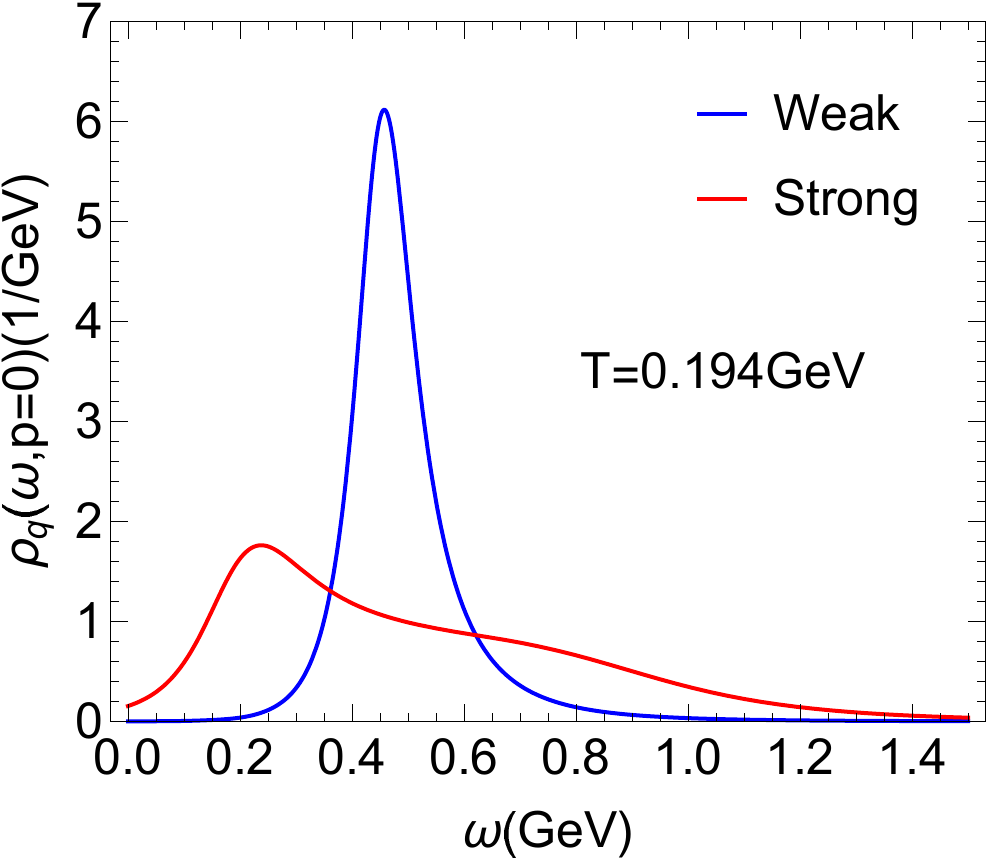}
\hspace{0.1cm} 
\includegraphics[width=0.31\textwidth]{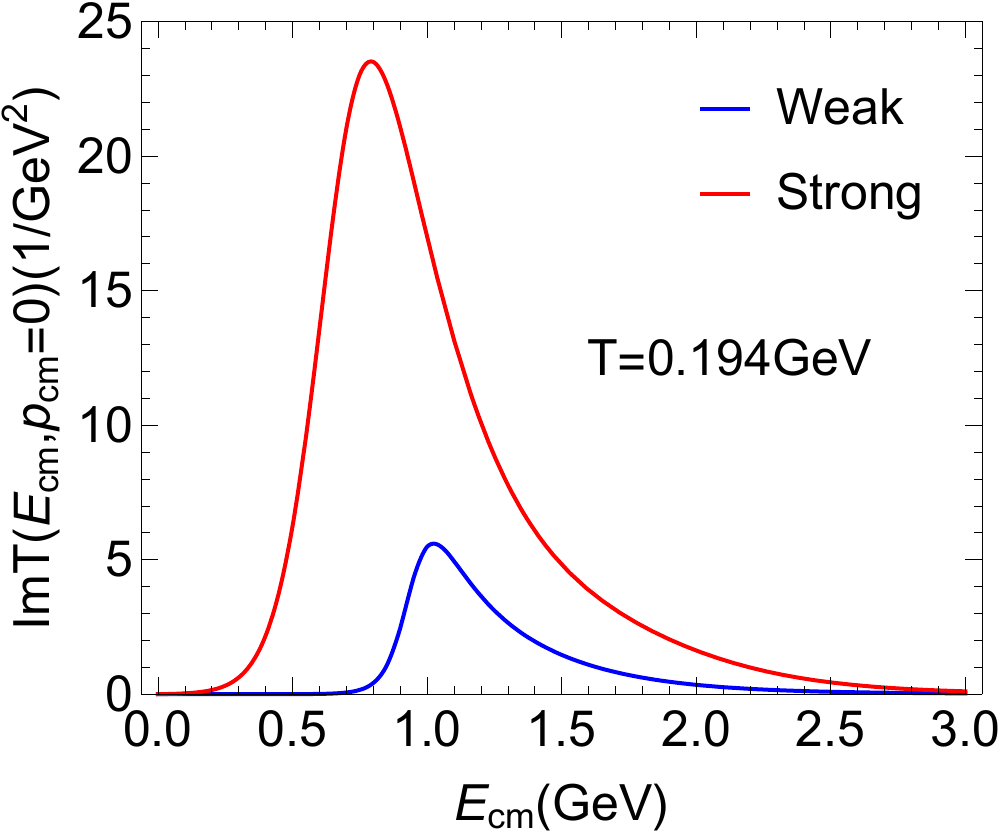} 
\caption{Thermodynamic $T$-matrix approach to the QGP implemented within the Luttinger-Ward-Baym formalism employing the two-body parton 
interaction potentials shown in Fig.~\ref{fig_FvsV} for the weakly-coupled scenario (WCS, blue lines) and the strongly-coupled scenario (SCS, red lines); 
the left panel shows the fit to the pressure computed in lQCD (total: dashed lines, $\Phi$ contribution: solid lines), while the middle and right panels show corresponding underlying (selfconsistent) one- and two-body spectral properties, \ie, the quark spectral functions and the imaginary part of the color-singlet 
$S$-wave $T$-matrix, respectively; figures taken from Ref.~\cite{Liu:2018ons}.}
\label{fig_eos-tmat}
\end{figure}

In Refs.~\cite{Liu:2017qah,Liu:2016ysz}, the $T$-matrix approach has been extended to the light-parton sector, formulated as a Hamiltonian approach. 
The same underlying potential has been used as following from the constraints from the HQ free energy and quarkonium correlators, including relativistic 
corrections and the inclusion of all possible color channels in the interactions of thermal partons among each other (and partial waves up to $l$=5). 
Light-parton $T$-matrices and selfenergies have been calculated selfconsistently, while the free energy of the QGP has been computed in the 
Luttinger-Ward-Baym (or two-particle irreducible) approach, schematically written as
\beq
\Omega(T) = \sum_j \mp d_j \int d\tilde{p} \ \left[\ln \left( -G_j^{-1}(\tilde{p}) \right)+D_j(\tilde{p}) \Sigma_j(\tilde{p}) \right] + \Phi(T) \ , 
\eeq
where $d\tilde{p}$ denotes a 3D momentum integration and a summation over Matsubara frequencies (in the imaginary time formalism).
The first term corresponds to the contribution of the thermal anti-/quarks and gluons with their fully dressed in-medium propagators, $G_j$, and 
selfenergies, $\Sigma_j$ (summed over their color, spin and flavor degeneracies, with the $\pm$ sign for bosons or fermions). The more challenging 
part is the two-body interaction contribution encoded in the Luttinger-Ward functional, $\Phi$, which contains an infinite summation over 
``skeleton diagrams" of increasing order, $\nu$,  in the in-medium interaction potential. Due to an extra combinatorial factor of $1/\nu$ (to avoid 
double-counting), it is not straightforwardly resummed for non-separable potentials (as are being used here). In 
Ref.~\cite{Liu:2016ysz} this problem was overcome with a matrix-log inversion technique leading to
\beq
\Phi(T) = \mp \frac{1}{2} \sum_j d_j   \int d\tilde{p} \ G_j(\tilde{p}) \  {\rm Log}(\Sigma_j(\tilde{p}))  \ .
\eeq
To fit the lQCD data for the pressure, an effective parton mass term is introduced (with appropriate color factors) that is associated with effects 
that are not explicitly treated in the current set-up (\eg, a gluon condensate). The results for the pressure are shown in the left panel of 
Fig.~\ref{fig_eos-tmat}. While both SCS and WCS can reproduce the lQCD data, the key difference is in the underlying spectral properties of the 
constituents of the system, especially at temperatures close to $\Tpc$. In the WCS, the quark spectral function (middle panel of Fig.~\ref{fig_eos-tmat})
consists of a well-defined quasiparticle peak at an energy close to the nominal thermal mass of about 0.5\,GeV, with a moderate spectral width of 
ca.~0.1\,GeV; at the same time, the color-singlet $q\bar q$ $T$-matrix (right panel of Fig.~\ref{fig_eos-tmat}) shows a rather weak resonance at a mass
of $\sim$1\,GeV whose thermodynamic contribution to the pressure is very small (blue line in the left panel). On the other hand, for the SCS, the quark 
spectral functions are essentially melted, with a width of 0.6\,GeV that is larger than the effective quark mass; in addition, a collective mode develops at low 
energy, which is also quite broad and thus has a rather limited thermodynamic weight. However, the color-singlet $S$-wave $T$-matrix develops a 
prominent resonance at a mass of about 0.8\,GeV, not far from the vacuum mass of the light-vector mesons (since spin-spin interactions are neglected, this
resonance is also present in the pseudoscalar channel). The thermodynamic contribution of the resonances (which also form in the color anti-triplet diquark
channel) are encoded in the $\Phi$ functional, whose relative contribution markedly rises with decreasing temperature reaching more than 50\% at 
$T$=194\,MeV. In this sense, the SCS develops a transition in the degrees of freedom from melting partons to pre-hadronic resonances.  

	\subsection{Hadronic Interactions}
    \label{ssec_had}

In a confined medium, heavy quarks diffuse within HF hadrons, in particular $D$-mesons and single-charm baryons and their excited states, through their 
interactions with the thermal abundance of (mostly light) hadrons in the heat bath. The study of the spectral properties of charm hadrons, such as in-medium modifications of masses and widths, have been carried out in various approaches, \eg, QCD sum rules~\cite{Hilger:2008jg,Suzuki:2015est} or quantum 
many-body theory~\cite{Fuchs:2004fh,Lutz:2005vx,Tolos:2007vh,Montana:2020lfi,Montana:2020vjg}. The studies of the transport (diffusion) properties 
of HF hadrons have been initiated more recently~\cite{Laine:2011is,He:2011yi,Ghosh:2011bw,Abreu:2011ic}. These efforts started with the 
calculation~\cite{Laine:2011is} of the $D$-meson diffusion coefficient in a pion gas at low temperatures using Born amplitudes from heavy-meson 
chiral perturbation theory. At a temperature of $T=100$\,MeV, where the hadronic medium is essentially a hot pion gas, the $D$-meson 
relaxation rate has been found to be $\gamma_D\simeq 0.05/{\rm fm}$ at zero momentum~\cite{Laine:2011is}. Similar results have also been 
obtained with Born amplitudes for $D$-meson scattering off various mesons and baryons~\cite{Ghosh:2011bw}. In Ref.~\cite{He:2011yi} a more 
phenomenological approach was adopted to calculate the same quantity by employing empirical scattering amplitudes (parameterized resonant amplitudes 
for $D$-$\pi$ scattering based on $D_0^*(2308)$ and $D_2^*(2460)$ resonances, and vacuum unitarized amplitudes with other pseudoscalar and 
vector mesons as well as baryons and antibaryons from existing effective model calculations). The resulting zero-momentum relaxation rate turned 
out to be a factor of $\sim 10$ smaller at $T=100$\,MeV.
When extrapolated to temperatures close to $\Tpc$, where contributions from other mesons and baryons become comparable, the zero-momentum 
$D$-meson relaxation rate increases to $\sim 0.06/\rm fm$. This translates into a pertinent spatial diffusion coefficient of $D_s(2\pi T)\simeq 6$, 
not far from the estimate from the QGP side within the $T$-matrix approach~\cite{Riek:2010fk,Liu:2018syc}, implying a minimum structure of the 
charm-quark diffusion coefficient across the phase transition region~\cite{He:2012df}. Finally, another calculation~\cite{Abreu:2011ic} using unitarized 
chiral $D$-$\pi$ interactions also featuring resonances led to a result of $\gamma_D\simeq 0.005/{\rm fm}$ at $T=100$\,MeV, in close agreement with 
the value found in Ref.~\cite{He:2011yi}. This highlights the importance of the unitarization procedure in constructing realistic interactions between 
$D$-mesons and pions -- without unitarization, the scattering amplitudes grow rapidly with energy and the calculated relaxation rate rises 
in a rather uncontrolled way. This approach was extended to include a complete set of pseudoscalar mesons and baryons (nucleons and 
$\Delta$'s)~\cite{Tolos:2013kva,Ozvenchuk:2014rpa} within the same unitarized formalism; the resulting $D$-meson relaxation rate and diffusion 
coefficient at high temperatures close to $\Tpc$ have been found to be similar to the results obtained in Ref.~\cite{He:2011yi}.

The diffusion properties of $B$ mesons in hadronic matter have since been also investigated, starting with Ref.~\cite{Das:2011vba} where heavy-light 
scattering length at threshold from heavy-meson chiral effective theory without resummation were adopted. The resulting $B$-meson relaxation rate 
($\gamma_B\simeq 0.001/{\rm fm}$ at $T=100$\,MeV) is only slightly smaller than the $D$-meson counterpart within the same 
framwork~\cite{Ghosh:2011bw}. In contrast, the unitarized calculations accounting for dynamical generation of resonances, first in a 
pseudosclar-meson gas~\cite{Abreu:2012et} and later on also including baryons~\cite{Torres-Rincon:2014ffa}, has yielded a value of 
$\gamma_B\simeq 0.003/{\rm fm}$ at $T=100$\,MeV (increasing by $\gtsim 50$\% at $T\geq 140$\,MeV). This helps to give a $B$-meson 
spatial diffusion coefficient $D_s=\gamma/m_HT$ (where $m_H$ is the heavy-meson mass) similar to that of the $D$-meson (especially at temperatures 
near $\Tpc$)~\cite{Abreu:2012et}. This supports the notion of using HF quarks/hadrons as a universal probe of the transport properties of the hot 
medium which should be independent of the mass of the probes. This notion has  been further confirmed between $\Lambda_c$ and $\Lambda_b$ 
in a calculation of the heavy baryons' diffusion coefficient within the same formalism~\cite{Tolos:2016slr,Das:2016llg}.

The above studies have all employed vacuum scattering amplitudes. In a recent work within the unitarized chiral effective approach
thermal corrections for the heavy-light interactions~\cite{Montana:2020lfi,Montana:2020vjg} have been  implemented  into an off-shell kinetic 
equation~\cite{Torres-Rincon:2021yga}. 
While the off-shell broadening effects have not been found to be significant, the Landau contribution arising from heavy-light interaction below the 
energy threshold as implicit in the thermal scattering amplitudes appears to be substantial, causing a factor of $2\sim3$ increase relative to the result 
from using vacuum amplitudes, for the $D$-meson relaxation rate at moderate temperatures $T=100$-$150$\,MeV~\cite{Torres-Rincon:2021yga}.


	\newpage

	\section{Transport Coefficients}
  \label{sec_trans}
In this chapter we discuss how the in-medium HQ interactions manifest themselves in various transport properties. In Sec.~\ref{ssec_Ap} we start 
with the thermal relaxation rate (or friction coefficient), which is directly calculated from the in-medium heavy-light scattering amplitudes. In particular, 
we focus on its momentum dependence, which reflects on the microscopic properties of the amplitudes and is also of considerable phenomenological 
importance. We also highlight the role of quantum effects in these calculations. In Sec.~\ref{ssec_Gam} we turn to the inelastic reaction rates of
quarkonia, which are closely related to the single HQ transport coefficients and again are pivotal to the phenomenology of quarkonium transport in
heavy-ion collisions. An additional scale in this context is the binding energy of the quarkonia, which causes a further interplay with the reaction rates.
In Sec.~\ref{ssec_Ds} we briefly review the status of the HF spatial diffusion coefficient, a widely discussed transport coefficient related to the 
zero-momentum limit of the friction coefficient, and return to the question of how the long-wavelength limit of HF diffusion
relates back to QGP properties, in particular the shear viscosity as a central transport parameter of hydrodynamic evolution models.

\subsection{Thermal Relaxation of Heavy Flavor}
\label{ssec_Ap}
\begin{figure}[t]
\centering
 \includegraphics[width=0.9\textwidth]{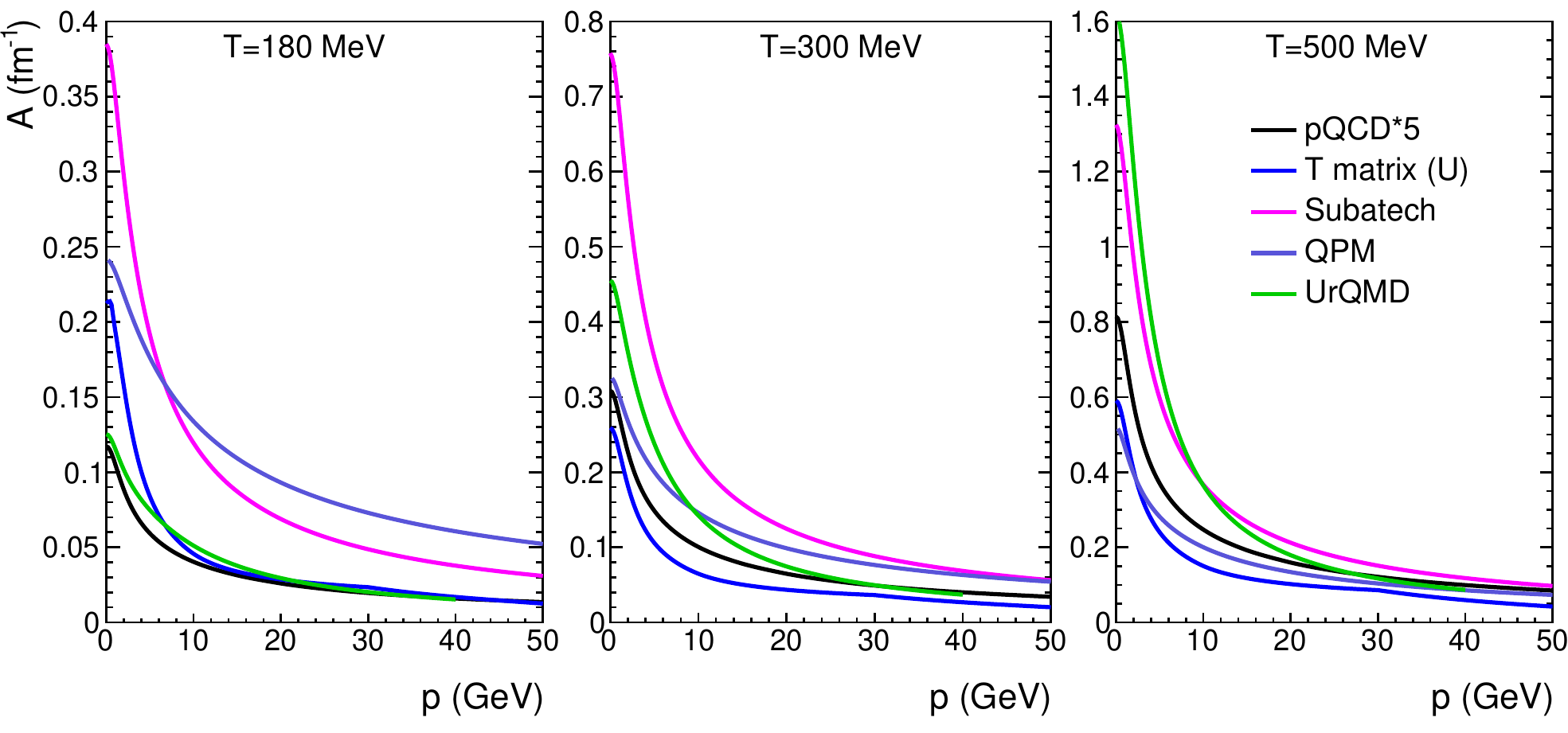} 
\caption{Friction coefficients for charm-quark diffusion in the QGP as a function of three-momentum for three different temperatures from various 
model calculations:
black lines: pQCD Born diagrams with $\alpha_s$=0.4, multiplied with and overall $K$-factor of 5, blue lines: $T$-matrix results using the 
internal-energy ($U$) as potential proxy~\cite{Riek:2010fk,Huggins:2012dj}, pink lines: pQCD with running coupling constant and reduced Debye 
mass~\cite{Gossiaux:2008jv,Peshier:2008bg}, purple lines: quasiparticle model with coupling constant fitted to the lQCD EOS~\cite{Plumari:2011mk}, 
and green lines: $D$-meson resonance model~\cite{vanHees:2004gq}; figure taken from Ref.~\cite{Rapp:2018qla}.}
\label{fig_Ap-models}
\end{figure}

Based on the original idea of Ref.~\cite{Svetitsky:1987gq}, HQ motion in a QGP at moderate temperatures is akin to a Brownian motion and as such
amenable to a description with a Fokker-Planck equation. As elaborated in more detail in Sec.~\ref{ssec_trans-app} below, this leads to a set of 
well-defined transport coefficients, in particular the friction coefficient, $A(p)$, and the transverse and longitudinal momentum diffusion coefficients, 
$B_0(p)$ and $B_1(p)$, respectively. In principle, these coefficients are related through an Einstein relation (as to ensure kinetic equilibrium in the 
long-time limit), and therefore we here focus on the friction coefficient which may also be interpreted as the thermal-relaxation rate (or inverse
thermal-relaxation time).  In Fig.~\ref{fig_Ap-models}, the results for charm quarks of several of the models discussed in the previous chapter 
(which are also frequently used in heavy-ion phenomenology as discussed in Sec.~\ref{ssec_hf-obs}) are compiled for three different temperatures 
roughly representative of the QGP temperatures reached at RHIC and the LHC.  At the smallest temperature ($T$=180\,MeV), the 
pQCD Born calculations with running coupling and reduced 
Debye mass (which we refer to as pQCD*) have the largest value at low momentum (driven by the large interaction strength for soft momentum 
transfers), but they also feature a strong fall-off with momentum. The in-medium $T$-matrix results (calculated with the internal energy, $U$, as the 
potential proxy) have a similarly strong fall-off, but are smaller in magnitude by a factor of 2-2.5; however, in this case the drop-off with momentum 
is caused by a transition from a resummed nonperturbative string force, prevalent at large distances, to a color-Coulomb force at small distance. 
The pQCD quasiparticle model results are comparable to the $T$-matrix  results at low momentum, but with a much harder momentum dependence, 
exceeding the pQCD* results for $p\gtsim$\,10\,GeV. This is probably caused by the rather large thermal-parton masses at low temperatures 
(recall Fig.~\ref{fig_qp} right), which remain more effective in transferring momentum to the charm quark even if the latter has a relatively 
high momentum.  Also shown is the effective $D$-meson resonance model, which is very comparable to a leading-order pQCD calculation with 
$\alpha_s=0.4$ and an additional $K$-factor of 5 (which we refer to as pQCD-5); this is presumably a coincidence. At an intermediate temperature of
$T$=300\,MeV, the pQCD* model still gives the largest friction coefficient, while the pQCD quasiparticle results now have a softer momentum dependence
and are only slightly larger than the pQCD-5 results. The friction coefficient from the $T$-matrix is now below the pQCD-5 result, as the resonant
interaction strength has significantly weakened.  On the other hand, in the resonance model, the assumption of undiminished resonance interactions 
leads to rather large friction coefficients which is even more extreme (and unrealistic)  at $T$=500\,MeV. At this temperature, the pQCD* friction 
coefficient remains the strongest, while $T$-matrix, pQCD quasiparticle and pQCD-5 rsults are all comparable. Also note that at high momentum 
the agreement between the different calculations is better than at low temperature.
Note that the temperature dependence for the zero-momentum friction coefficient is approximately quadratic for the pQCD-5 results, a little stronger than 
linear for pQCD*,  and weaker than linear for the pQCD quasiparticle and $T$-matrix models at low temperature (due to the drop in coupling constant and
 loss of resonance interactions, respectively) and approximately linear at higher temperature.

\begin{figure}[t]
\centering
 \begin{minipage}{0.47\textwidth}
 \includegraphics[width=0.97\textwidth]{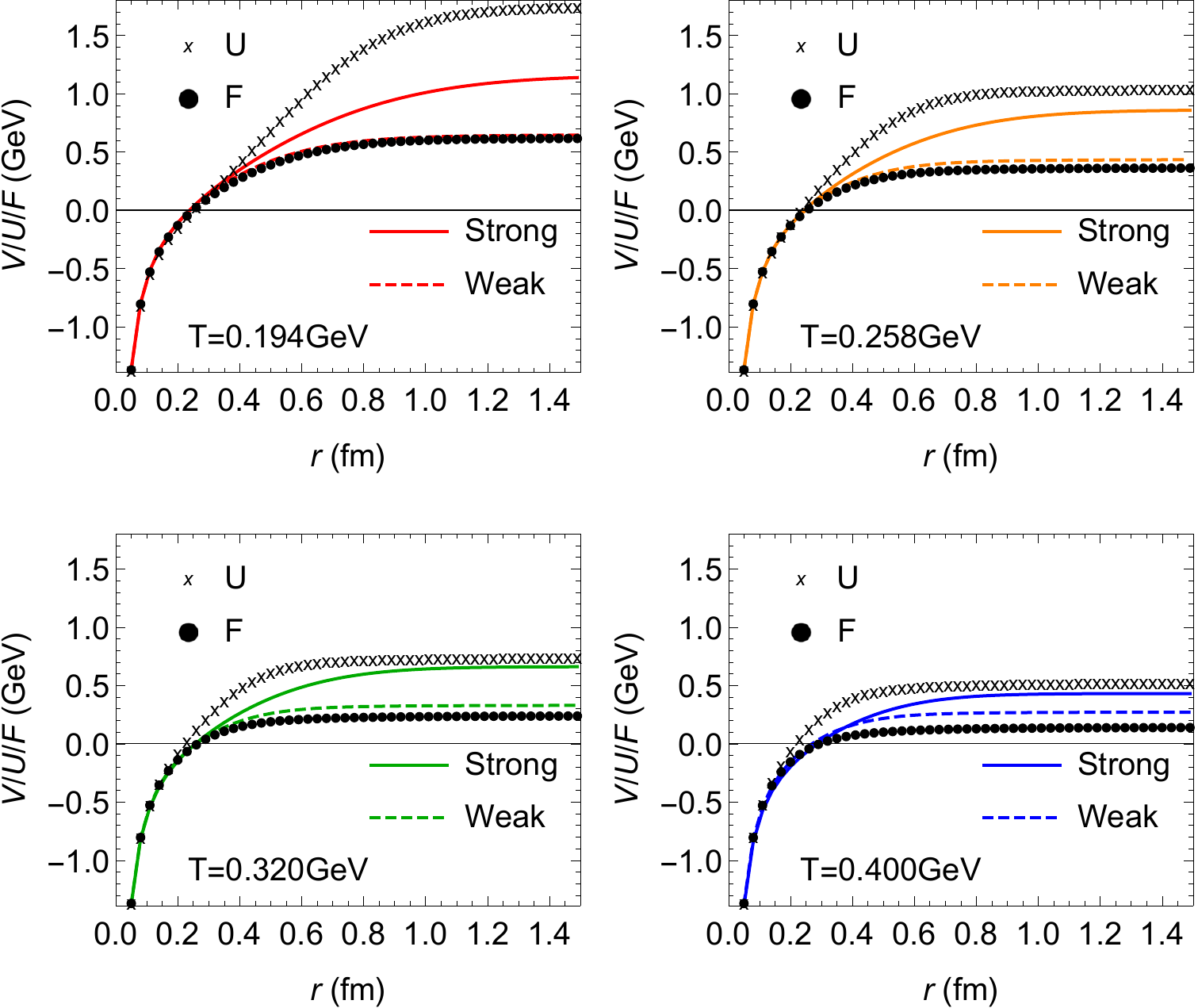}
\end{minipage} 
\hspace{0.3cm}
\begin{minipage}{0.47\textwidth}
\includegraphics[width=0.97\textwidth]{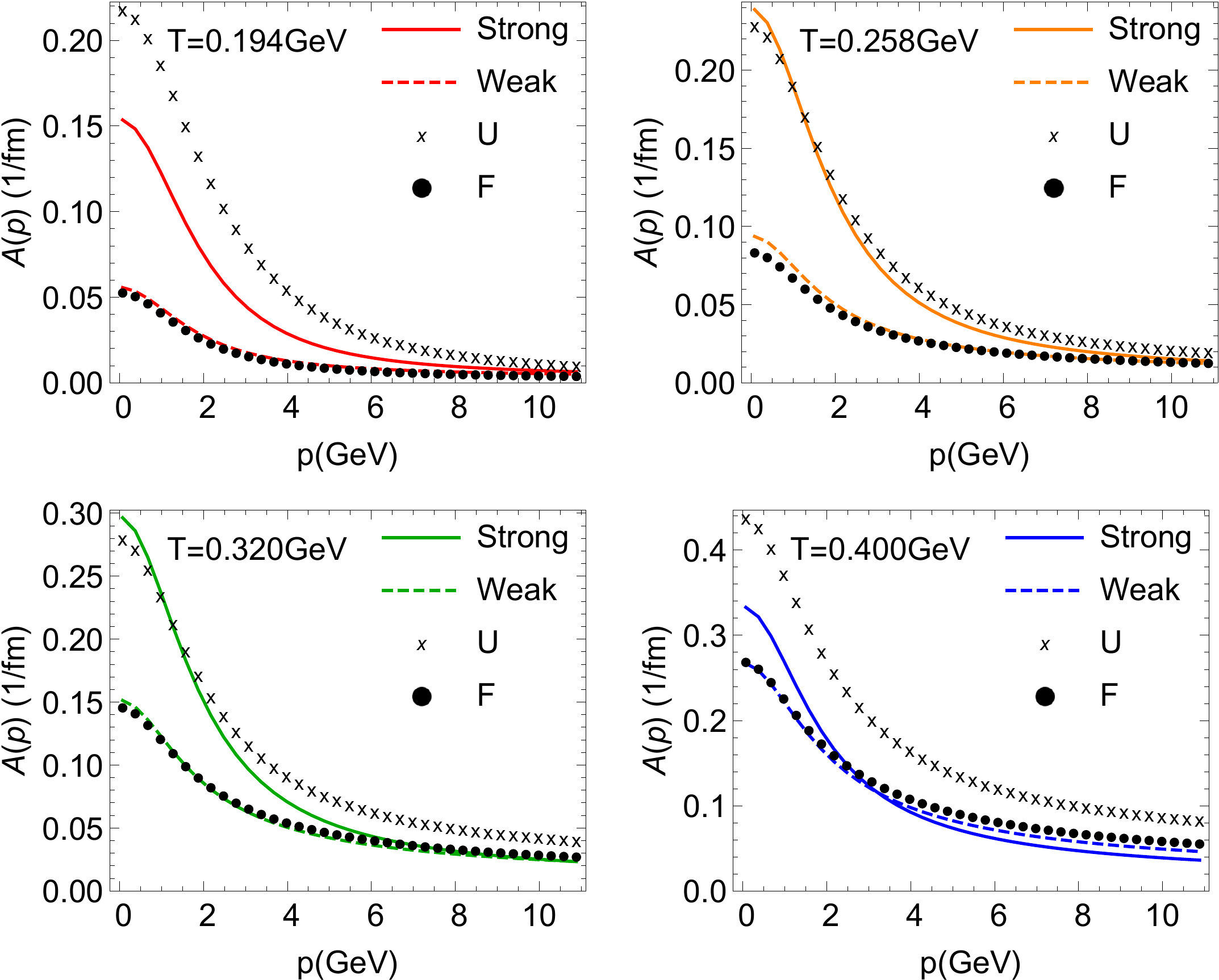}
\end{minipage}
\caption{The left-hand four panels show the 4 different input potentials that are utilized to compute the heavy-light $T$-matrices for charm-quark 
scattering of thermal partons in quasiparticle approximation (in all color and spin-flavor channels, including partial waves up to $l$=5) that are
further used to compute the pertinent friction coefficients in the QGP shown in the right four panels (dots, crosses, dashed lines and solid lines  are for the 
free-energy, internal energy, weakly-coupled and strongly-coupled potentials, respectively); figures are taken from Ref.~\cite{Liu:2018syc}.}
\label{fig_Ap-tmat-qp}
\end{figure}
To investigate the dependence of the friction coefficient on the underlying interactions and the manifestation of nonperturbative effects more explicitly, 
various inputs to the $T$-matrix calculations have been scrutinized in Ref.~\cite{Liu:2018syc}. Specifically, different input potentials have been considered
(cf.~the four left panels of Fig.~\ref{fig_Ap-tmat-qp}) and employed to calculate the heavy-light $T$-matrices using the same quasiparticle medium for 
the QGP, with approximately constant quark masses of about 0.4\,GeV and gluon masses smoothly increasing from 0.8\,GeV at $T$=400\,MeV to 1.2\,GeV at 
$T$=200\,MeV. However, the in-medium charm-quark masses, while based on the same bare-quark mass, are governed by the infinite-distance limits 
of the respective potentials, ranging from 1.35-1.5\,GeV at $T$=400\,MeV to 1.6-2.1\,GeV at $T$=200\,MeV, with the largest values from the 
$U$-potential and the smallest from the $F$-potential. The resulting friction coefficients are shown in the four right panels of Fig.~\ref{fig_Ap-tmat-qp}.
Not surprisingly, the proximity of the free-energy and the WCS potential lead to very similar transport coefficients for these two potentials throughout; the 
internal-energy and SCS potential generate much stronger effects, especially at low momentum and at smaller temperatures. Given the appreciable 
difference of $U$ and $V_{\rm SCS}$ at distances of $r$$\simeq$0.5-1\,fm at small temperatures, the difference in the low-momentum friction coefficient 
is less pronounced than may have been expected. It turns out, though, that the force (derivative of the potential) is very similar between $U$ and 
$V_{\rm SCS}$ at distances of around 1\,fm at $T$=194\,MeV, and even slightly larger for $V_{\rm SCS}$ for $r\gtsim0.7$\,fm at $T$=258\,MeV. The 
long-range parts of the force are therefore instrumental for the zero-momentum limit of $A(p)$. These calculations corroborate  that the zero-momentum 
limit of the friction coefficient (which also determines the spatial diffusion coefficient) indeed reflects the long-range properties 
of the interaction. In part, this is so because for a long-range force the heavy quark can interact with a larger number of thermal partons in its vicinity.

\begin{figure}[t]
\centering
\begin{minipage}{0.65\textwidth}
\includegraphics[width=0.97\textwidth]{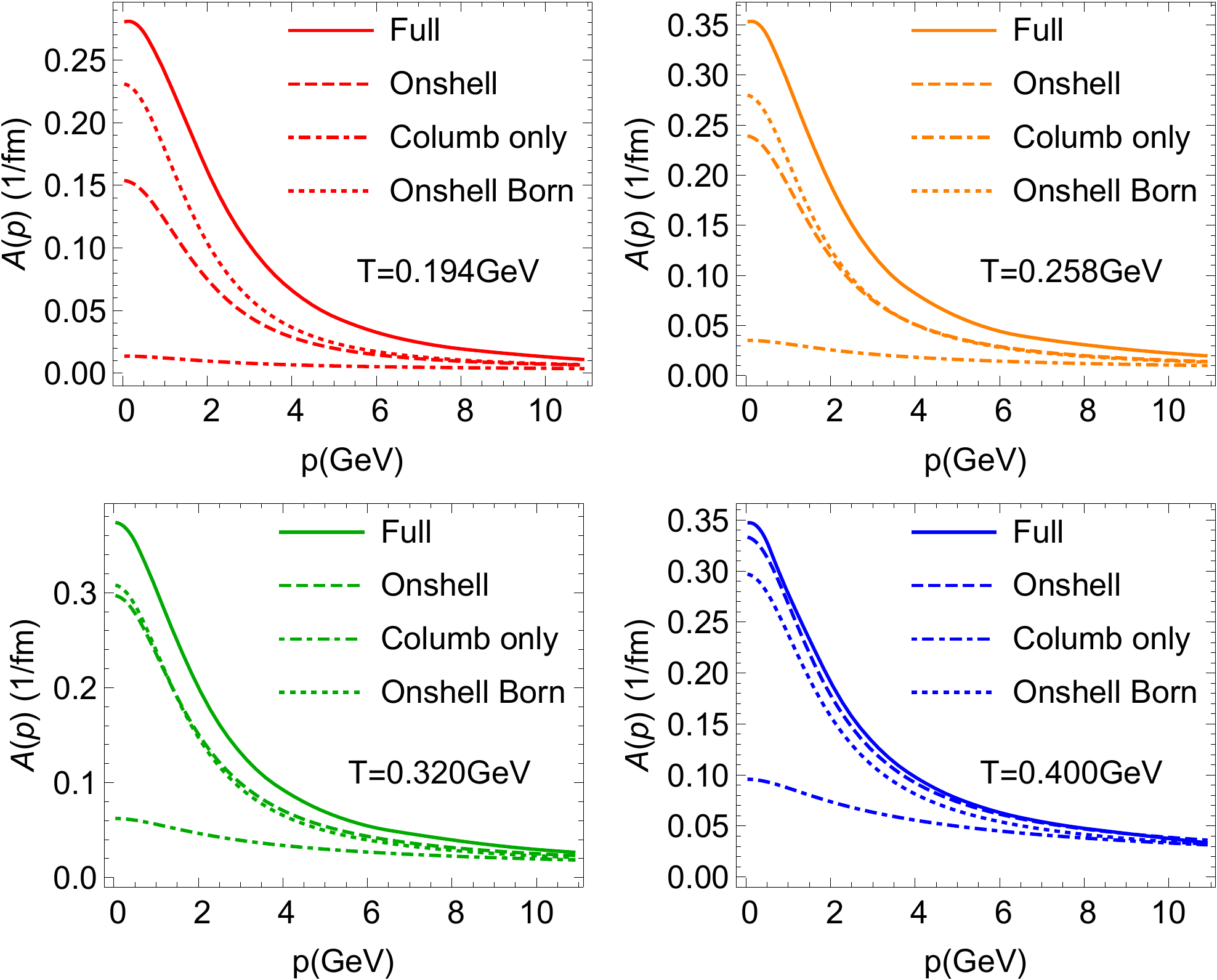}
\end{minipage}
\caption{Charm-quark friction coefficients in the QGP from the $T$-matrix approach in the SCS. The different curves in the four panels for diffferent 
temperatures include an increasing level of nonperturbative effects, from one-gluon (pQCD) exchange Born terms only (dash-dotted lines), via 
including the full Born term with nonperturbative string contributions (dotted lines) and a fully resummed $T$-matrix with all incoming and outoging 
partons put on-shell (dashed lines) to the full result which accounts for the broad spectral functions of the thermal partons as well as the outgoing 
charm-quark; figures taken from Ref.~\cite{Liu:2018syc}.}
\label{fig_Ap-tmat-scs}
\end{figure}
Another issue of interest to better understand the microscopic mechanisms of HQ transport in the QGP is the manifestation of both
nonperturbative and quantum effects (which are rather closely related as will be seen in the following; \eg,  a large interaction strength generates 
large collisional widths implying broad spectral functions).  Toward this end, four cases have been set up in Ref.~\cite{Liu:2018syc} within the SCS, 
by systematically switching off various components in the calculation of the charm-quark friction coefficient. 
In the first one, the string interactions in the input potential have been switched off (by putting the string tension in Eq.~(\ref{Va}) to zero), leaving
only the color-Coulomb interactions (with $\alpha_s$=0.27). Compared to the full result, a reduction of the friction coefficient by up to a factor of 
$\sim$10 at low momentum and small temperatures has been found, cf.~Fig.~\ref{fig_Ap-tmat-scs}. On the other hand, for momenta above 10\,GeV, 
and especially at the higher temperatures, the results of the Coulomb-only calculation are close to the full ones, \ie, the nonperturbative interactions have 
ceased. In the second case, the ladder resummations of the $T$-matrix are switched off, leaving only the Born terms as the scattering amplitudes.
This leads to surprisingly large friction coefficients, which even at low momenta and temperatures are within ca.~20\% of the full results. Clearly, 
the string Born term is responsible for this result. In the third case, the full $T$-matrices are calculated, including both color-Coulomb and string 
interactions, but the thermal partons, as in the previous two cases, are still assumed to be quasiparticles, \ie, the friction coefficients are evaluated 
with  $\delta$-functions for the parton spectral functions enforcing the on-shell condition for the parton energies at given momentum.  In this case, 
a {\it reduction} of the friction coefficients has been found, relative to the Born-case, which is most pronounced again for low temperatures and 
momenta. The reason for this is that the ladder resummation in the $T$-matrix generates bound states, especially at low temperatures, which are 
not accessible in on-shell heavy-light scattering kinematics. 
Finally, case 4 is the full result, where the broad spectral functions of the in-medium thermal partons, as well as for the outgoing charm-quark, are
accounted for. This is critical for the charm quarks to access the large resonant interaction strength of the broad sub-threshold pre-hadronic bound 
states, recall Fig.~\ref{fig_c-spec}. In this way the quantum uncertainty in the energy spectra of both the constituents and their scattering amplitudes 
plays a key role in producing large friction coefficients at low momenta and temperatures.

\subsection{Quarkonium Dissociation Rates}
\label{ssec_Gam}
The central quantity in the description of quarkonium (${\cal Q}$) transport in the QCD medium is the inelastic reaction rate of the various bound states. 
To leading order in  the strong coupling constant, the rate in the QGP is given by gluo-dissociation processes, 
$g+{\cal Q} \to Q+ \bar{Q}$~\cite{Bhanot:1979vb,Kharzeev:1994pz} (more recently also referred to as singlet-to-octet transition~\cite{Brambilla:2011sg}). However, in practice, it turns out that this mechanism is often not efficient, especially for excited states and if in-medium reductions of the binding energy 
are accounted for~\cite{Grandchamp:2001pf}. The reason is that for a weakly bound state, gluo-absorption occurs on a near on-shell heavy quark, for 
which the phase space is very limited (it vanishes if both incoming and outgoing quarks are on-shell). Instead, inelastic scattering processes on the heavy 
quarks in the bound state take over, \ie, $p +  {\cal Q} \to p+ Q+ \bar{Q}$ involving all thermal partons $p=q, \bar q, g$. 
The temperature where the dominant rate contribution transits from gluo-dissociation to inelastic parton scattering is 
parametrically of the order of the binding energy, $T \sim E_B(T)$, but more quantitatively, for realistic couplings and bound-state kinematics, it is rather 
$T \simeq E_B(T)/2$~\cite{Zhao:2010nk,Du:2017qkv}. More importantly, in the temperature regime where the transition occurs, the rates are small on 
an absolute scale, around 10-20\,MeV, \ie, the pertinent quarkonium lifetimes of 10-20\,fm/$c$ are much larger than the fireball evolution time spent
in this temperature regime in a heavy-ion collision. Thus, the overall relevance of gluo-dissociation is quite limited.  
In the following, we therefore focus on the inelastic parton scattering mechanisms.
Thus far, they have been mostly implemented using a perturbative coupling to the bound state both in momentum~\cite{Grandchamp:2001pf} and 
coordinate space~\cite{Strickland:2011aa,Krouppa:2017jlg}. In principle, the full dynamics of the $2\to3$ process requires a three-body phase. However, 
taking advantage of the HQ mass, one can factorize the process into ``quasifree" scattering of the thermal partons on one of the heavy quarks and 
incorporate the quarkonium binding energy, $E_B$, through the off-shellness of the incoming heavy quark while enforcing four-momentum conservation 
(further corrections are of higher order in $E_B/m_Q$). In this approximation, the reaction rate can be directly written in terms of the half off-shell
heavy-light scattering amplitudes, ${\cal M}_{p\tilde{Q}\to pQ}$~\cite{Du:2019xqq}, 
\beq
\Gamma^{\rm qf}_{\cal Q}(p_{\cal Q};T) = \frac{2}{E_{\cal Q}} \sum\limits_{p=q,\bar q, g}  \int d^3\bar{p}_i d^3\bar{p}_f d^3\bar{p}_{Q,f}
\overline{|{\cal M}_{p\tilde{Q}\to pQ}|^2} (2\pi)^4 \delta^{(4)}(P_{\rm in}-P_{\rm out}) d_p f_p(E_{p_i}) [1\pm f_p(E_{p_f})]
\label{Gam-qf-1}
\eeq
where we defined the Lorentz-invariant phase space elements as $d^3\bar{p} \equiv d^3p/(2\pi)^3 2E_p$, $f_p$ are thermal-parton distribution 
functions, $d_p$ is the spin-color-flavor degeneracy of the incoming thermal parton, and an overall factor of 2 accounts for the scattering off both 
the heavy quark and antiquark in $\cQ$. This expression can be written in a more compact form using the inelastic quasifree $2\to2$ cross section, 
\begin{equation}
\Gamma^{\rm qf}_{\cQ}(p_{\cQ};T)= 2  \sum\limits_{p=q,\bar q, g} \int \frac{d^3p_i}{(2\pi)^3} \sigma(s) v_{\rm rel}  
d_p f_p(E_{p_i}) [1\pm f_p(E_{p_f})] \ ,
\label{Gam-qf-2}
\end{equation}
although the definition of a cross section can be problematic in medium. Employing quantum many-body theory, the quarkonium width from 
quasifree dissociation
can be obtained in a more rigorous form in terms of the imaginary part of the quarkonium selfenergy, 
\begin{equation}
{\rm Im} \Sigma_\cQ(p_{\cQ};T)= 2 \sum\limits_{p=q,\bar q, g} \int \frac{d^3p_i}{(2\pi)^3} {\rm Im} {\cal M}_{pQ\to pQ} \  
[f_p(E_{p_i}) \mp f(E_{p_i}+E_{p_{Q,i}})] \ ,
\label{Gam-qf-3}
\end{equation}
where  ${\rm Im} {\cal M}_{pQ\to pQ}$ denotes the imaginary part of the {\em forward}-scattering amplitude (which in the vacuum is related to the total
cross section via the optical theorem); the quarkonium width follows as $\Gamma_\cQ = - {\rm Im}\Sigma_\cQ/E_\cQ$. Note that in the NLO calculation 
of quasifree dissociation, the amplitude corresponds to the LO heavy-light Born amplitudes displayed in Fig.~\ref{fig_HQpQCD}, which do not possess an 
imaginary part, and thus require the use of Eq.~(\ref{Gam-qf-1}) or (\ref{Gam-qf-2}).
However, Eq.~(\ref{Gam-qf-3}) can be straightforwardly evaluated from the resummed in-medium $T$-matrices, although such a calculation has not 
been done yet. Alternatively, one can extract the quarkonium widths directly from the bound-state widths in the pertinent spectral functions, recall 
Fig.~\ref{fig_corr}. That figure also illustrates the importance of interference effects in the scattering off the heavy quark and antiquark in the bound 
state. These are, a priori, not included in Eqs.~(\ref{Gam-qf-1})-(\ref{Gam-qf-3}). In Ref.~\cite{Du:2017qkv} a pertinent interference factor, 
$(1-{\rm e}^{i\vec q \cdot\vec r})$ has been implemented based on the original work of Ref.~\cite{Laine:2006ns}, where $\vec q^2$ is the momentum 
transfer in the heavy-light quark scattering amplitude. For example, for a vacuum $\Upsilon(2S)$ binding energy of 0.5 GeV at $T$=300\,MeV, the 
interference effect reduces the quasifree dissociation rate by almost a factor of $\sim$5 (cf.~Fig.~2 in Ref.~\cite{Du:2017qkv}), which is roughly 
comparable to what is found for the in-medium $\Upsilon(1S)$ spectral function at $T$=320\,MeV in the upper right panels of Fig.~\ref{fig_corr}.

\begin{figure}[!t]
\centering
 \includegraphics[width=0.335\textwidth]{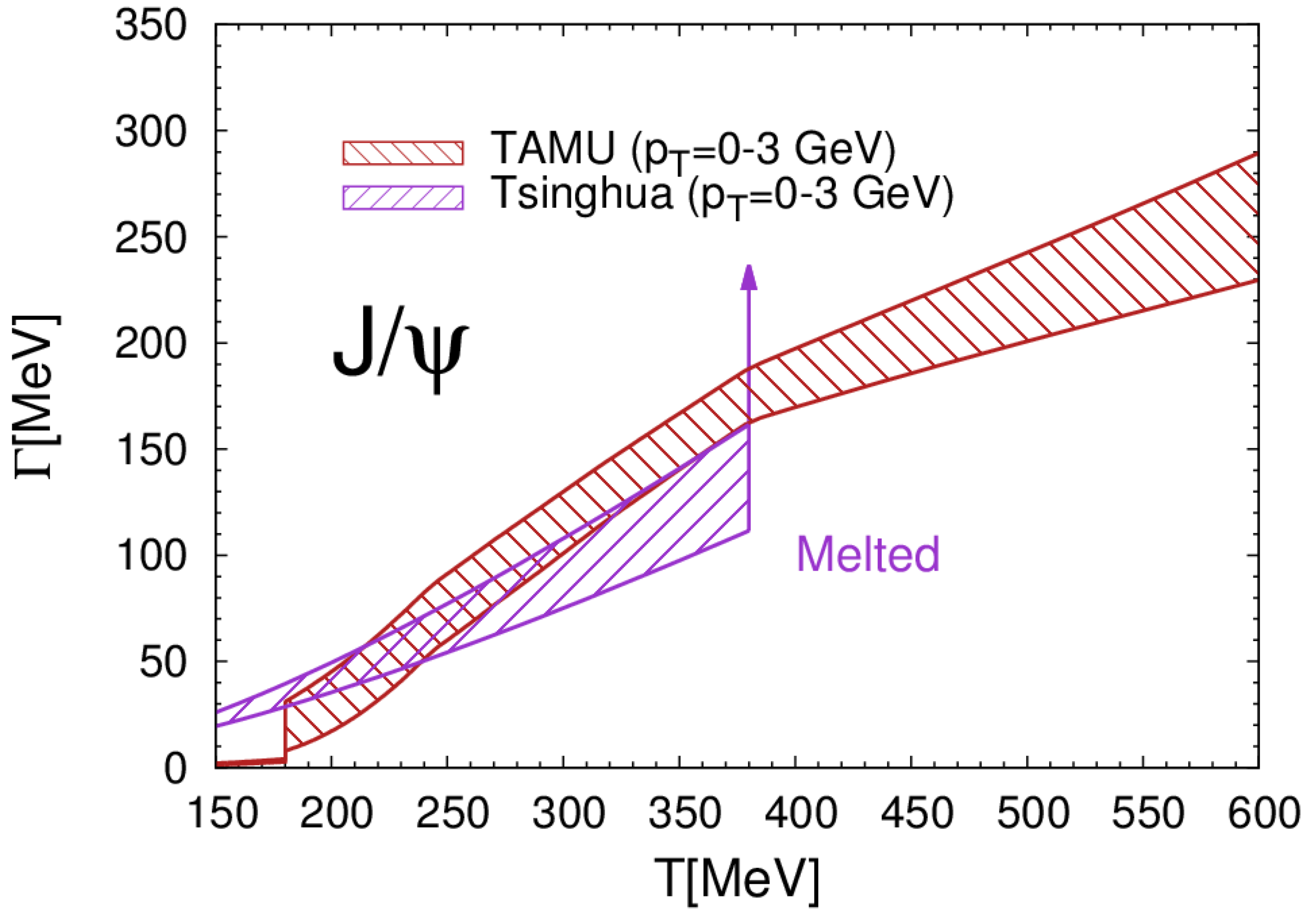}
\hspace{-0.28cm} 
 \includegraphics[width=0.335\textwidth]{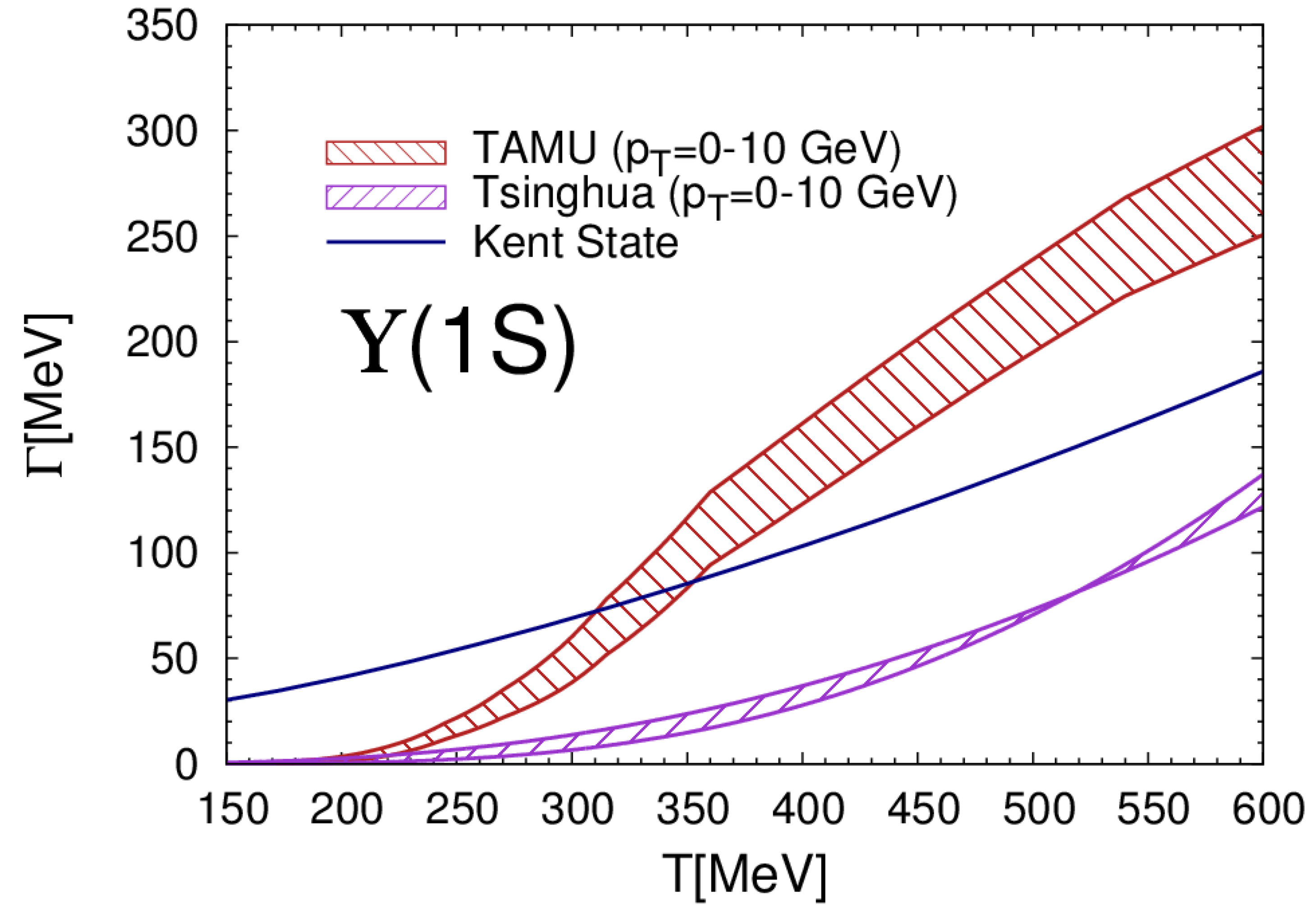}
\hspace{-0.28cm}
\includegraphics[width=0.333\textwidth]{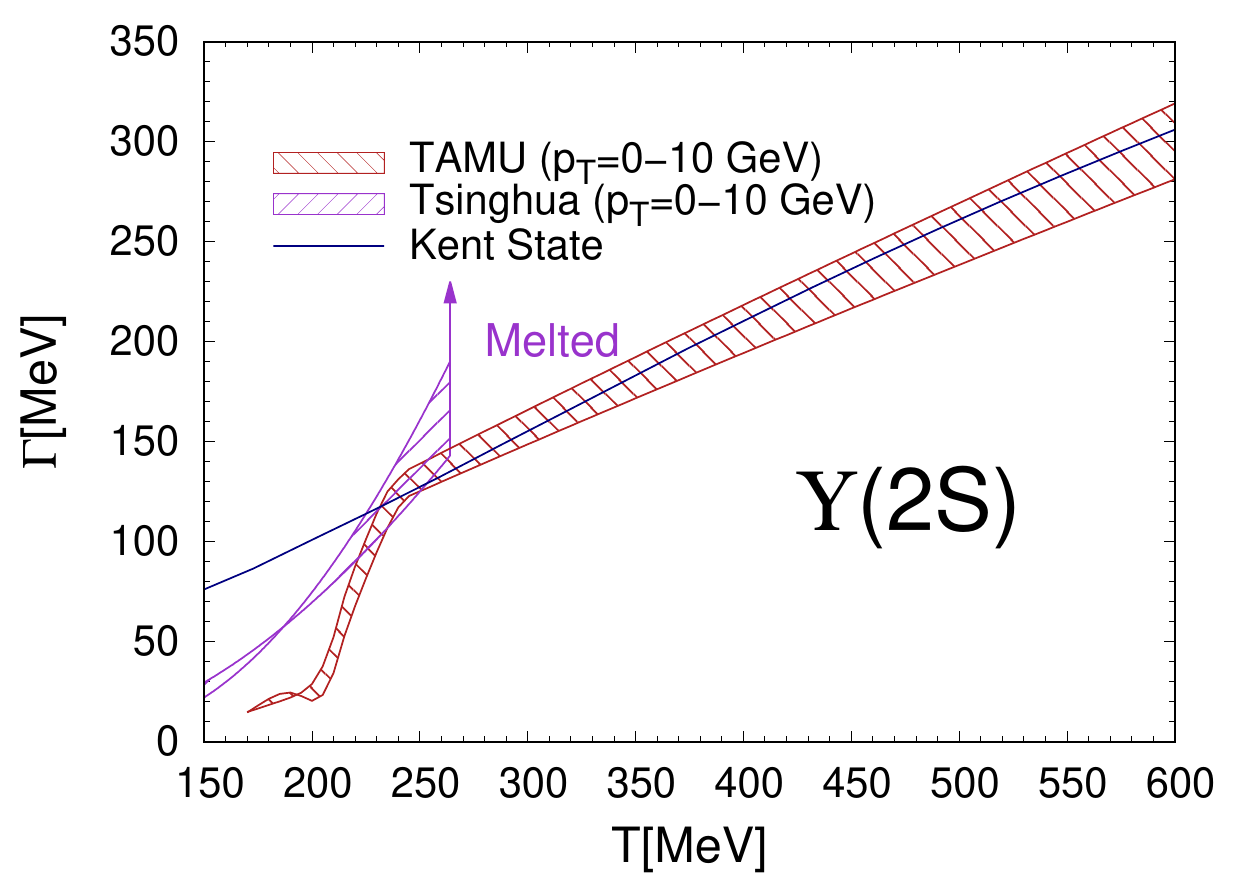} 
\caption{Dissociation rates of $J/\psi$ (left panel), $\Upsilon(1S)$ (middle panel) and $\Upsilon(2S)$ (right panel) in the QGP as calculated from the 
Kent State (blue lines), Tsinghua (purple bands) and TAMU (maroon  bands) groups. For the latter two, the bands reflect the indicated ranges in the
3-momentum of the quarkonium state; figures taken from Ref.~\cite{Rapp:2017chc}.}
\label{fig_gam-psi-ups}
\end{figure}
In Fig.~\ref{fig_gam-psi-ups} we show examples of charmonium and bottomonium dissociation rates from three different groups that have been used 
in the description of experimental data in heavy-ion collisions at RHIC and the LHC. The Kent State results~\cite{Strickland:2011aa} utilize an in-medium 
HQ potential in coordinate space with an imaginary part that includes interference effects. The TAMU results~\cite{Zhao:2010nk,Du:2017qkv} are based
on quasifree dissociation rates with LO heavy-light scattering amplitudes employing in-medium binding energies from the $T$-matrix approach with an 
internal-energy potential~\cite{Riek:2010fk}; interference effects are included for the bottomonium states. 
The Tsinghua results~\cite{Zhou:2014kka,Liu:2010ej} utilize gluo-dissociation reactions with vacuum binding energies for $\Upsilon$ and $J/\psi$ states, 
augmented with a geometric scaling for excited charmonia with an in-medium radius, and dissociation temperatures taken from potential model 
calculations with an internal-energy potential~\cite{Satz:2005hx}. 
Overall, the rates are in rather reasonable agreement, especially for the $J/\psi$ and 
$\Upsilon(2S)$\footnote{We found that the authors of Ref.~\cite{Rapp:2017chc} made a mistake of a factor of 10 in plotting the Tsinghua results
for the $\Upsilon(2S)$ rate in their paper which has been corrected here.}, although the underlying assumptions differ considerably. 
A better understanding of the ingredients and their applicability is thus an 
important task for future studies, in particular also with regard to nonperturbative effects.

\subsection{Spatial Diffusion Coefficient and Shear Viscosity}
\label{ssec_Ds}

An important objective of HF probes of QCD matter is to put their findings into a broader context of the research on the QGP and its 
hadronization. In particular, the HF diffusion coefficient characterizes the long-wavelength properties of the transport of the heavy-flavor quantum number 
through QCD matter. It can be defined through the zero-momentum value of the drag coefficient discussed in Sec.~\ref{ssec_Ap}, 
\beq
{\cal D}_s = \frac{T}{A(p=0) m_Q} \ .
\label{Ds}
\eeq 
Note that this definition implies that the leading HQ mass dependence in $A\propto 1/m_Q$ is divided out. This is one of the reasons that the HF diffusion 
coefficient contains generic information on the QCD medium and is therefore believed to be closely related to its other transport coefficients, \eg,  the shear 
viscosity or electric conductivity. To facilitate their comparison and  interpretation as a (temperature-dependent) interaction strength of the medium, they 
are commonly scaled to dimensionless quantities,  \ie, by the  inverse ``thermal wavelength" of the medium ($1/2\pi T$) for the HF diffusion coefficient, 
$\cD_s(2\pi T)$, by the entropy density ($s$) for the shear viscosity, $\eta/s$, or by temperature for the electric conductivity, $\sigma_{\rm el}/T$. 
In perturbation theory, these quantities are proportional to the inverse coupling constant squared, $1/\alpha_s^2$. It might therefore be expected 
that such a connection persists in the nonperturbative regime, \ie, that the microscopic in-medium interactions that drive HF diffusion also govern the 
transport of energy-momentum, electric charge, and other conserved quantities. This is further supported by the general definition of transport coefficients 
as the zero-energy limit of pertinent spectral functions (see, \eg, Ref.~\cite{Aarts:2002cc}), where the width of the transport peak (after dividing by the 
energy) essentially reflects the collisional broadening of the carriers of the pertinent quantum number (and the transport coefficient itself is essentially proportional to the inverse width).  An example of comparing the HQ diffusion from both the the zero-momentum limit of the friction coefficient, 
Eq.~(\ref{Ds}), to the zero mode of the quarkonium spectral function has been elaborated in Ref.~\cite{Riek:2010py}.

\begin{figure}[t]
\centering
\begin{minipage}{0.6\textwidth}
\includegraphics[width=0.97\textwidth]{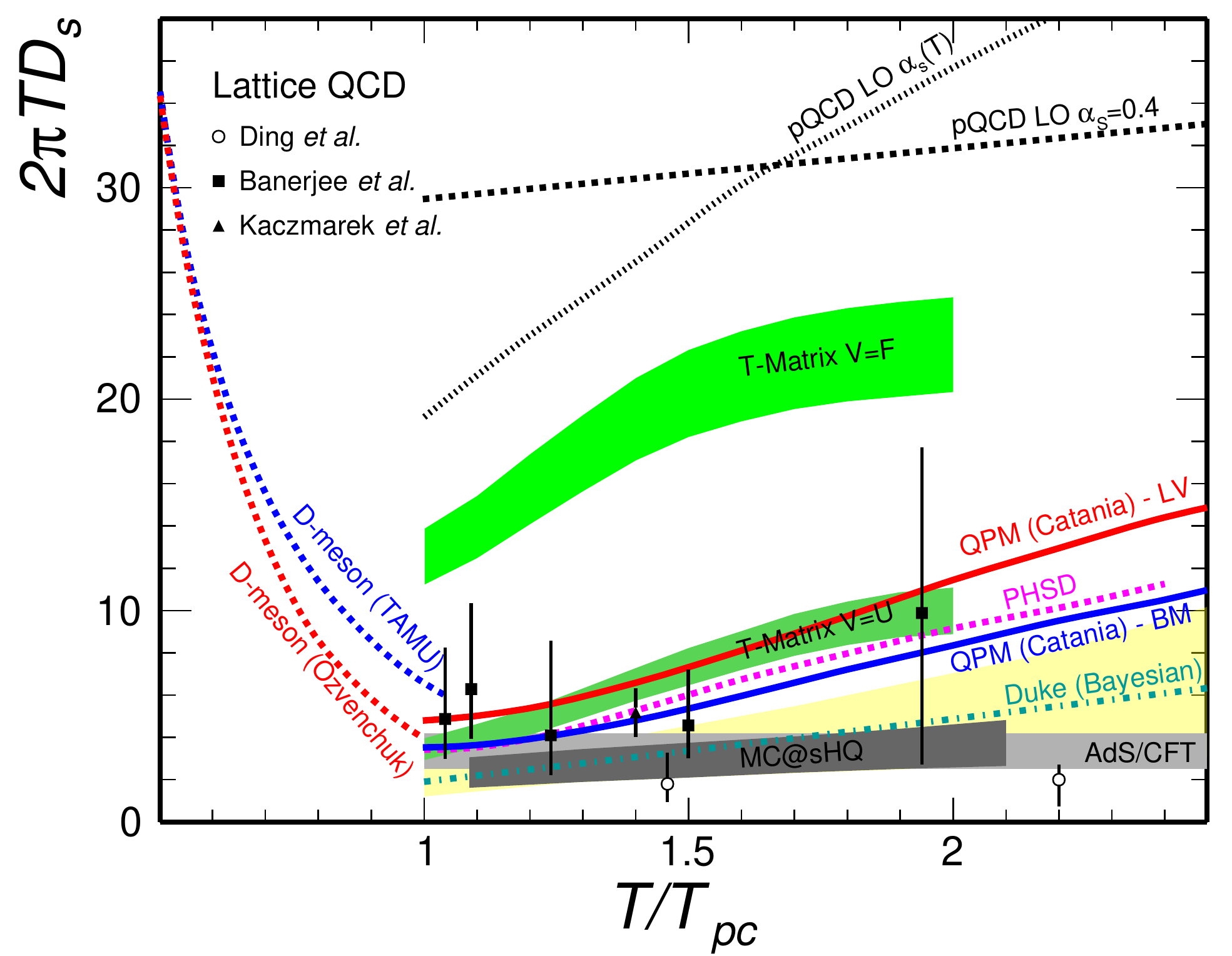}
\end{minipage}
\caption{A recent compilation of charm-quark diffusion coefficients in the QGP from various approaches: quenched lQCD~\cite{Banerjee:2011ra,Ding:2012sp,Kaczmarek:2014jga}, pQCD Born diagrams with either fixed~\cite{Svetitsky:1987gq,Rapp:2009my} or temperature-dependent 
coupling~\cite{Moore:2004tg}, QPM using using either a Langevin or Boltzmann-transport extraction~\cite{Das:2015ana}, DQPM~\cite{Song:2015sfa}, 
$T$-matrix approach with either free- or internal-energy potential~\cite{Riek:2010fk}, pQCD*~\cite{Gossiaux:2008jv,Andronic:2015wma}, 
AdS/CFT correspondence~\cite{Horowitz:2015dta}, Bayesian fits to HF data~\cite{Xu:2017obm}, and hadronic calculations from 
Refs.~\cite{He:2011yi,Ozvenchuk:2014rpa} discussed in Sec.~\ref{ssec_had};
 figure taken from Ref.~\cite{Dong:2019byy}.}
\label{fig_Ds}
\end{figure}
An overview of recent calculations of the temperature dependence of the HF diffusion coefficient in various theoretical approaches is shown in 
Fig.~\ref{fig_Ds}, adopted from the rather recent review in Ref.~\cite{Dong:2019byy}. From current phenomenological applications to charm-quark
observables in heavy-ion collisions, in particular $D$-meson nuclear modification factors and elliptic flow as discussed in Sec.~\ref{ssec_hf-obs},
the values of $\cD_s(2\pi T)$ have been estimated to be in a range of 1.5-4.5 in the QGP near $\Tpc$ , with most model calculations featuring a 
gradualincrease with temperature suggesting a decrease in the ``coupling strength" of the medium (especially the large magnitude observed for 
the elliptic flow requires a large coupling strength close to $\Tpc$, cf.~Sec.~\ref{ssec_hf-obs}).  This is about a factor of $\sim$10 smaller 
than the results of pQCD calculations with a fixed Debye mass and a coupling constant of $\alpha_s$=0.4. It is also much smaller than 
the results from the  thermodynamic $T$-matrix when using a free-energy potential from lQCD data (upper green band in Fig.~\ref{fig_Ds}). 
On the other hand, when using the internal-energy potential, augmented with a moderate $K$ factor of $\sim$1.5, a rather good description 
of the $D$-meson observables at the LHC has been  found~\cite{He:2019vgs}, cf.~\ref{ssec_hf-obs} (part of the $K$ factor may be resolved 
when going from the $U$-potential to the SCS potential, as discussed in Sec.~\ref{ssec_Ap}, recall 
Figs.~\ref{fig_Ap-tmat-qp} and\ref{fig_Ap-tmat-scs}). It is important to note that the momentum dependence of the friction coefficient
plays an important role in the phenomenological extraction of the diffusion coefficient. For example, the Subatech approach (labeled ``MC@sHQ" in 
Fig.~\ref{fig_Ds}) has a rather soft 3-momentum dependence in $A(p)$ (\ie, falling off rather quickly), relative to the QPM, 
recall~Fig.~\ref{fig_Ap-models}, which leads to an extraction of a smaller $\cD_s$ than for the QPM.  A more systematic investigation of this correlation 
has been reported in Ref.~\cite{HF-ect21}. Also shown in Fig.~\ref{fig_Ds} are results for $\cD_s$ in the hadronic phase, \ie, for $D$-mesons. 
While these calculations carry an uncertainty that is probably a bit larger than for charm quarks in the QGP (cf.~the discussion in Sec.~\ref{ssec_had}), 
the results are quite suggestive for both a minimum structure in the vicinity of $\Tpc$ and a continuous transition from 
the confined to the deconfined medium (as to be expected for a cross-over transition). The minimum feature, in particular, would further corroborate
that the relevant interactions are closely related to the phase change, \ie, forces that ultimately lead to confining the quarks and gluons into hadrons 
(effects due to chiral symmetry breaking are expected to be less relevant for heavy quarks).

\begin{figure}[t]
\centering
 \begin{minipage}{0.47\textwidth}
 \includegraphics[width=0.97\textwidth,height=0.8\textwidth]{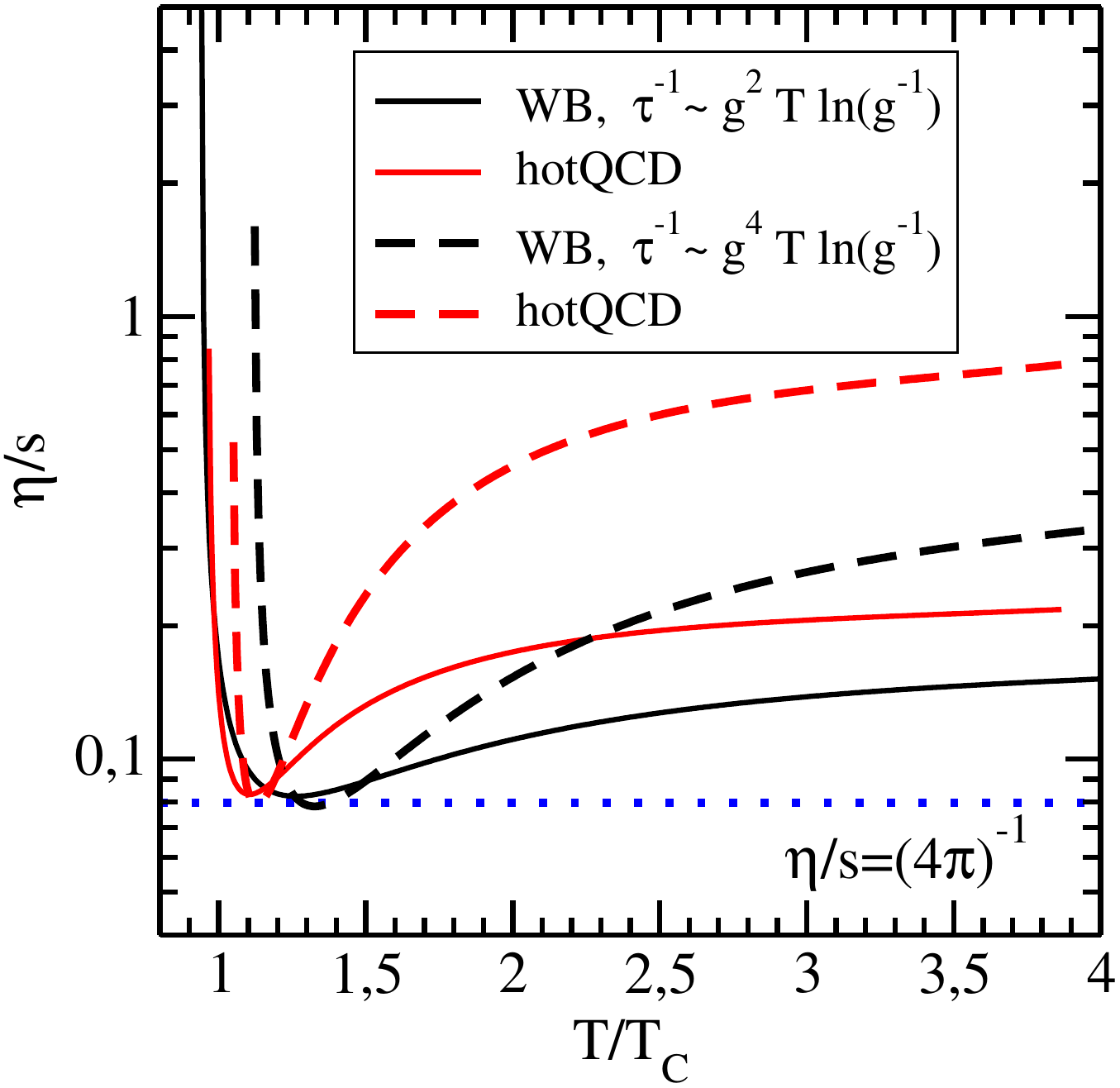}
\end{minipage} 
\hspace{0.3cm}
\begin{minipage}{0.47\textwidth}
\includegraphics[width=0.97\textwidth]{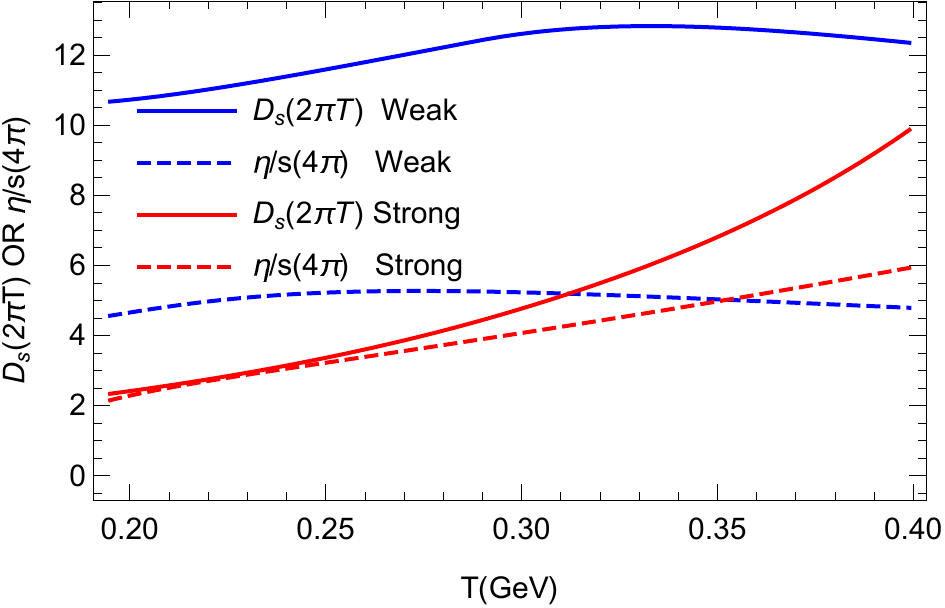}
\end{minipage}
\caption{Ratio of shear viscosity over entropy density in the quasiparticle model (left panel; figure taken from Ref.~\cite{Plumari:2011mk} and the 
thermodynamic $T$-matrix approach (right panel; figure taken from Ref.~\cite{Liu:2016ysz}).
}
\label{fig_eta-s}
\end{figure}
As stated above, it is of great interest to study the relations of the HF transport coefficient to other transport properties of the QCD medium. 
Here, we will focus on the approaches that can compute the shear viscosity (and EoS) within the same framework as $\cD_s$, paralleling the
discussion in Sec.~\ref{ssec_med}.
For the QPM~\cite{Plumari:2011mk}, the coupling constant extracted from the fit to the EoS has been injected into pQCD ansatz for the 
viscosity~\cite{Arnold:2003zc}, 
\beq
\eta = \frac{\eta_1 T^3}{g^4 \ln\left( \frac{\mu_*}{gT} \right)}  \ ,
\eeq
where the parameters $\eta_1$ and $\mu_*$ have been to constrained to recover the high-temperature pQCD limit and requiring the minimum
of the resulting $\eta/s (T)$ to reach $1/4\pi$. The predictive power then lies in the temperature dependence of this quantity, shown by the dashed lines
in the left panel of Fig.~\ref{fig_eta-s}. Alternatively, an effective $1/g^2$ dependence has been considered (motivated by a relaxation time that scales with the collisional width~\cite{Peshier:2004bv,Khvorostukhin:2010aj}), which leads to a significantly weaker temperature dependence (solid lines in the left
panel of Fig.~\ref{fig_eta-s}). In the $T$-matrix approach, the viscosity has been computed using the Kubo formula~\cite{Liu:2016ysz},
\beq
\eta=\lim\limits_{\omega\rightarrow 0}\sum_i\frac{\pi d_i}{\omega} \int \frac{d^3\textbf{p}d\lambda}{(2\pi)^3} \frac{p_x^2p_y^2}{\omega^2_i(p)}
\rho_i(\omega+\lambda,p)\rho_i(\lambda,p)  [f_i(\lambda)-f_i(\omega+\lambda)]  \  , 
\eeq
which directly employs the thermal-parton spectral functions from the selfconsistent solutions of the WCS or SCS (recall Fig.~\ref{fig_eos-tmat}).
The pertinent $\eta/s$ ratio is plotted in units of $1/4\pi$ in the right panel of Fig.~\ref{fig_eta-s}.  In the WCS,
the $\eta/s$ ratio reaches down to twice the conjectured lower bound at low QGP temperatures, rising by a factor of $\sim$3 at $T$=400\,MeV (or 
$\sim$2.5$\Tpc$). In the WCS, the ratio is about a factor of 2 larger at low $T$, but does not rise much with temperature. This is quite different from
the (scaled) HQ diffusion coefficient, which differs by a factor of 5 between the WCS and SCS at low temperature, while at $T$=400\,MeV they are 
comparable. This suggests that the absolute value of $\cD_s(2\pi T)$ is a better discriminator between a strongly and a weakly coupled system than
 $\eta/s$.  In Ref.~\cite{Rapp:2009my}, the double ratio,  $[2\pi T\cD_s)] / [\eta/s]$, has been proposed as a quantitative
measure of the strong-coupling nature of the QGP, varying between values of 2.5 for a perturbative scenario and as low as 1 from the strong-coupling
limit of conformal field theory~\cite{Policastro:2002se}. Evaluating this ratio for the WCS $T$-matrix results shown in Fig.~\ref{fig_eta-s} right, 
one finds $\sim$1 for the lowest $T$ and approximately 2.5 at the highest $T$, while for the WCS the ratio is approximately constant 
at 2.5~\cite{Liu:2016ysz}.


\newpage

	\section{Open Heavy-Flavor Phenomenology}
  \label{sec_open}
In this chapter we discuss the transport of charm and bottom quarks in hot QCD matter and its applications to URHICs at RHIC and the LHC. 
In Sec.~\ref{ssec_trans-app}, we will briefly review the Fokker-Plank/Langevin approach to implement the diffusion of HF particles based on
transport coefficients as discussed in Sec.~\ref{ssec_Ap}. In particular, we will recall comparisons between Langevin and Boltzmann 
simulations of HQ motion in an expanding QGP, and elaborate on the role of quantum effects that go beyond the semiclassical Boltzmann 
framework but can still be included in the transport coefficients of the  Langevin approach. In Sec.~\ref{ssec_hf-obs} we turn to a 
quantitative discussion of open HF observables and their comparisons to theoretical model calculations, focusing on recent LHC 
measurements of $D$-mesons, non-prompt $D$-mesons arising from $B$-hadron decays, $D_s$ mesons and $\Lambda_c$ baryons.

\subsection{Transport Approaches}
\label{ssec_trans-app}

In the context of the transport parameters discussed in Sec.~\ref{ssec_Ap}, we have already referred to the Fokker-Planck equation as a suitable 
framework to define these coefficients. As is well-known, the Fokker-Planck equation can be derived from the relativistic Boltzmann 
equation~\cite{DeGroot:1980dk,Cercignani:2002} for the HQ distribution function, $f_Q(x,\bvec{p})$, 
\begin{equation}
\label{trans.1}
\frac{1}{E_{\bvec{p}}} p^{\mu} \partial_{\mu} f_Q(x,\bvec{p}) =\int \dd^3 \bvec{k} [w(x,\bvec{p}+\bvec{k},\bvec{k})
f_Q(x,\bvec{p}+\bvec{k}) - w(x,\bvec{p},\bvec{k}) f_Q(x,\bvec{p}) ]\ ,
\end{equation}
in the limit of  small momentum transfers from the medium to the heavy quark, $k \sim T \ll m_Q$.
As first noted in Ref.~\cite{Svetitsky:1987gq}, this is well satisfied for the motion of a heavy quark in the QGP due to the HQ mass 
being much larger than the typical temperatures under conditions relevant for high-energy heavy-ion collisions. 
In the above equation, $x=(t,\bvec{x})$ denotes the HQ space-time coordinate and $E_{\bvec{p}}=\sqrt{m_Q^2 + \bvec{p}^2}$ its on-shell energy. 
The collision term is given by the transition rate for a heavy quark of momentum $\bvec{p}$ being scattered to momentum 
$\bvec{p}-\bvec{k}$ through collisions with the light quarks and gluons in the QGP,
\begin{equation}
\label{trans.4}
w(x,\bvec{p},\bvec{k}) = \frac{1}{64 \pi^2} \int \frac{\dd^3
  \bvec{q}}{(2 \pi)^3} \frac{1}{g_Q} \sum |\mathcal{M}_{iQ}|^2 \frac{1}{E_{\bvec{p}}
  E_{\bvec{q}} E_{\bvec{p}-\bvec{k}} E_{\bvec{q}+\bvec{k}}}
  \delta(E_{\bvec{p}} + E_{\bvec{q}} - E_{\bvec{p}-\bvec{k}} -
  E_{\bvec{q}+\bvec{k}}) f_X(x,\bvec{q}) \  .
\end{equation}
Here, $\mathcal{M}_{iQ}$ is the invariant amplitude for the HQ elastic scattering off $i \in \{q,g \}$, and the sum
is over spin, flavor, and color in the initial and final states.
Expanding the integrand up to $2^{\text{nd}}$ order in $\bvec{k}$, one obtains
\begin{equation}
\label{trans.6}
\frac{1}{E_{\bvec{p}}} p^{\mu} \partial_\mu f_Q(x,\bvec{p}) =
\frac{\partial}{\partial p_i} [A_i(x,\bvec{p}) f_Q(x,\bvec{p})] +
\frac{\partial^2}{\partial p_i \partial p_j} [B_{ij}(x,\bvec{p})
f_Q(x,\bvec{p})] \ .
\end{equation}
The momentum drag and diffusion coefficients, $A_i$ and $B_{ij}$, respectively, are defined in the
local rest frame of the medium as
\begin{equation}
\label{trans.7}
A_i(x,\bvec{p}) = \int \dd^3 \bvec{k} w(x,\bvec{p},\bvec{k}) k_i \; , \quad
B_{ij} = \frac{1}{2} \int \dd^3 \bvec{k} w(x,\bvec{p},\bvec{k}) k_i k_j \ .
\end{equation}
For an isotropic medium in local thermal
equilibrium, one has $A_i(x,\bvec{p}) = A(x,p) p_i$ and $B_{ij}(x,\bvec{p}) = B_0(x,p)
P_{ij}^{\perp}(\bvec{p}) + B_1(x,p) P_{ij}^{\parallel}(\bvec{p})$
with $p=|\bvec{p}|$ and the projection operators onto the longitudinal and transverse direction of $\bvec{p}$: 
$P_{ij}^{\parallel}=p_i p_j/p^2$ and $P_{ij}^{\perp}=\delta_{ij} - P_{ij}^{\parallel}(\bvec{p})$.  This
defines the drag and diffusion coefficients, $A$, $B_0$, and $B_1$, discussed in Sec.~\ref{ssec_Ap}.
As in the non-relativistic case~\cite{risken1989fpe}, the drag coefficient is related to the diffusion 
coefficients by an Einstein dissipation-fluctuation relation, $B=TE_pA$, which ensures that in the
long-time limit the phase-space distribution function converges to the
Maxwell-Boltzmann distribution, \ie, it properly describes the equilibration of the heavy quark with the 
ambient medium. The generalization to momentum-dependent coefficients and relativistic motion of the heavy
quarks can be found in Refs.~\cite{Rapp:2009my,Dunkel:2008ngc,Beraudo:2009pe}.

In practice, the Fokker-Planck equation for the HQ phase-space distribution function
is simulated using an equivalent Langevin equation of the form
\begin{equation}
\label{trans.14}
\dd x_j = \frac{p_j}{E_p} \dd t, \quad \dd p_j =-\Gamma(x,\bvec{p}) p_j \dd t +
\sqrt{\dd t} C_{jk}(x,\bvec{k}+\xi \dd \bvec{p}) \rho_k(t) \ , 
\end{equation}
in the local rest frame of the medium, where $\rho_k(t)$ is normal-distributed white noise,
$\erw{\rho_j(t_1)\rho_k(t_2)}=\delta_{jk} \delta(t_1-t_2)$, and
$\xi \in [0,1]$, where $\xi$=0, 1/2 and 1 correspond to the pre-point Ito, mid-point Stratonovic, and post-point Ito
realizations, respectively, of the stochastic integral~\cite{Rapp:2009my}. The relation
to the transport coefficients is given by $C_{jk} = \sqrt{2 B_0} P_{jk}^{\perp} + \sqrt{2 B_1} P_{jk}^{\parallel}$ and 
$A p_j=\Gamma p_j - \xi C_{lk} \frac{\partial C_{jk}}{\partial p_l}$.
\begin{figure}[t]
\centering
\includegraphics[width=0.3 \linewidth]{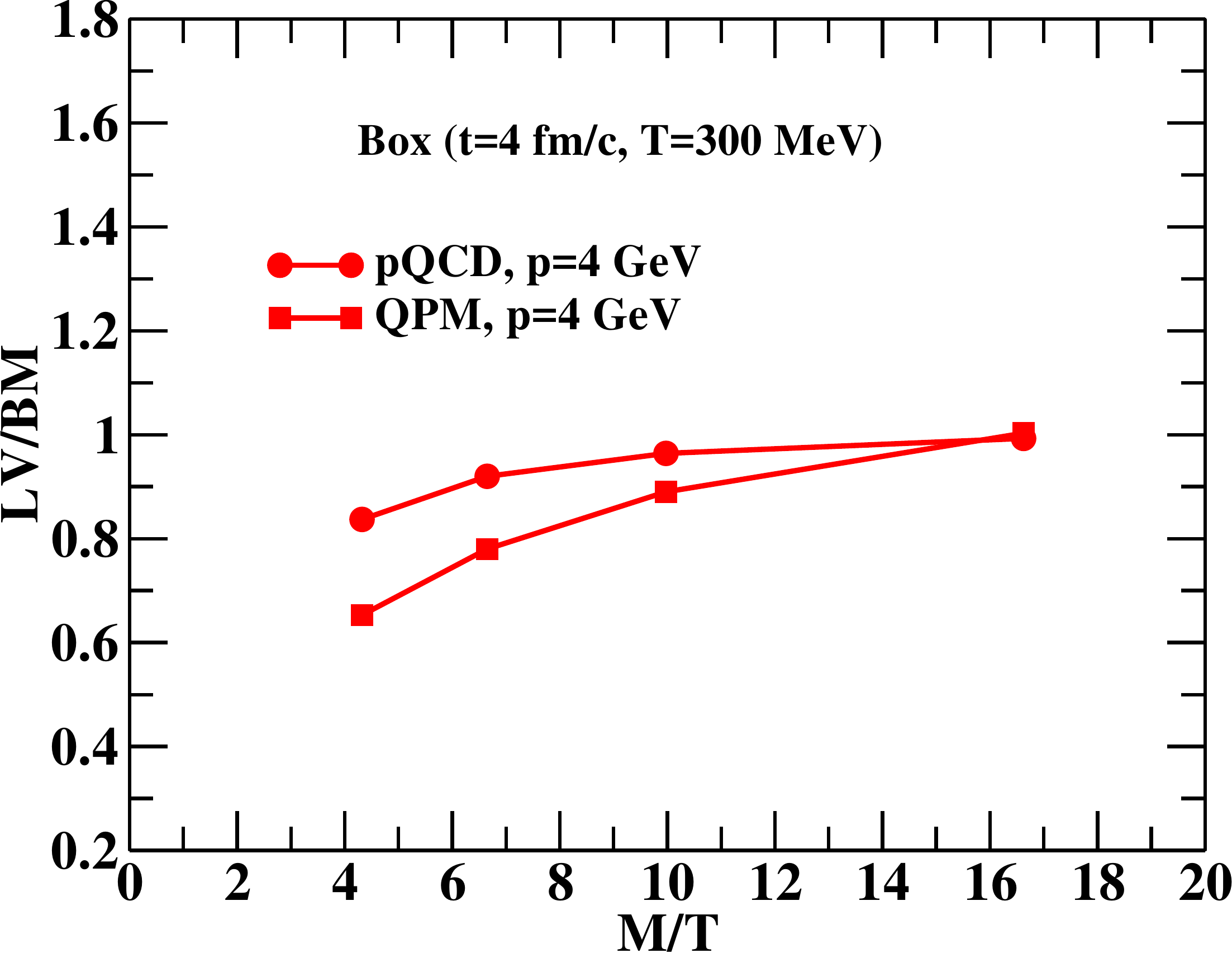}
\hspace*{5mm}
\includegraphics[width=0.3 \linewidth]{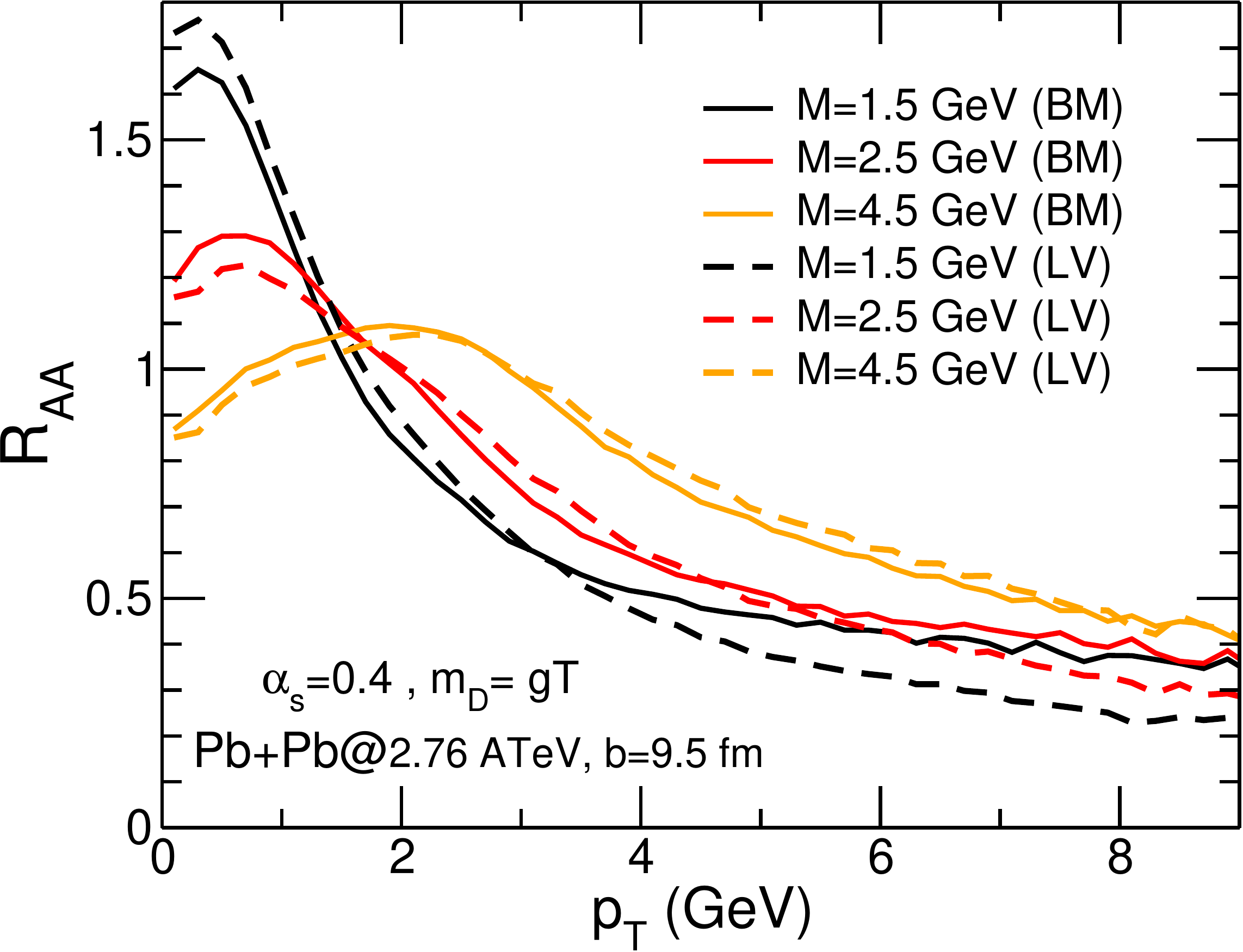}
\hspace*{5mm}
\includegraphics[width=0.3 \linewidth]{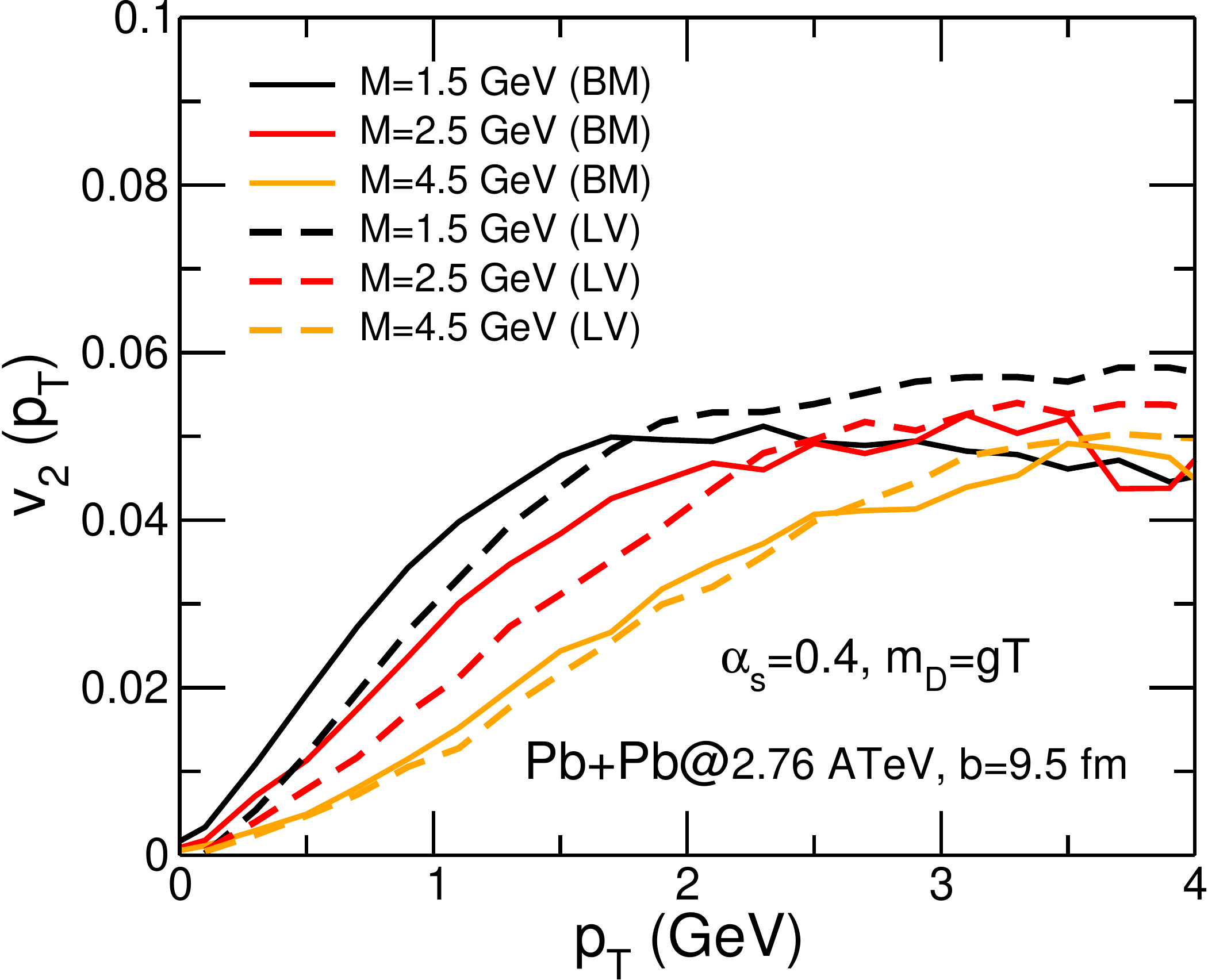}
\caption{Comparison between Boltzmann and a Langevin simulation for heavy quarks of 
different mass. Left panel: The ratio of the nuclear modification factor, $\RAA$, for HQ
  diffusion in a box at temperature, $T=300 \; \MeV$, at a HQ  momentum of $p=4 \;\GeV$. 
Middle and right panels: Nuclear
  modification factor, $\RAA$, and elliptic flow, $v_2$, from 
  simulations of a semi-central Pb+Pb collision at
  $\sqrt{s_{NN}}=2.76 \; \TeV$ in Boltzmann transport
  (solid lines) in comparison to a Langevin description for the heavy
  quarks. Figure taken from Ref.~\cite{Rapp:2018qla}.}
\label{fig_Boltz-Lang}
\end{figure}
The dissipation-fluctuation relation is not guaranteed to be satisfied within microscopic model calculations for HQ scattering in the medium, 
as discussed, \eg,  in Ref.~\cite{Rapp:2018qla}. However, from direct comparisons
between numerical realizations of the Boltzmann transport and the Langevin/Fokker-Planck equations it turns out that a 
good strategy  is to use $\Gamma=A$ and $B_0$ from the microscopic model and set $B_1=A ET$ in the post-point Ito 
realization of the Langevin process, Eq.~(\ref{trans.14})~\cite{Das:2013kea,Rapp:2018qla}. 
In the latter works,  quantitative comparisons of Boltzmann and Langevin simulations have been carried out for heavy quarks of different
masses using transport coefficients evaluated from two microscopic models: 
(a) pQCD Born diagrams with a Debye screening mass $m_{\text{D}}= g T$, a coupling constant $\alpha_{\mathrm{s}}=g^2/4 \pi=0.4$, 
and an additional $K$-factor of with massless light quarks and gluons, and 
(b) the QPM discussed in Secs.~\ref{ssec_med} and \ref{ssec_Ap} with massive partons of $m_g=0.69 \; \GeV$ and $m_q=0.46 \;
\GeV$.  
First, in a box of thermal QGP with fixed temperature $T$=300\,MeV, the nuclear modification factor has been evaluated at a HQ momentum
of $p$=4\,GeV and plotted as a function of mass in the left panel of Fig.~\ref{fig_Boltz-Lang}. For the case of massless thermal partons, the 
deviations are quite small, below 10\% for HQ masses of 1.8\,GeV and higher, \ie, for $m_Q/T\gtsim6$, while for the QPM they are somewhat
larger due to the larger momentum transfers imparted on the heavy quark in their scattering off the masssive  thermal quasiparticles, caused 
by more isotropic cross sections compared to the forward-peaked ones in the massless pQCD model. 
However, for bottom-quarks, both Boltzmann and Langevin approaches lead to virtually identical results, for both microscopic models.
In the middle and right panel of Fig.~\ref{fig_Boltz-Lang} the comparison is carried out for a more realistic situation where an expanding
fireball for semicentral Pb-Pb(2.76\,TeV) is simulated using a transport model for the bulk medium from the QPM model but with 
HQ transport coefficients evaluated with the pQCD Born diagrams in a massless QGP.  For the nuclear modification factor, $R_{AA}$, the 
agreement is quite good for momenta up to $p_T\simeq 4$\,GeV for the three considered HQ masses, but
at higher $p_T$, the lightest HQ mass result shows significant deviations (\eg, $\sim$30\% at $p_T$=7\,GeV), with the Langevin simulation 
over-predicting the suppression relative to the Boltzmann one. 
For the $v_2$, significant deviations appear at small $p_T \ltsim m_Q/2$ for both $m_Q$=1.5 and 2.5\,GeV, and  again for momenta 
$p_T\gtsim 2m_Q$, while for a bottom-like quark mass close agreement over the entire $p_T$ range is observed.

Concerning the calculation of the drag and diffusion coefficients we note that for the scattering matrix elements, $\mathcal{M}_{iQ}$, in
Eq.~(\ref{trans.4})  the corresponding transition rates from nonperturbative many-body quantum-field theoretical calculations like
the $T$-matrix approach discussed in Sec.~\ref{ssec_tmat} can be used to take into account ``off-shell effects'', particularly the
substantial broadening of the spectral functions in the strongly coupled QGP.
As discussed already in Sec.~\ref{ssec_Ap}, a large impact of the off-shell dynamics of the light constituents of the medium has 
been found~\cite{Liu:2018syc}, which is in large part due to the nonperturbative string interaction and the dynamical formation of
heavy-light-quark resonances. Even though the resonances generally lie below the nominal heavy-light-quark mass threshold,  
the width of the resonances and the inclusion of the width of the spectral functions of the thermal partons (as well as the outgoing 
charm quark) open the sub-threshold phase space and therefore give access to the the resonant interaction strength.
The widths of both the light and heavy quarks, as well as gluons are taken into account selfconsistently in the Luttinger-Ward-Baym 
(or two-particle-irreducible (2PI)) formalism. The importance of the selfconsistent approach also lies in its consistency with the equation 
of state of the medium, which by construction of the model is constrained to fit the corresponding lQCD results~\cite{Liu:2017qah} 
(cf.~Fig.~\ref{fig_eos-tmat}). 

\begin{figure}[!t]
\centering
\includegraphics[width=0.3 \linewidth]{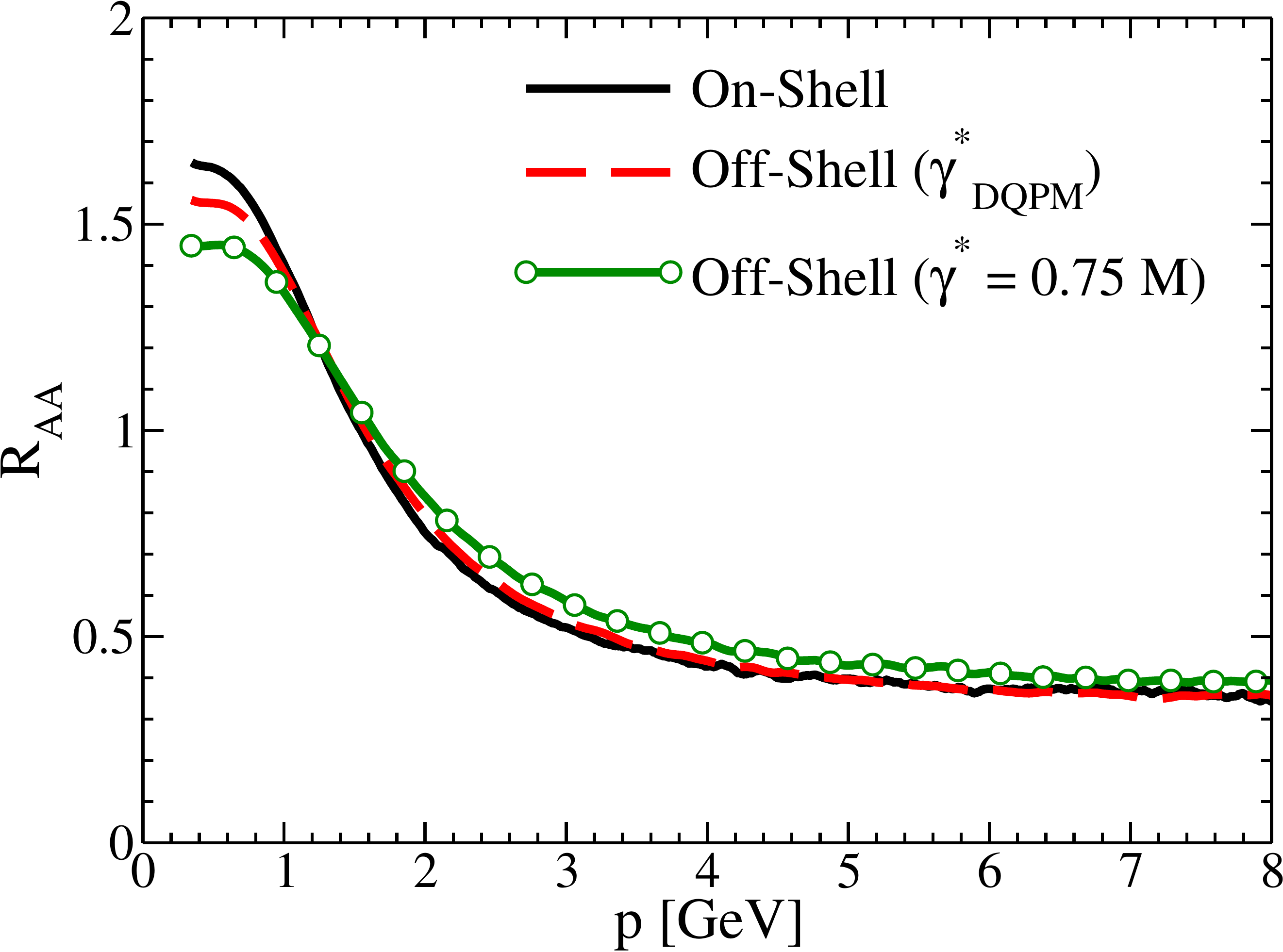}
\hfill
\includegraphics[width=0.335 \linewidth]{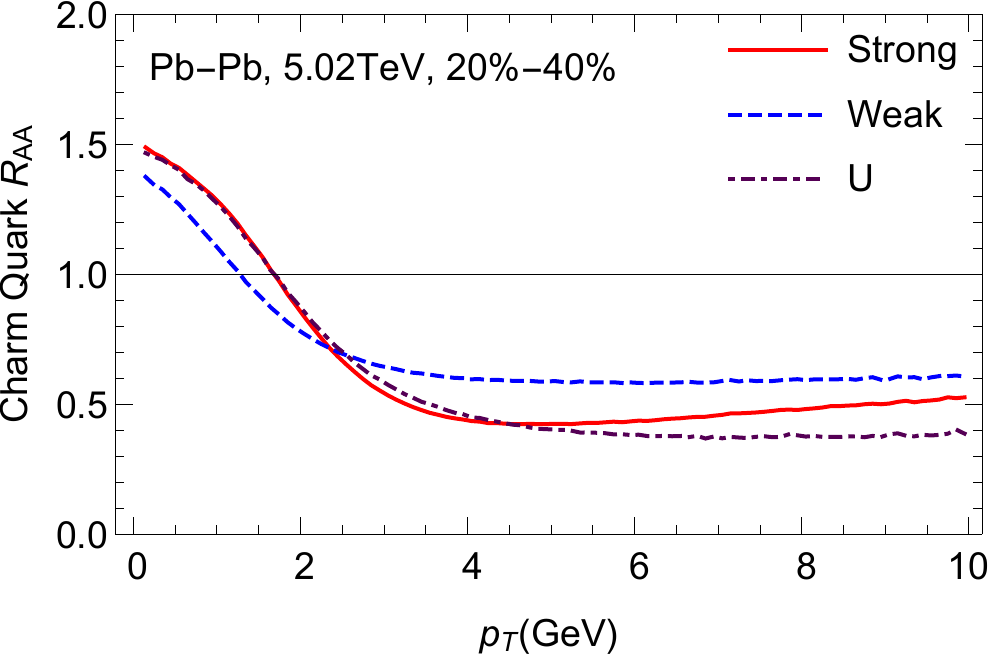}
\hfill
\includegraphics[width=0.335 \linewidth]{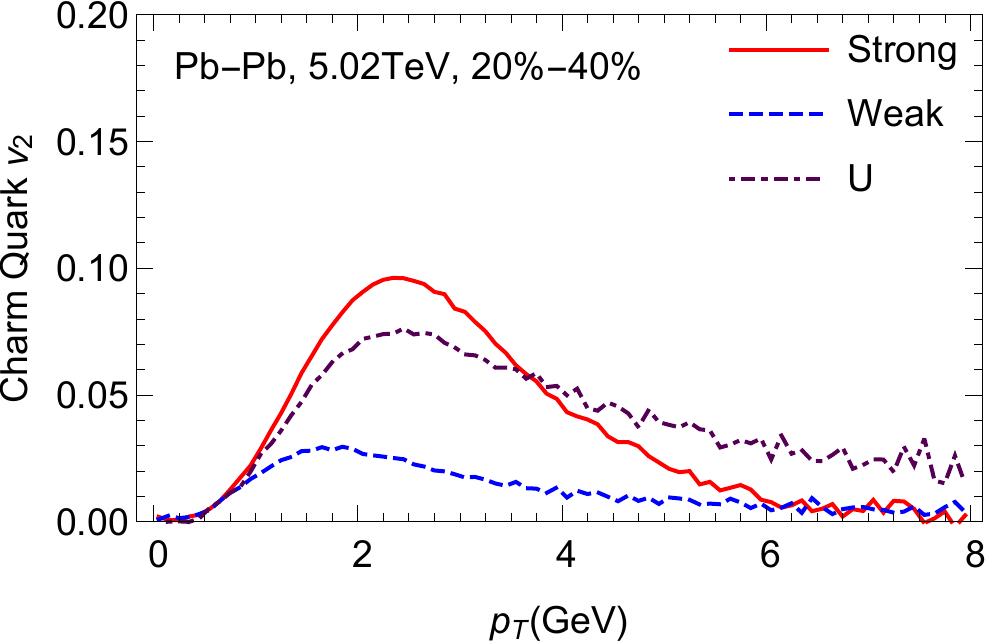}
\caption{Left panel: comparison of the nuclear modification factor, $\RAA$, of charm quarks in an expanding QGP  using  transport 
coefficients from a QPM with on-shell vs.~off-shell kinematics for HQ scattering off thermal partons, where the latter are approximated 
with Breit-Wigner distributions for two different values for their width; figure taken from Ref.~\cite{Sambataro:2020pge}.
Middle and right panel: $\RAA$ and $v_2$, respectively, of $c$-quarks from a Langevin simulation with transport coefficients from a
self-consistent 2PI calculation of the transport coefficients within a weakly- and strongly-coupled scenario, as well as for the internal-energy 
potential within a quasiparticle QGP; figures taken from Ref.~\cite{Liu:2018syc}).}
\label{fig_offshell}
\end{figure}

The significance of these features of the 2PI  approach for HQ transport is has been studies in a more schematic implementation of off-shell
effects conducted in Ref.~\cite{Sambataro:2020pge}, where finite widths for the thermal partons have been introduced into the QPM 
via Breit-Wigner mass distributions in the evaluation of the transport coefficients (using pQCD scattering matrix elements).
It has been found that the width effects only have a small impact on the HQ transport coefficients, actually leading to a slight reduction in the
friction coefficient, $A(p)$. The pertinent results for the nuclear modification factor of charm quarks also show little change even for
thermal-parton widths as large as 75\% of their mass, see left panel of Fig.~\ref{fig_offshell}.  
The middle and right panels of Fig.~\ref{fig_offshell} show the $\RAA$ and $v_2$ from Langevin simulations of $c$-quarks in an 
expanding QGP of Pb-Pb collisions at the LHC with drag and diffusion coefficients evaluated from the selfconsistent 2PI $T$-matrix 
approach~\cite{Liu:2018syc} in the strong- and weak-coupling scenarios (SCS and WCS), recall Sec.~\ref{ssec_med},  as well as taking 
the HQ-internal-energy potential for the heavy-light $T$-matrices. In particular the elliptic flow exhibits an excellent sensitivity to the 
underlying HQ interactions in the QGP.

In addition to a consistent calculation of the  HQ-medium interactions and the QGP EoS, nonperturbative many-body approaches
are promising for establishing further consistency toward the kinetic recombination of heavy with light quarks into hadrons 
(which is not present in, \eg,  in instantaneous coalescence models). The in-medium bound states generated by the two-body
$T$-matrix naturally provide for the formation of (pre-) hadronic states, thus putting the interactions underlying the diffusion 
and hadronziation processes on the same footing. Also for the latter, this will require the use of off-shell quark spectral 
functions, to ensure four-momentum conservation and the correct long-time limits of kinetic and chemical equilibrium.  



\subsection{Observables}
\label{ssec_hf-obs}



The two most common observables to characterize the medium effects on open heavy-flavor probes in AA collisions are the nuclear modification factor,
 $R_{\rm AA}$,  and the elliptic flow, $v_2(p_T)$.  The former is defined as the ratio of the $p_T$-differential particle yield in nucleus-nucleus 
collisions over the production cross section in $pp$ collisions (at the same center-of-mass energy) scaled by the nuclear overlap function for a given
centrality class; it characterizes the spectral modifications due to interactions in the fireball formed in AA collisions. The elliptic-flow coefficient, 
$v_2=\langle {\rm cos}(2\phi)\rangle$ (where $\phi$ is the azimuthal emission angle of the measured particle relative to the reaction plane of the AA
collision) characterizes angular anisotropies that can be induced, \eg, through the collectively expanding fireball medium (usually at relatively low $p_T$)  
or different path lengths through the medium (usually for high-$p_T$ particles). Precision of measurements
of these observables for $D$ mesons have matured considerably in recent years, to the level that comparisons to theoretical model predictions allow 
for stringent constraints on the underlying transport coefficients (in particular, the determination of the charm-diffusion coefficient $\cD_s$, which is a 
central goal of using HF particles as probes of the QGP properties).

\begin{figure}[!t]
\begin{center}
\begin{minipage}{0.8\linewidth}
\vspace{-0.5cm}
\hspace{-0.4cm}
\includegraphics[width=1.0\textwidth]{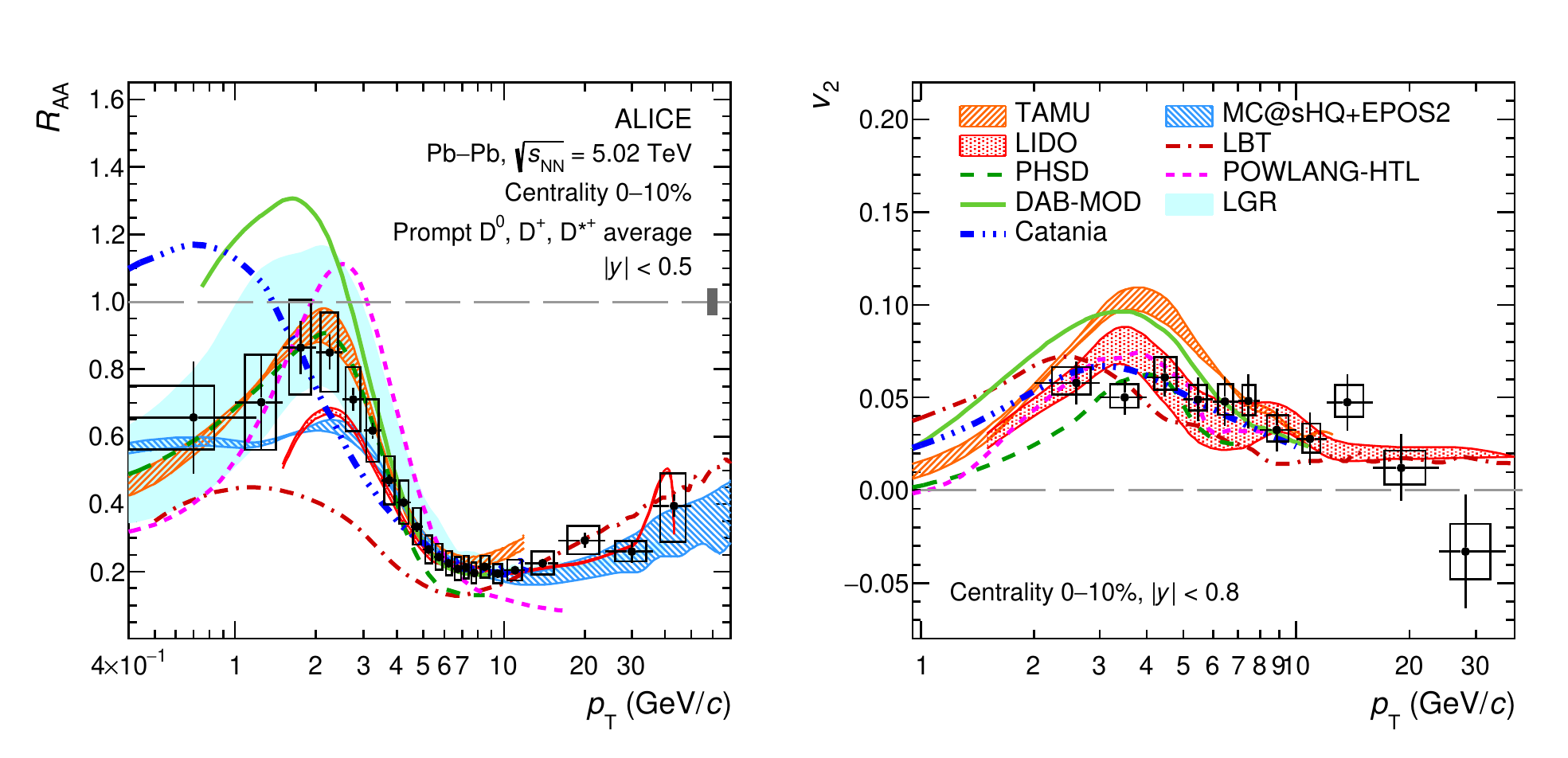}
\end{minipage}
\vspace{-0.25cm}

\begin{minipage}{0.8\linewidth}
\vspace{-0.5cm}
\hspace{-0.4cm}
\includegraphics[width=1.0\textwidth]{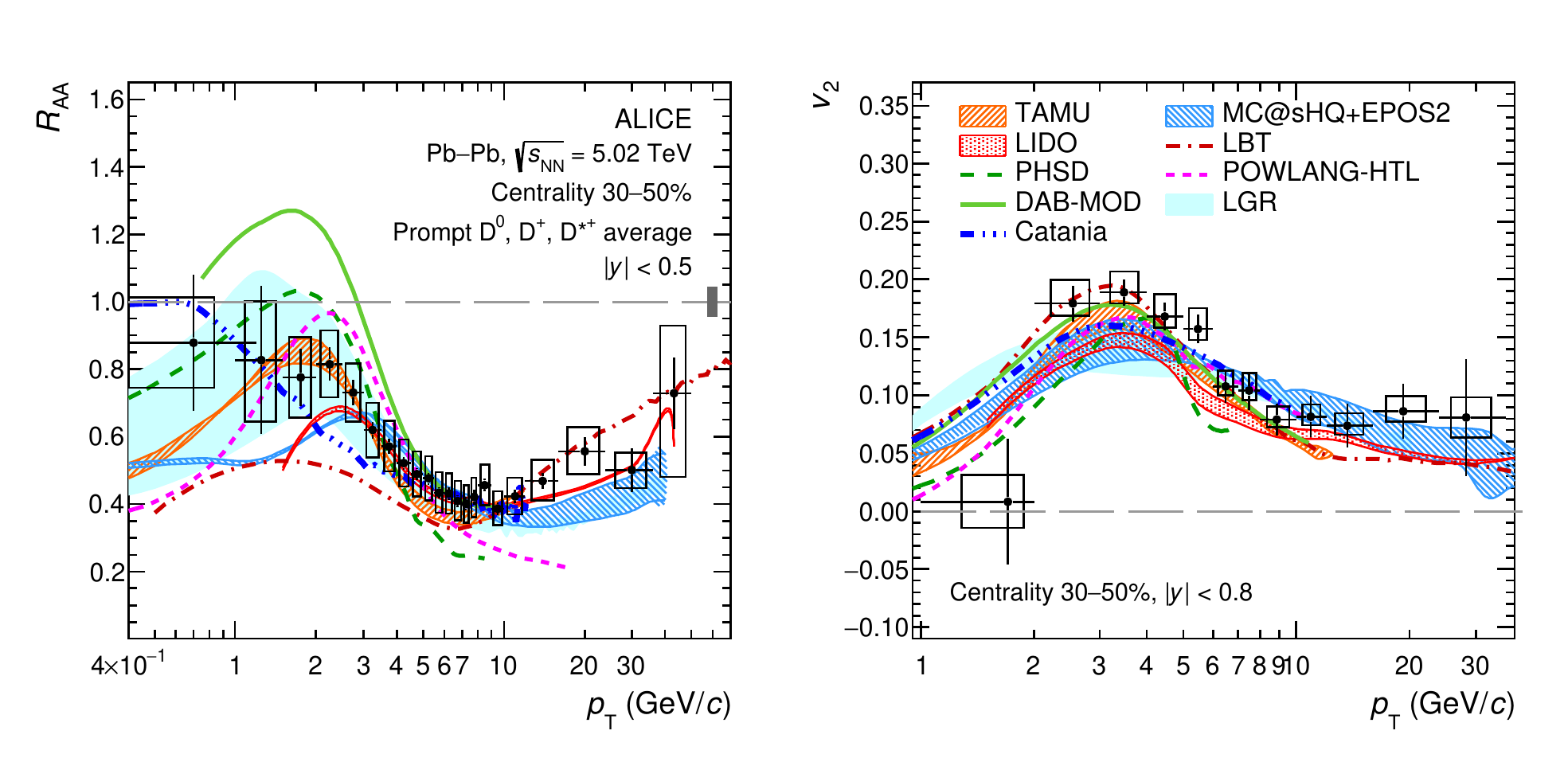}
\end{minipage}
\end{center}
\vspace{-0.2cm}
\caption{Nuclear modification factor, $\RAA$ (left), and elliptic flow, $v_2$ (right), averaged over prompt $D^0$, $D^+$, and $D^{*+}$ mesons in 
the 0-10\% (top) and 30-50\% (bottom) centrality classes in 5.02 TeV Pb-Pb collisions compared with predictions of theoretical models. Figure taken from Ref.~\cite{ALICE:2021rxa}.}
\label{fig_D-RAA-v2}
\end{figure}
As a recent example, we show in Fig.~\ref{fig_D-RAA-v2}, an ALICE compilation of their $\RAA$ and $v_2$ data averaged over 
prompt $D$ and $D^*$ mesons in central and semicentral 5.02\,TeV Pb-Pb collisions, in comparison to various theoretical model 
predictions~\cite{ALICE:2021rxa}. A suppression of prompt $D$ mesons by up to a factor of $\sim$5 (2.5) is observed in the $R_{\rm AA}$ around 
$p_T\sim 7$-10\,GeV in 0-10 (30-50)\% central collisions, which is captured by most of the transport models that simulate charm diffusion with 
elastic and/or radiative interactions and compute charm hadronization with coalescence and fragmentation processes in a hydrodynamically expanding medium~\cite{He:2019vgs,Beraudo:2014boa,Beraudo:2017gxw,Song:2015sfa,Scardina:2017ipo,Plumari:2019hzp,Nahrgang:2013xaa,Katz:2019fkc,Cao:2016gvr,Cao:2017hhk,Li:2019lex,Ke:2018jem}. Beyond this regime of maximum suppression, \ie, for $p_T>10$\,GeV, where gluon radiation is
expected to dominate the energy loss, both $R_{\rm AA}$ and $v_2$ are fairly well described by pQCD based jet-medium interaction 
models~\cite{Xu:2014tda,Xu:2015bbz,Shi:2018izg,Stojku:2020tuk,Kang:2016ofv}. At low and intermediate $p_T<6$\,GeV, most transport models 
give a reasonable description of the $D$-meson $v_2$ but show significant deviations from the measured $R_{\rm AA}$. The low-$p_T$ $R_{\rm AA}$ 
data remain significantly below unity indicating a nuclear-shadowing suppression of charm production (stronger in central than in semicentral collisions),
 consistent with estimates using nuclear PDFs~\cite{Lansberg:2016deg,Kusina:2017gkz,Kusina:2020dki,Eskola:2009uj,Eskola:2016oht}. In central collisions, 
the data suggest a ``bump" structure around $p_T\sim2$\,GeV, which is commonly associated with the collective flow that charm quarks pick up via a 
strong drag from the expanding medium. Another important effect in developing the flow bump is caused by the recombination of near-thermalized 
charm quarks with flowing light anti-/quarks from the medium, which also adds significant $v_2$ to the formed $D$-meson in the intermediate 
$p_T$ regime and is essential for a quantitative description of the observed $v_2$. One also notes that further ingredients, such as differing features 
in the hydrodynamic evolution models adopted for the transport calculations, as well as interactions in the hadronic phase after hadronization (which 
are taken into account in some approaches) affect the predictions of the observables.

\begin{table*}[!t]
\begin{center}
\begin{tabular}{lcccccc}
\hline\noalign{\smallskip}
~Model~~~~                               & $~~~~~~~~~~~~~~~~~~\chi^2/\mathrm{ndf}$  \\
                                                    & $R_{\rm AA}$    & $v_2$    \\
\noalign{\smallskip}\hline\noalign{\smallskip}

Catania~\cite{Scardina:2017ipo,Plumari:2019hzp}		& 143.8$/$30    &  14.0/8 \\
DAB-MOD~\cite{Katz:2019fkc}							& 234.1$/$30    &  9.8/6  \\
LBT~\cite{Cao:2016gvr,Cao:2017hhk}				    & 411.8$/$30    &  15.8/12 \\
LIDO~\cite{Ke:2018jem}								& 46.4$/$26     &  62.0/11 \\
LGR~\cite{Li:2019lex}								& 9.2$/$30      &  15.5/11  \\
MC@sHQ+EPOS2~\cite{Nahrgang:2013xaa}				& 56.6$/$30     &  5.7/12   \\
PHSD~\cite{Song:2015sfa}							& 294.7$/$30    &  19.6/11 \\
POWLANG-HTL~\cite{Beraudo:2014boa,Beraudo:2017gxw}	& 468.6$/$30    &  13.5/8 \\
TAMU~\cite{He:2019vgs}								& 30.2$/$30     &  8.15/9  \\

\noalign{\smallskip}\hline
\end{tabular}
\end{center}
\caption{Summary of the $\chi^2$ per degree-of-freedom values for theoretical-model calculations (listed in the left column) in comparison to ALICE 
data~\cite{ALICE:2021rxa,ALICE:2020iug} for the average $D$-meson $R_{\rm AA}$ in 0-10\% and 30-50\% central Pb-Pb(5.02\,TeV) collisions  
combined (middle column) and for the $v_2$ in 30-50\% central collisions (right column). Values taken from Refs.~\cite{ALICE:2021rxa,ALICE:2020iug}.}
\label{tab_chi2}
\end{table*}
Simultaneous comparisons of $R_{\rm AA}$ and $v_2$ measurements of $D$ mesons with transport model predictions, in particular in the low- and 
intermediate-$p_T$ regime where the sensitivity to charm-quark diffusion and hadronization is the highest,  provide rather stringent constraints on the 
charm-quark interactions with the medium. To  obtain a conservative estimate of the charm quark's spatial diffusion coefficient, in particular to encompass uncertainties due to different model implementations of charm-quark recombination (\eg,  resonance recombination~\cite{He:2019vgs} vs.~instantaneous coalescence~\cite{Scardina:2017ipo,Plumari:2019hzp,Cao:2016gvr,Cao:2017hhk}), the ALICE collaboration applied a mild requirement for data-to-model consistency, namely $\chi^2/\mathrm{ndf}<5$ for $R_{\rm AA}$ for $0<p_T<8$\,GeV~\cite{ALICE:2021rxa} and $\chi^2/\mathrm{ndf}<2$ for $v_2$ in the $p_T$ ranges provided by the different models~\cite{ALICE:2020iug}. This requirement was met by the TAMU~\cite{He:2019vgs}, MC@sHQ+EPOS2~\cite{Nahrgang:2013xaa}, LIDO~\cite{Ke:2018jem}, LGR~\cite{Li:2019lex}, and Catania~\cite{Scardina:2017ipo,Plumari:2019hzp} 
models, cf.~Tab.~\ref{tab_chi2}. The charm-quark spatial diffusion coefficient used in these models amounts to a range of $1.5<\mathcal{D}_s(2\pi T)<4.5$ 
near $\Tpc$ (recall Fig.~\ref{fig_Ds}), and represents the current state-of-the-art extraction of the transport coefficient of the QGP from HF probes.
In particular, it is in agreement with the properties of a strongly coupled QGP as discussed in Sec.~\ref{ssec_Ds}.

\begin{figure}[!t]
\centering
\includegraphics[width=0.65\textwidth]{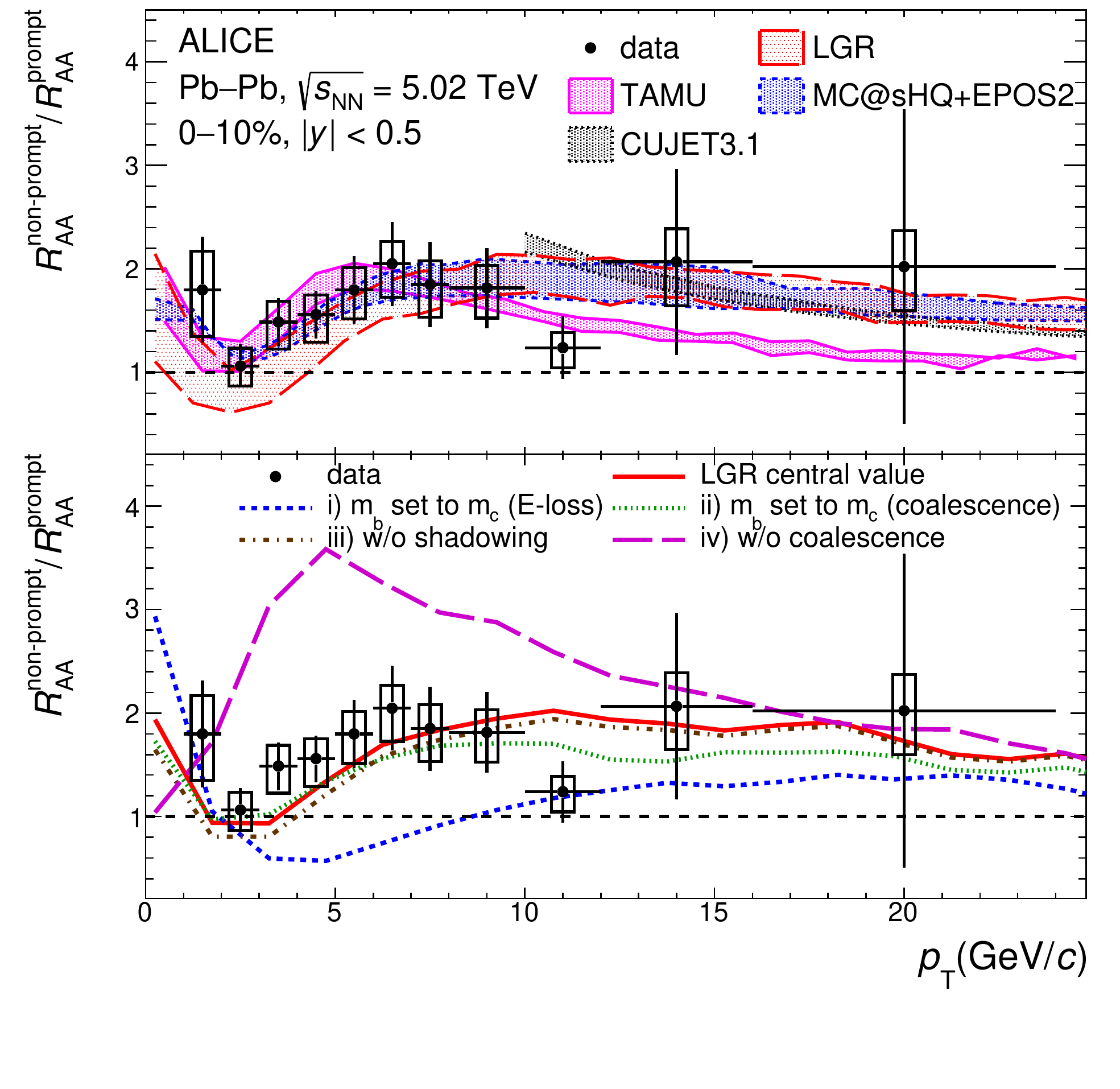}
\vspace{-0.6cm}
\caption{Double ratio of the $\RAA$ of non-prompt over prompt $D^0$-mesons as a function of $p_T$ measured by ALICE~\cite{ALICE:2021rxa} in 
 0-10\% central Pb-Pb(5.02\,TeV) collisions, compared to model predictions~\cite{He:2014cla,Nahrgang:2013xaa,Li:2019lex,Li:2020umn,Shi:2018izg} 
(upper panel), and to different modifications of the LGR calculations (bottom). Figure taken from Ref.~\cite{ALICE:2022tji}.}
\label{fig_RAA-nonprompt-D0}
\end{figure}
Bottom quarks, with a mass of about three times that of charm quarks, are expected to have a correspondingly longer thermal relaxation time and thus 
exhibit  less thermalization and pertinent spectrum modifications. The larger mass of bottom quarks also extends the reliability of the Langevin 
implementation of diffusion processes in the expanding QGP to significantly larger momentum~\cite{Das:2013kea} (recall Sec.~\ref{ssec_trans-app}), 
rendering them cleaner ``Brownian markers". At high $p_T$, where energy loss is presumably dominated by radiative processes, bottom quarks
are expected to lose less energy than charm quarks because of a stronger ``dead cone" effect, which suppresses gluon radiation at angles smaller than 
$m_Q/E$ relative to the quark's direction of motion~\cite{Dokshitzer:2001zm,Armesto:2003jh,ALICE:2021aqk}. Therefore, comparisons of the bottom- 
and charm-hadron observables offer means to disentangle the different manifestations of these effects and further constrain the HQ diffusion coefficient in 
the medium. Figure~\ref{fig_RAA-nonprompt-D0} shows the ALICE measurement of the double ratio of the $R_{\rm AA}$ of 
so-called non-prompt $D^0$ mesons (which arise from weak decays of bottom hadrons and thus reflect primary bottom spectrum modifications) 
to that of prompt $D^0$ mesons in 0-10\% central 5.02 TeV Pb-Pb collisions, in comparison with different model predictions~\cite{ALICE:2022tji}. 
The trends in the $p_T$ dependence are fairly well described by several transport models that implement charm and bottom diffusion and hadronization 
in QGP~\cite{He:2014cla,Nahrgang:2013xaa,Li:2019lex,Li:2020umn}. A non-prompt $D^0$-meson $R_{\rm AA}\geq 1$ at low 
$p_T\leq 2$\,GeV~\cite{ALICE:2022tji} indicates little or no shadowing on bottom production at mid-rapidity, whereas significant charm shadowing 
results in an $R_{\rm AA}$ below unity for prompt $D^0$ mesons~\cite{ALICE:2021rxa}. Going to $p_T\sim2$-3\,GeV, the prompt $\RAA$ 
is characterized by a pronounced ``flow bump" resulting from near-thermalized charm quarks and their recombination with flowing light quarks, 
which is much less pronounced (or even absent) for the non-prompt $\RAA$. Taken together, these observations help explain the minimum at 
$p_T\sim2$-3\,GeV in the $R_{\rm AA}^{\rm nonprompt}/R_{\rm AA}^{\rm prompt}$ ratio followed by a monotonous increase toward 
$p_T=0$. For $p_T\simeq 4-10$\,GeV, weaker suppression of bottom quarks than charm quarks leads to a rise of the ratio to close to two, reflecting 
less radiative energy loss of bottom than charm in this $p_T$ regime. Eventually the double ratio is expected to undergo a slow decrease toward unity at 
high $p_T$ where mass differences become irrelevant. In the lower panel of Fig.~\ref{fig_RAA-nonprompt-D0} the ratio from the scenario of 
the LGR model is shown without coalescence effects, and turns out to be too much enhanced compared to the data, corroborating that the 
minimum structure is mainly due to the formation of prompt $D^0$ mesons via charm-quark coalescence.

\begin{figure}[!t]
\centering
\begin{minipage}{0.8\linewidth}
\includegraphics[width=1.0\textwidth]{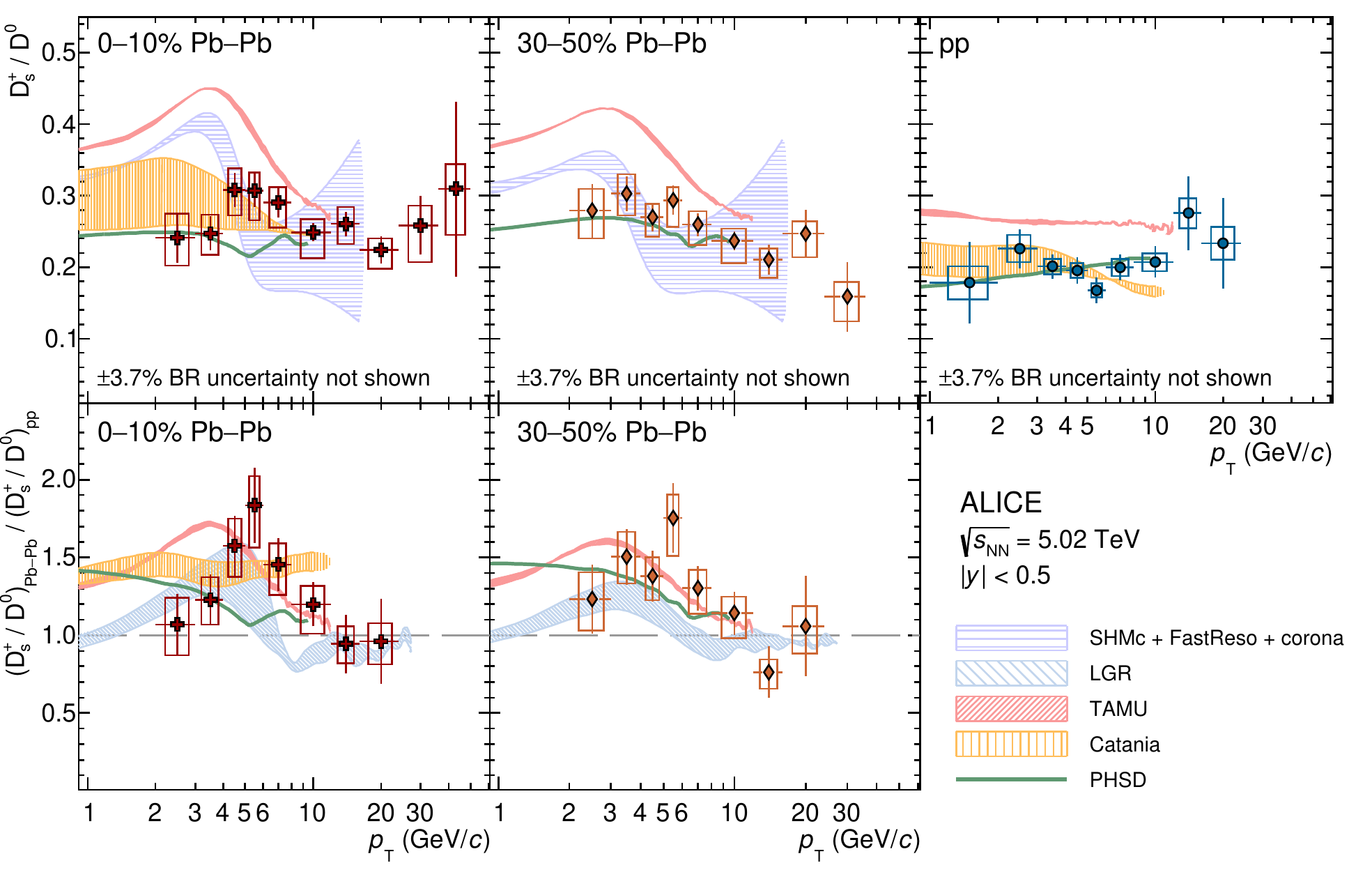}
\end{minipage}
\vspace{-0.1cm}
\caption{Upper panels: $D_s^+/D^0$ $p_T$-differential production ratios in 0-10\% (left) and 30-50\% (middle) Pb-Pb and  $pp$  collisions (right) 
at the same center-of-mass energy of 5.02\,TeV, compared with theoretical calculations based on charm-quark transport in a hydrodynamically 
expanding 
QGP~\cite{He:2019vgs,Li:2019lex,Song:2015sfa,Plumari:2017ntm,Song:2015ykw,Minissale:2020bif,He:2019tik} and on statistical 
hadronization~\cite{Andronic:2021erx}. Lower panels: double ratio of $D_s^+/D^0$ in Pb-Pb collisions divided by those in $pp$ collisions, 
for 0-10\% (left) and 30-50\% (middle) centrality, compared with theoretical calculations. Figure taken from Ref.~\cite{ALICE:2021kfc}.}
\label{fig_Ds-D0}
\end{figure}
Transverse-momentum dependent ratios between different charm-hadron species provide a measure of the charm hadro-chemistry and its 
possible in-medium modifications from $pp$ to $AA$ collisions, and have recently attracted intense interest~\cite{Dong:2019byy}. 
These observables are sensitive probes of charm-quark hadronization mechanisms, as a the same underlying charm-quark distribution function 
from its transport through the deconfined medium is converted into different hadrons (much like what has been found in the light- and strange-flavor 
sector~\cite{Braun-Munzinger:2003pwq}, the charm chemistry is not expected to change in the hadronic phase). An enhancement of the yield of the 
ground-state charm-strange meson, $D_s^+$, relative to that of nonstrange $D$ mesons at low and intermediate $p_T$, has been predicted for Au-Au 
collisions at RHIC~\cite{Kuznetsova:2006bh,He:2012df}, as a consequence of recombination of near-thermalized charm quarks with strange quarks in 
a  chemically equilibrated QGP in which strangeness production is much enhanced with respect to $pp$ collisions. This has been confirmed by the 
recent measurement of the $D_s^+/D^0$ ratio in 5.02 TeV $pp$ and Pb-Pb collisions~\cite{ALICE:2021kfc}, as shown in Fig.~\ref{fig_Ds-D0} 
(STAR published similar results in Au-Au collisions at RHIC~\cite{STAR:2021tte}). The $D_s^+/D^0$ ratios in $pp$ and Pb-Pb collisions are fairly 
described by the Catania~\cite{Plumari:2017ntm,Minissale:2020bif} and PHSD model~\cite{Song:2015ykw}.
The TAMU model~\cite{He:2019vgs}, which computes charm-quark hadronization with resonance recombination, overestimates the ratio in both $pp$ 
and Pb-Pb collisions by a similar amount;  in the latter, 
the $D_s/D^0$ ratio develops a peak around $p_T\sim3$-4\,GeV which is caused by collective flow of the heavier strange quark in the recombination 
of $D_s^+$ mesons compared to light quarks in the recombination of $D$ mesons. This peak migrates also into the double ratio and appears to 
be supported by the data; it is also reported by the SHMc model~\cite{Andronic:2021erx} that approximates the charm-meson spectra with a 
hydrodynamic blastwave ansatz normalized to yields from the inclusive statistical hadronization model calculations.

\begin{figure}[!t]
\centering
\begin{minipage}{0.8\linewidth}
\includegraphics[width=1.0\textwidth]{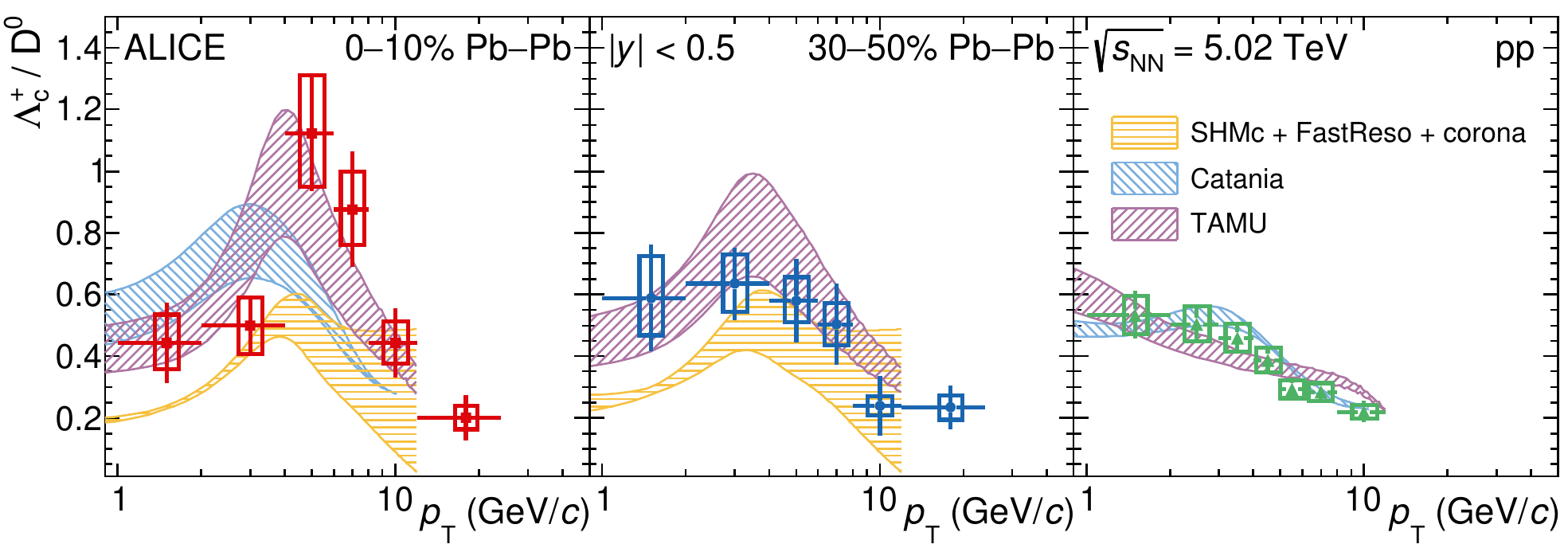}
\end{minipage}
\vspace{-0.0cm}
\caption{The $\Lambda_c^+/D^0$ yield ratio as a function of $p_T$ in 5.02 TeV 0-10\% (left panel) and 30-50\% (middle panel) central Pb-Pb 
and $pp$ (right panel) collisions compared with predictions of different theoretical 
calculations~\cite{Plumari:2017ntm,He:2019vgs,Andronic:2021erx,Minissale:2020bif,He:2012df}. Figure taken from Ref.~\cite{ALICE:2021bib}.}
\label{fig_Lc-D0}
\end{figure}
An enhancement of the baryon-to-meson ratio at intermediate $p_T\sim3-$6\,GeV in heavy-ion collisions with respect to $pp$ 
collisions~\cite{PHENIX:2003tvk,STAR:2006pcq,ALICE:2019hno,ALICE:2013cdo} is a well established phenomenon for light and strange hadrons, and 
has been commonly interpreted as being due to quark coalescence processes~\cite{Greco:2003xt,Fries:2003vb,Hwa:2003bn,Molnar:2003ff,Fries:2008hs}. 
In principle, this can also be regarded as a remnant of the radial-flow effect that pushes heavier baryons (containing three valence quarks) further out 
in $p_T$ than lighter mesons (containing two valence quarks).  In the charm sector, the $\Lambda_c/D^0$ ratio exhibits a significant enhancement 
already in high-energy 
$pp$~\cite{ALICE:2020wfu,ALICE:2020wfu} and $p$-Pb collisions~\cite{ALICE:2020wfu} (see the right panel of Fig.~\ref{fig_Lc-D0}) over the 
value of $\sim$0.1 measured in $e^+e^-$ and $ep$ collisions (essentially independent of $p_T$).
This implies a considerable change of charm-quark fragmentation fractions into different kinds of charm hadrons~\cite{ALICE:2021dhb},
and has been explained by three models implementing different charm-baryon enhancement mechanisms: a color reconnection beyond leading order 
with junctions fragmenting into baryons~\cite{Bierlich:2015rha}, statistical hadronization augmented with excited charm-baryon states beyond the listings 
of the particle data group~\cite{He:2019tik,Chen:2020drg}, and instantaneous coalescence in a deconfined fireball postulated to be formed in $pp$ 
collisions~\cite{Minissale:2020bif}. In anaolgy to the $p/\pi$~\cite{ALICE:2019hno} and $\Lambda/K_s^0$~\cite{ALICE:2013cdo} ratios, the
$\Lambda_c/D^0$ ratio is enhanced from $pp$ to Pb-Pb collisions at intermediate $p_T\simeq4-8$\,GeV, cf.~the ALICE measurements shown in the left and 
middle panels of Fig.~\ref{fig_Lc-D0} (a similar enhancement has also been observed in 200\,GeV Au-Au collisions at RHIC~\cite{STAR:2019ank}), 
with indications for a larger effect in the 0-10\% central collisions and a hint of a depletion at low $p_T$. This is consistent with a larger radial
flow of the expanding fireball, which, from light-hadron production, one knows to be stronger in more central collisions. At high $p_T>10$\,GeV, 
the ratio is comparable to the measurement by CMS in minimum-bias Pb-Pb collisions~\cite{CMS:2019uws} and tends to approach the value measured 
in $pp$ collisions. 
Three theoretical calculations are compared to the $\Lambda_c/D^0$ data in Fig.~\ref{fig_Lc-D0}. The results of the Catania model, based on 
instantaneous coalescence~\cite{Plumari:2017ntm} of charm- and light-quark spectra that are both based on a combination of thermal blastwave spectra 
and quenched primordial spectra and supplemented with independent fragmentation at high $p_T$, tend to overestimate (underestimate) the 
$\Lambda_c/D^0$ in central collisions at low (intermediate) $p_T$;  the SHMc, assuming hydrodynamic blastwave spectra normalized to statistical 
hadronization yields (including all charm hadrons in the PDG list)~\cite{Andronic:2021erx}, tends to underestimate the data, indicative of an overall lack of 
charm-baryon production. The TAMU results, utilizing the resonance recombination model and fragmentation for hadronizing transported $c$-quark spectra 
on a hydrodynamic hypersurface, describe the shape and magnitude of the ratio in both 0-10\% and 30-50\% collisions fairly well~\cite{He:2019vgs}.
In particular, this approach takes into account space-momentum correlations between charm- and light-quark phase space distributions developed 
through Langevin transport and hydrodynamic flow, respectively, which turn out to harden the charm-baryon spectrum more than the charm-meson one. 
Also, utilizing the same augmented baryon spectrum relative to the PDG as used in the $pp$ calculations~\cite{He:2019tik} leads to substantially larger
feeddown contributions relative to the SHMc, and thus predicts an integrated $\Lambda_c/D^0$ value in Pb-Pb collisions that is compatible with the one 
in $pp$, 
as dictated by the {\it relative} chemical equilibrium among different charm-hadron species~\cite{He:2019vgs}. This suggests that the $\Lambda_c/D^0$ enhancement at intermediate $p_T$ is primarily due to a kinematic redistribution of charm baryons and mesons in momentum space (where the
former experience a larger collective-flow effect), in analogy to the case of $\Lambda/K_s^0$~\cite{ALICE:2013cdo}, 
rather than additional charm-baryon production channels opening up in Pb-Pb collisions.

\newpage

	\section{Quarkonium Phenomenology}
  \label{sec_onium}
In this chapter we discuss heavy-quarkonium transport in hot QCD matter. In Sec.~\ref{ssec_QQ-kin} give a brief review of the kinetic approaches 
that have been used in the past to describe the dissociation and regeneration of charmonia and bottomonia in heavy-ion collisions; this includes the 
semiclassical Boltzmann and rate equations with their pertinent transport parameters discussed in Sec.~\ref{ssec_Gam}, and a short discussion of 
quantum approaches represented by a Lindblad-type equation for the quarkonium density matrix. In Sec.~\ref{ssec_QQ-obs} we first give an overview
of the excitation functions of quarkonium production from SPS via RHIC to LHC energies, followed by a discussion of more differential charmonium and
bottomonium observables at the LHC in comparison to model calculations. In both contexts, our focus will be on the features that have a close connection
to the open HF sector, specifically the role of regeneration reactions for $J/\psi$ which directly relate to the charm-quark diffusion properties, and
the role of the in-medium QCD force in the observed pattern of bottomonium suppression.

\subsection{Kinetics}
\label{ssec_QQ-kin}

To date, the phenomenology of quarkonia kinetics to describe their dissociation and regeneration in URHICs has mostly been carried out in 
semiclassical transport approaches.
Starting point is the Boltzmann equation for the quarkonium phase space distribution function, $f_{\cQ}$, which can be written in compact form as
\beq
p^\mu \partial_\mu f_{\cQ} = -E_{\cQ}(p) \Gamma_{\cQ} f_{\cQ} +  \beta \  ,
\label{Boltz}
\eeq
where $p^\mu= (E_{\cQ}(p),\vec{p})$ denotes the (on-shell) f-moourmentum of the quarkonium state $\cQ$. The first and second term on the 
right-hand-side of Eq.~(\ref{Boltz}) denote the loss term with dissociation rate $\Gamma_{\cQ}$ and the gain term $\beta$, respectively. The latter is 
only operative below the temperature where the state $\cQ$ can survive, usually referred to as the dissociation temperature, $T_{\rm diss}^{\cQ}$.
This notion is not strictly defined; in practice, different criteria for $T_{\rm diss}^{\cQ}$ have been employed, \eg, as the temperature where the 
binding energy of $\cQ$ vanishes, or when the dissociation rate (determining the width of the state) equals the binding energy. A more rigorous 
treatment of this issue requires a quantum transport approach which we will return to below.
The dissociation rate, $\Gamma_{\cQ}$,  has been discussed in Sec.~\ref{ssec_Gam}. It does not directly involve the individual HQ distribution functions
(except for final-state Pauli blocking which can be safely neglected for heavy quarks).  The explicit form of the gain term (much like for the dissociation rate, 
recall, \eg, Eq.~(\ref{Gam-qf-1})) depends on the type of process, \eg, leading-order gluo-dissociation ($2\to2$) or inelastic parton scattering ($3\to 2$). 
In the latter case one has
\beq
\beta =\sum\limits_p d_p d_Q d_{\bar{Q}} \int d\Pi_3 \overline{|{\cal M}_{p{Q}\bar{Q}\to p\cQ}|^2} f_p(p_i)) f_Q(p_Q) f_{\bar{Q}}(p_{\bar Q})  
 [1\pm f_p(p_f)]  \  ,
\eeq
where $d\Pi_3$ denotes the Lorentz-invariant phase space (including an energy-momentum-conserving $\delta$-function) for the incoming 
heavy quark, antiquark, and thermal parton ($p$), as well as the outgoing light parton (all being on-shell). The scattering matrix elements are exactly the 
same as for the inverse process of dissociation, with an initial-state average and a final-state summation which ensures detailed balance (approximation
methods, in particular the quasifree approximation discussed in Sec.~\ref{ssec_Gam}, also apply here).  Of critical importance for a quantitative evaluation
of the quarkonium regeneration rate are the HQ distribution functions, $f_{Q,\bar{Q}}$. In the context of a heavy-ion collision, they deviate from both
thermal and chemical equilibrium. Assessing the deviation from thermal equilibrium is precisely the problem of HQ transport through the fireball of
a heavy-ion collision as discussed in Sec.~\ref{sec_open}, with pertinent transport coefficients discussed in Sec.~\ref{ssec_Ap}. For quarkonium
dissociation, this figures as a space- and time-dependent HQ distribution function whose momentum distribution  gradually evolves from its initial production
toward local thermal equilibrium in the expanding fireball. As emphasized before, the heavy-light scattering matrices governing this transport are related
to the ones in the gain term, especially when evaluated in the quasifree approximation. Rather few calculations exist to date which have implemented the
off-equilibrium HQ kinetics in quarkonium regeneration processes (see, \eg, Refs.~\cite{Song:2012at,Blaizot:2017ypk,Yao:2020xzw} for weak-coupling 
approaches, or Ref.~\cite{He:2021zej}). The chemical off-equilibrium of the HQ distributions stems from the fact that HQ pair production is strongly 
suppressed at the typical temperatures encountered in heavy-ion collisions, as they are much smaller than the mass threshold of 2$m_Q$. Therefore, the 
number of heavy quarks in the fireball is expected to be determined by their hard production in primordial nucleon-nucleon collisions upon first impact of
the nuclei, and approximately conserved thereafter. This can be accounted for by a HQ fugacity, $\gamma_Q$, which depends on the initial HQ production
cross section, the fireball volume and temperature, and on the HQ mass (or more generally, the available HF states in the ambient medium, \eg, all 
HF hadron states in confined matter). This problem becomes more transparent when integrating the Boltzmann equation over space and three-momentum
(neglecting surface terms) to obtain a kinetic rate equation for the time evolution of the quarkonium number, $N_{\cQ}$, 
\beq
\frac{dN_{\cQ}}{d\tau} = -\Gamma_{\cQ} \left[N_{\cQ} -N_{\cQ}^{\rm eq}\right] \  .
\label{rate}
\eeq
This equation is now characterized by two transport parameters, the (momentum-averaged) inelastic reaction rate, $\Gamma_{\cQ}$, as discussed before 
(which still depends on temperature through the time dependence), and the quarkonium equilibrium limit, $N_{\cQ}^{\rm eq}$, emerging from the gain 
term. It can be written as
\beq
N_{\cQ}^{\rm eq}(T,\gamma_Q) = V_{\rm FB} \ d_{\cQ} \ \gamma_Q(T)^2
\int \frac{d^3p}{(2\pi)^3} \ f^B(m_{\cQ},T) \ ,
\label{Neq}
\eeq
with the spin degeneracy $d_{\cQ}$  of state $\cQ$, its Bose distribution function, $f^B$, and the fireball volume, $V_{\rm FB}$. The HQ 
fugacity factor, $\gamma_Q$ , is obtained from matching the number of $Q\bar Q$ states in the fireball to the total number of HF states in thermal
equilibrium (which in the QGP would simply be the number of heavy quarks plus antiquarks). 
Non-thermal HQ distributions can further affect the quarkonium equilibrium limit, typically leading to a reduction as the harder momentum spectra from 
off-equilibrium HQ distributions reduce  the phase space overlap for bound-state formation~\cite{Song:2012at}. In the TAMU transport approach, 
\eg, this has been accounted for by a thermal relaxation factor~\cite{Grandchamp:2002wp}, ${\cal R}(\tau) =1-\exp(\tau/\tau_Q)$ where $\tau_Q$ 
is the HQ thermal relaxation  time discussed in Sec.~\ref{ssec_Ap}.  The equilibrium limit of quarkonia has first been introduced in the context of the statistical 
hadronization model (SHM)~\cite{Gazdzicki:1999rk,Braun-Munzinger:2000csl,Andronic:2006ky,Braun-Munzinger:2009dzl}, where charmonium
production has been evaluated at the pseudocritical temperature using the emerging spectrum of hadrons containing charm to compute the charm-quark 
fugacity. 


While semiclassical approaches have been fairly successful in describing and predicting charmonium and bottomonium observables at RHIC and the LHC, 
some of the underlying approximations remain to be scrutinized. In general, the presence of a strongly coupled QGP, with transport coefficients 
near conjectured lower bounds by quantum mechanics, clearly call for more rigorous investigations of how off-shell effects in the presence of large 
collisional widths affect the microscopic description of observables. As indicated above, a key question concerning quarkonia is their fate as they approach 
their dissociation temperature, where the binding energy becomes a small scale that is susceptible to quantum effects. This is one of the main motivations 
for the development of quantum-transport approaches, in particular those involving bound states, which are usually formulated in the language of 
open-quantum systems highlighting the coupling to the $Q\bar Q$ continuum, cf.~\eg, 
Refs.~\cite{Blaizot:2015hya,Katz:2015qja,Yao:2018nmy,Rothkopf:2019ipj,Miura:2019ssi,Akamatsu:2020ypb,Strickland:2021cox} 
and references therein. 
Specifically, instead of solving for the (differential) number of quarkonium states, one simulates the evolution of the reduced quarkonium density matrix, 
$\varrho_{\cQ}$, which can usually be cast into a Lindblad form, 
\beq
\frac{d\varrho_{\cQ}}{dt} = -i[H,\varrho_{\cQ}] + \sum\limits_n L_n \varrho_{\cQ} L_n^\dagger - \frac{1}{2} \{L_n^\dagger L_n, \varrho_{\cQ}\}  \  ,  
 \eeq
where the sum is over all quarkonium states including the $Q\bar{Q}$ continuum. The Lindblad operator contains the information on the transition rates
in the form of amplitudes (rather than amplitudes squared or cross sections as in the Boltzmann equation), an essential feature to account for quantum 
effects. In particular, this incorporates transitions and the interference between different states in the time evolution of the wave packet, which is not 
treated in semiclassical frameworks, especially for overlapping states and threshold effects in the spectral distribution, \ie, for small binding energies and 
large widths. In this way, the quantum treatment also avoids the notion of a dissociation temperature, as this information is encoded in the time evolving wave functions (or density matrix).
Regeneration reactions are included in the Lindblad equation, but their practical implementation is rather challenging, especially for the case of multiple
HQ pairs in the system. In addition, most quantum-transport approaches to date involve weak-coupling methods (either for the bound-state dynamics 
or the medium description, or both). Therefore, applications to URHICs have focused on the bottomonium sector, which will be discussed in Sec.~\ref{sssec_bottomonia}.


\subsection{Observables}
\label{ssec_QQ-obs}


We are now in position to discuss quantitative comparisons of model calculations to charmonium and bottomonium data in 
URHICs at the SPS, RHIC, and the LHC. We will focus on theoretical approaches that have been discussed in the preceding sections, thereby 
trying to reconnect to the fundamental question of how the in-medium QCD interactions manifest themselves in the experimental observables.

\begin{figure}[!t]
\begin{minipage}[c]{0.5\linewidth}
\includegraphics[width=0.97\textwidth]{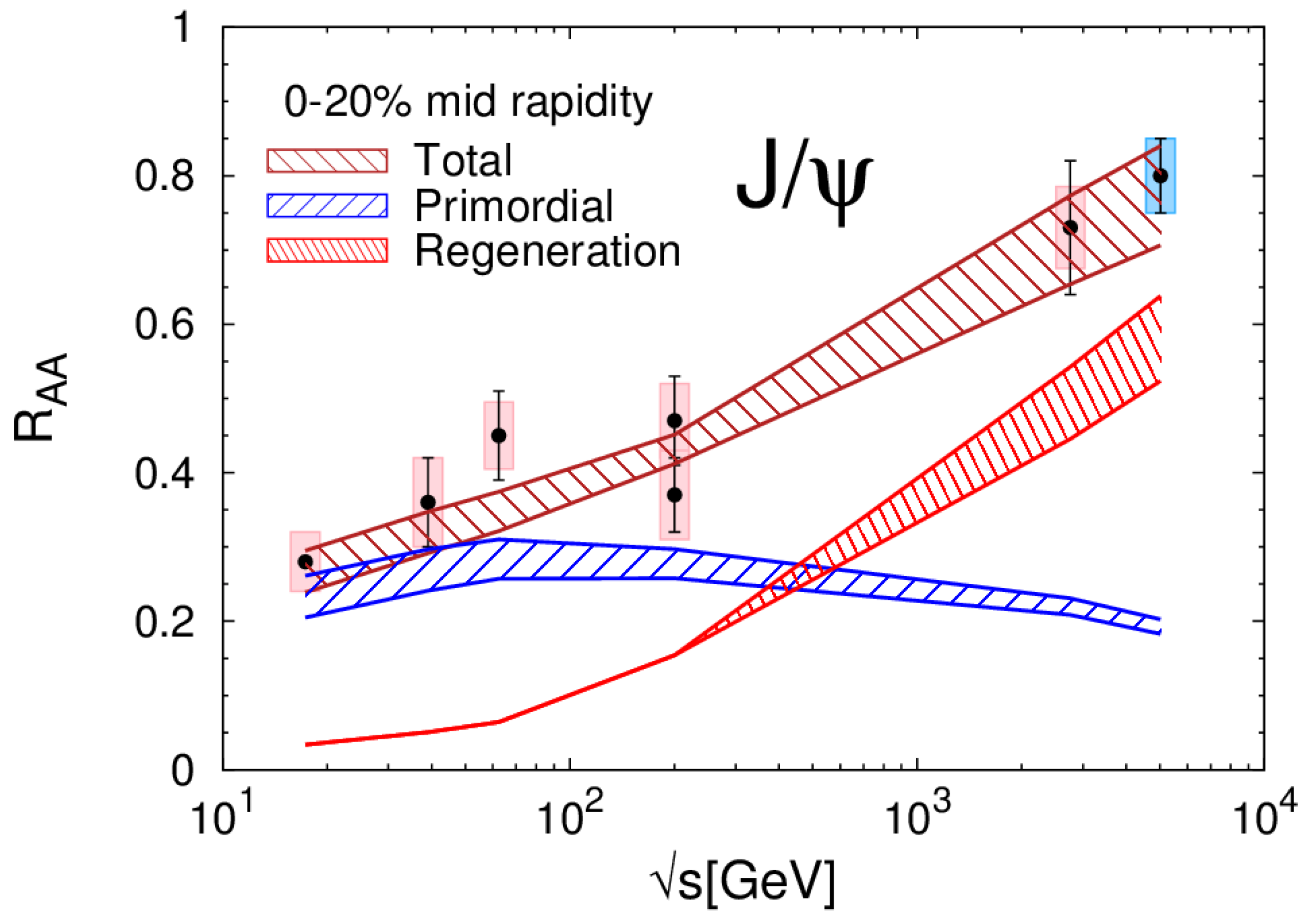}
\end{minipage}
\begin{minipage}[c]{0.5\linewidth}
\includegraphics[width=0.99\textwidth]{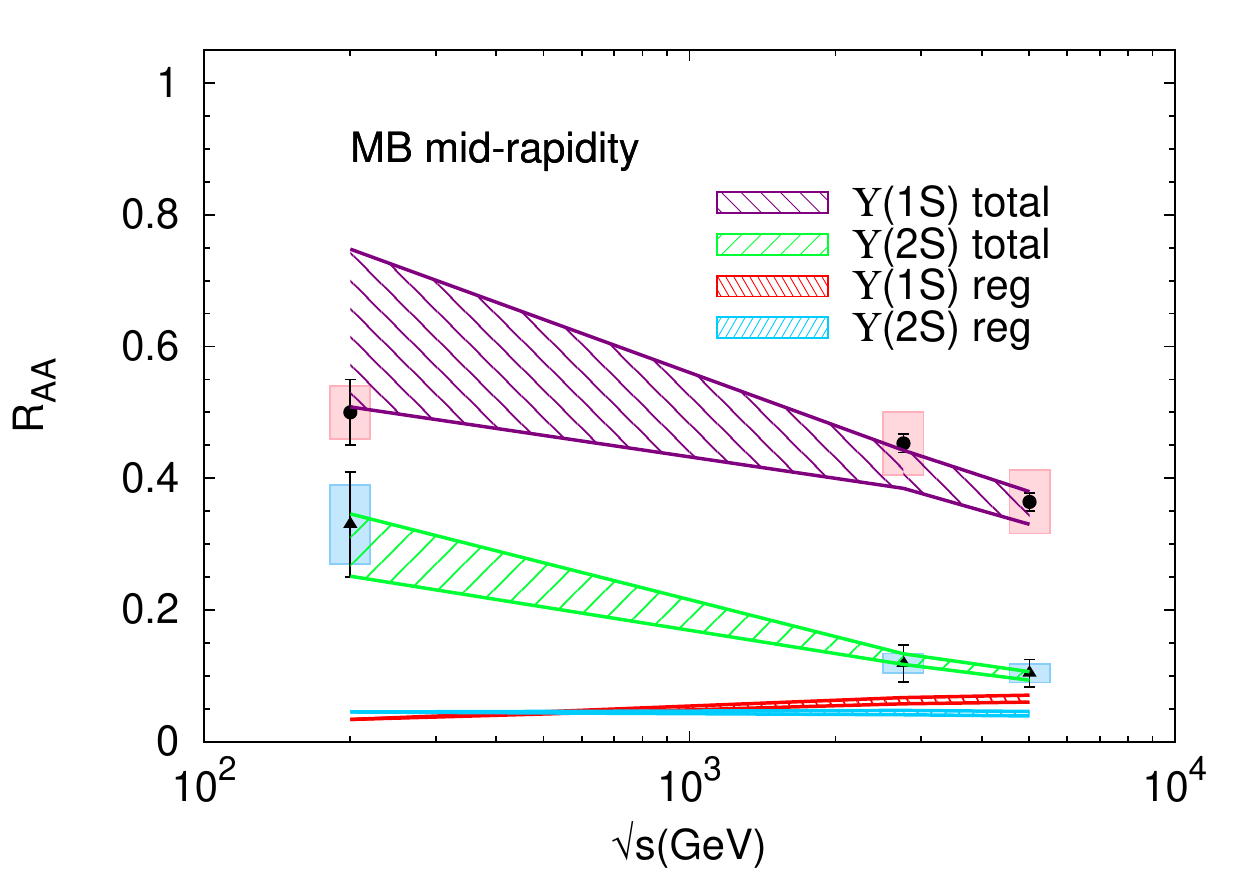}
\end{minipage}
\caption{Excitation function of the nuclear modification factor for inclusive-$J/\psi$ production in central Au-Au collisions at RHIC and in Pb-Pb collisions 
at the SPS and LHC (left panel), and for $\Upsilon (1S)$ and $\Upsilon(2S)$ production in minimum-bias collisions Au-Au and Pb-Pb collisions at RHIC 
and the LHC, respectively (right panel); TAMU transport model calculations are compared to data from the NA50, STAR, ALICE and CMS collaborations; 
figures are taken from Ref.~\cite{Rapp:2017chc}.}
\label{fig_excit-onia}
\end{figure}
To illustrate the production systematics of quarkonia over a broad range of collision energies, $\sqrt{s_{\rm NN}}$, in a compact form, we start with 
pertinent excitation functions of the nuclear modification factor, $R_{\rm AA}(\sqrt{s_{\rm NN}})$. The left panel of Fig.~\ref{fig_excit-onia} shows 
inclusive-$J/\psi$ production at mid-rapidity in central Pb-Pb and Au-Au collisions at the SPS (17 GeV) via RHIC (39, 62, 200 GeV) to the LHC (2.76 
and 5.02 TeV), comparing data from the NA50~\cite{NA50:2004sgj}, PHENIX~\cite{PHENIX:2006gsi}, STAR~\cite{STAR:2013eve,STAR:2016utm} 
and ALICE~\cite{ALICE:2013osk,JimenezBustamante:2017wxc,ALICE:2019nrq} collaborations to TAMU transport model 
calculations~\cite{Grandchamp:2003uw,Zhao:2011cv}
using the rate-equation approach with in-medium binding energies and dissociation rates as discussed in Sec.~\ref{ssec_Gam}. At SPS energy, 
regeneration is small and the $R_{\rm AA}\sim 0.3$ directly reflects the suppression of primordially produced $J/\psi$'s. A little more than 
half of the total suppression is actually caused by so-called cold-nuclear-matter effects, which at SPS energies are essentially attributed to
a nuclear absorption on the incoming nucleons with a dissociation cross section of $\sigma_{\rm abs}^{J/\psi}\simeq 7.5$\,mb as
extracted from $p$A collisions~\cite{NA60:2010wey}. The hot-medium effect, with initial temperatures of $T_0\simeq230$\,MeV (as 
extracted from thermal dilepton spectra~\cite{Rapp:2014hha}), is mostly driven by the suppression of excited states ($\chi_c$ and $\psi'$) which 
contribute up to $\sim$35-40\% to the inclusive $J/\psi$ yield in $pp$ collisions. This means that at the SPS the direct-$J/\psi$ yield is rather 
little affected, which implies that the $J/\psi$ should be able to survive, with a rather small width, up to temperatures of at least $\sim$230\,MeV or so
(cf.~Fig.~\ref{fig_gam-psi-ups}). The hot-medium suppression of the primordial $J/\psi$ becomes significantly stronger at RHIC energies as a 
consequence of the increasing temperature of the medium together with a subsiding nuclear absorption with increasing $\sqrt{s}$,
and reaches a factor of 5 at the LHC. At the same time, due to the increasing number of charm quarks and their near-thermalization, $J/\psi$ 
regeneration grows substantially and becomes the dominant source in central Pb-Pb collisions at the LHC. The situation is very different 
for the $\Upsilon$ states, where CMS~\cite{CMS:2012gvv,CMS:2016rpc} and STAR~\cite{Ye:2017fwv} data for minimum bias (MB) Pb-Pb and Au-Au
collisions, respectively,  are compared to the same theoretical approach for bottomonia, cf.~right panel of Fig.~\ref{fig_excit-onia}. The $\Upsilon(2S)$, 
albeit with a comparable binding energy to the $J/\psi$ in vacuum, shows a substantial  suppression already at RHIC energy, which becomes even 
stronger at the LHC. The main reason for the stark difference in the $J/\psi$ and $\Upsilon(2S)$ excitation functions is, of course, due to 
regeneration processes, with a small number of $b\bar b$ pairs but also the smallness of the $\Upsilon$ equilibrium limit including the slower 
$b$-quark thermalization time compared to $c$-quarks. The ground-state $\Upsilon(1S)$ is much more stable than the $\Upsilon(2S)$, especially 
given the ca.~50\% feeddown contributions in $pp$ collisions. This suggests that there is rather little direct-$\Upsilon(1S)$ suppression at RHIC, with
initial temperatures near $T_0 \simeq300$\,MeV in the fireball (for MB collisions), where the dissociation rate from Fig.~\ref{fig_gam-psi-ups} 
is still relatively small (around 50\,MeV).

In the following two sections, we will focus on two topics where quarkonia are rather directly related to the open HF sector, namely the role of
regeneration processes for charmonia (Sec.~\ref{sssec_charmonia}), and the extraction of the in-medium HQ potential from bottomonium 
suppression (Sec.~\ref{sssec_bottomonia}).   


\subsubsection{Charmonia}
\label{sssec_charmonia}
\begin{figure}[!t]
\begin{minipage}[c]{0.49\linewidth}
\includegraphics[width=1.0\textwidth]{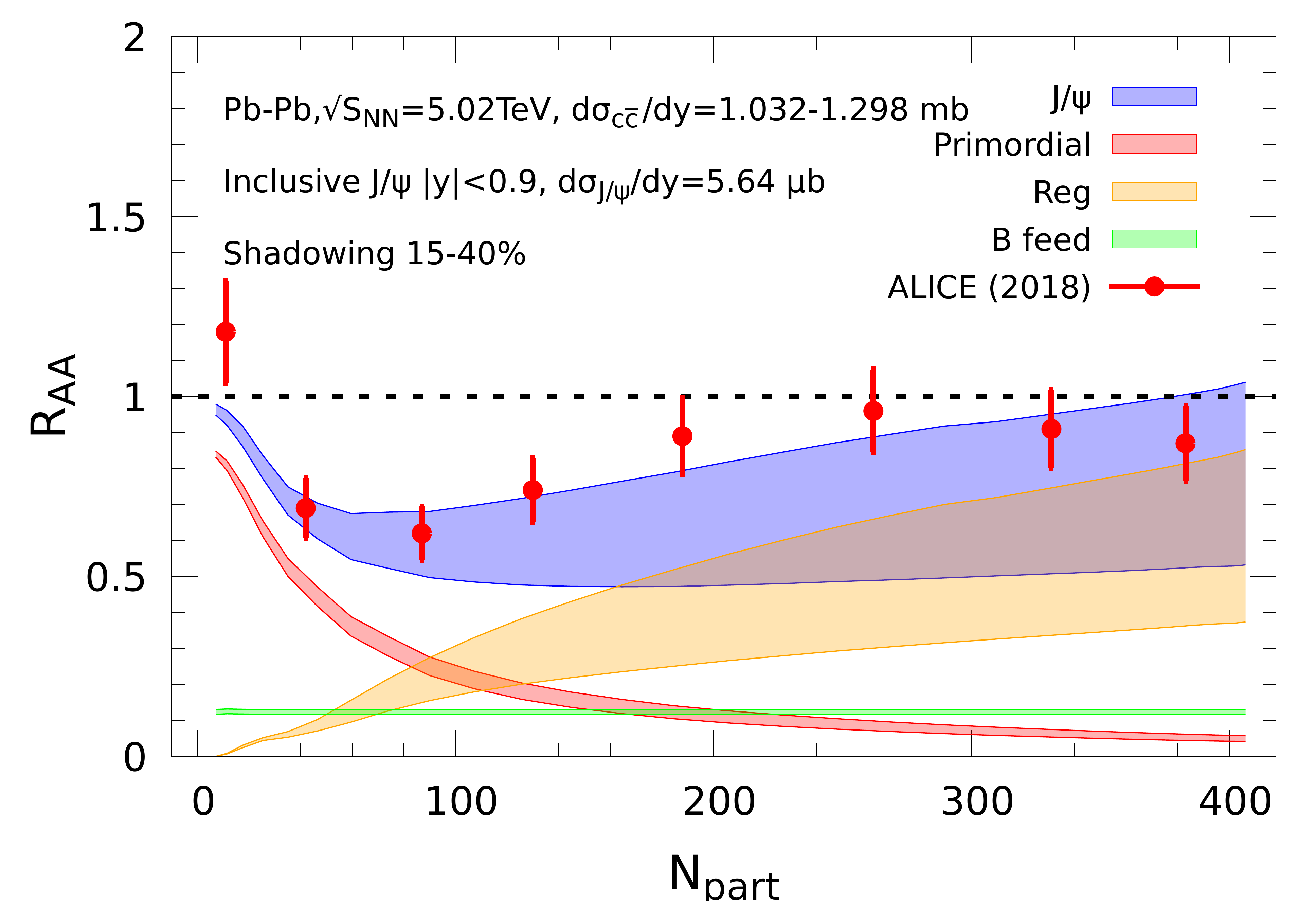}
\end{minipage}
\begin{minipage}[c]{0.49\linewidth}
\includegraphics[width=1.0\textwidth]{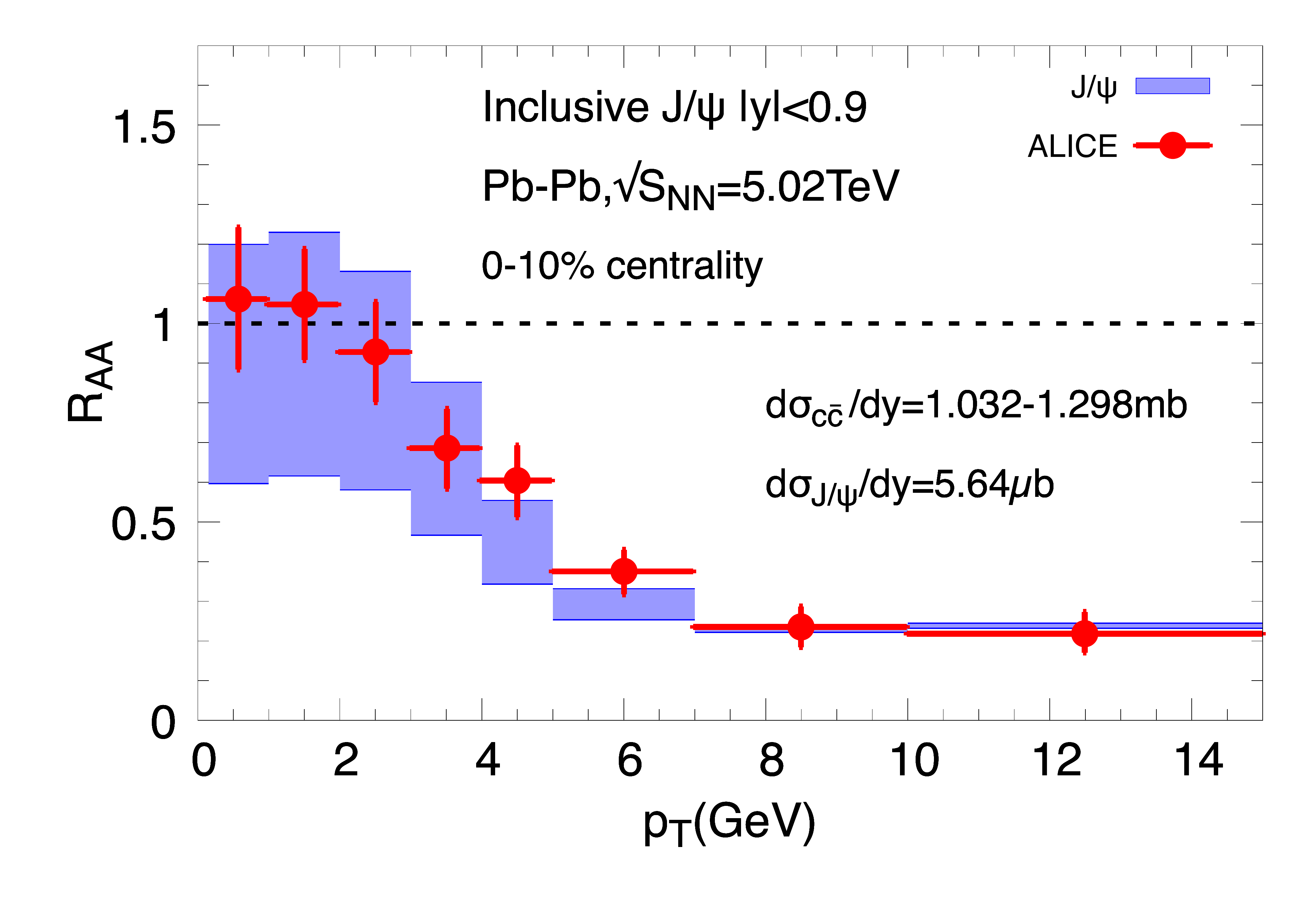}
\end{minipage}
\caption{Centrality (left panel) and transverse-momentum (right panel) dependence of the inclusive-$J/\psi$ $\RAA$ in Pb-Pb(5.02\,TeV) collisions 
measured by the ALICE collaboration, compared to TAMU transport model calculations with the most recent inputs for the $c\bar c$ cross section and 
nuclear shadowing.}
\label{fig_jpsi-raa-502}
\end{figure}
Quantitative predictions for the regeneration contribution to charmonia are rather sensitive to the number of $c\bar{c}$ pairs in the fireball (recall 
Eq.~(\ref{Neq})), and as such to the $c\bar{c}$ production cross section in $pp$ collisions and its suppression due to nuclear shadowing in AA collisions.
In Fig.~\ref{fig_jpsi-raa-502} we show results of the TAMU transport approach employing an updated charm cross section according to recent ALICE
data~\cite{ALICE:2021dhb}, 
and a range of nuclear shadowing of up to 15-40\% in central Pb-Pb collisions, in comparison to mid-rapidity ALICE data for inclusive $J/\psi$
production~\cite{ALICE:2019nrq,Bai:2020svs}. In the left panel, the centrality dependence of the calculated yields exhibits an increasing 
regeneration yield that overcompensates the strong suppression of primordial $J/\psi$ already in rather peripheral collisions. 
The right panel further corroborates the properties of the regeneration contribution being characterized by a concentration at low 
$p_T\ltsim_{J/\psi}$~\cite{ALICE:2016flj,ALICE:2019lga,ALICE:2019nrq} . This, in particular, implies a decrease of the ratio 
$r_{\rm AA}\equiv\langle p_T^2\rangle_{\rm AA}/\langle p_T^2\rangle_{pp}$, introduced by the Tsinghua group~\cite{Zhou:2009vz}, with centrality 
and also with collision energy, from $\sim 1.5$ at SPS~\cite{NA50:2000mfb} via $\sim 1$ at RHIC~\cite{PHENIX:2006gsi} to $\sim 0.5$ at the 
LHC~\cite{ALICE:2015nvt}, indicating the transition from primordial production with Cronin effect at the SPS to regeneration from a nearly thermal 
source. 
Due to the large elliptic flow of charm quarks required to describe the open HF observables discussed in Sec.~\ref{ssec_hf-obs}, one furthermore 
expects a sizable $v_2$ for the regenerated $J/\psi$ which has indeed been observed~\cite{ALICE:2017quq,ALICE:2020pvw}, reaching values of
$\sim 0.1$ in semicentral Pb-Pb collisions at 5.02\,TeV~\cite{ALICE:2017quq,ALICE:2020pvw}. 
Transport model calculations, which have approximated the regenerated $J/\psi$-$p_T$ spectrum with a thermal blastwave 
expression~\cite{Zhou:2014kka,Du:2015wha}, correctly predicted this effect up to momenta  of $p_T\simeq4$\,GeV, but fall well below the data
at higher $p_T$, which has cast some doubts on the role of regeneration (sometimes referred to as the ``$J/\psi$ $v_2$ puzzle").

\begin{figure}[!t]
\begin{minipage}[c]{0.5\linewidth}
\vspace{-0.5cm}
\hspace{-0.4cm}
\includegraphics[width=1.1\textwidth]{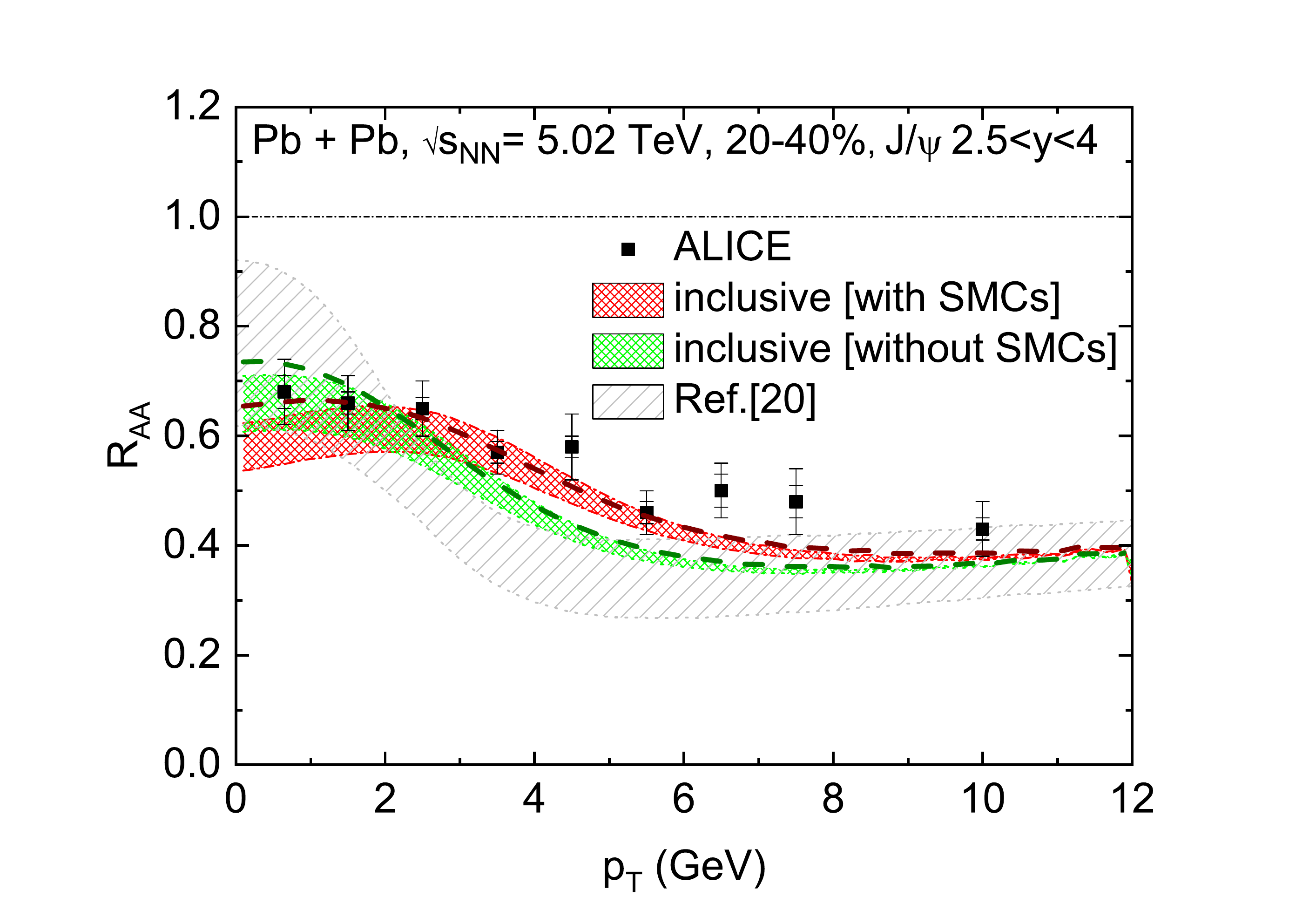}
\end{minipage}
\begin{minipage}[c]{0.5\linewidth}
\vspace{-0.5cm}
\hspace{-0.4cm}
\includegraphics[width=1.1\textwidth]{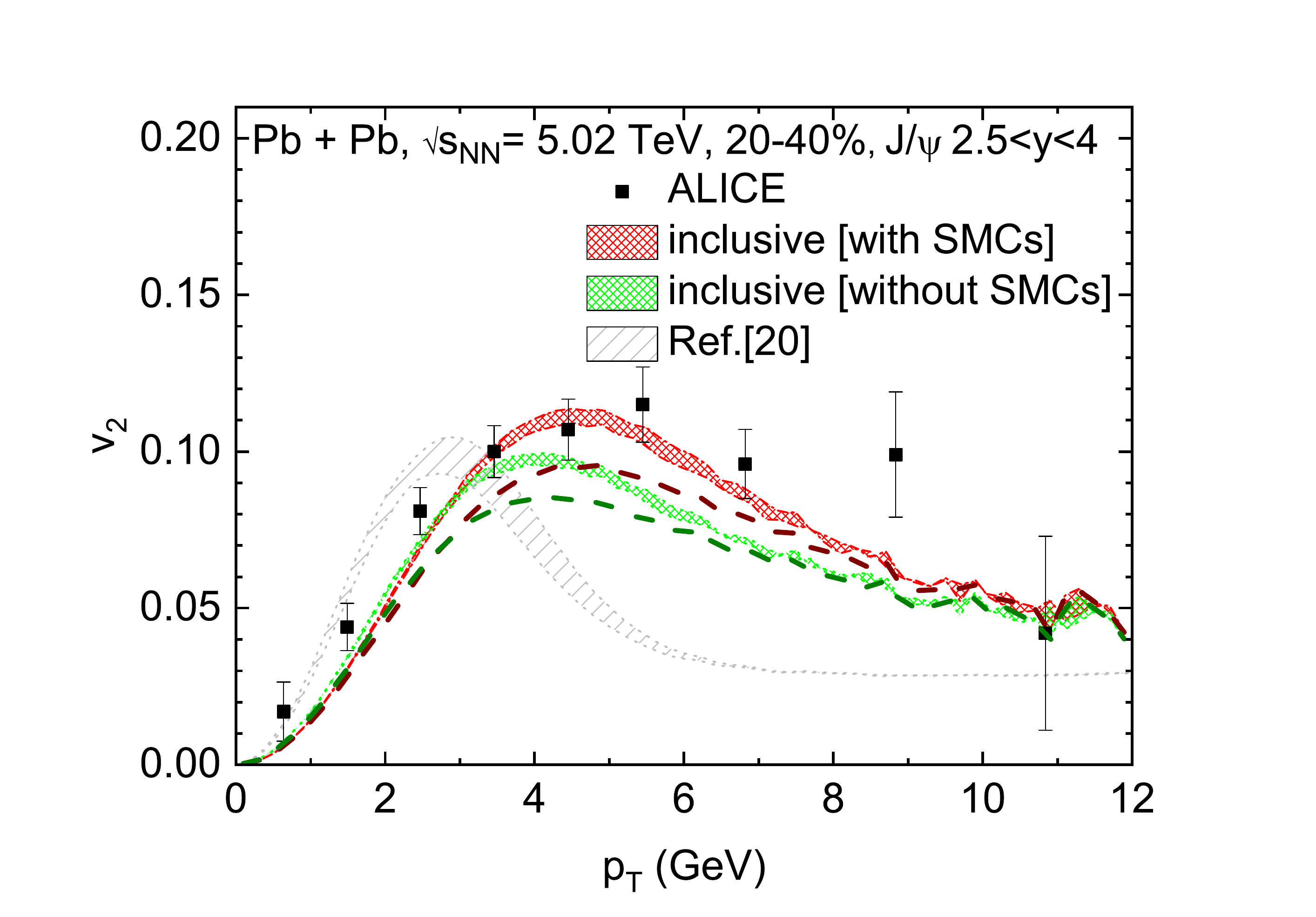}
\end{minipage}
\vspace{-0.3cm}
\caption{Inclusive-$J/\psi$ $R_{AA}$ (left) and $v_2$ (right) vs.~transverse momentum in semicentral 5\,TeV Pb-Pb collisions at the LHC as 
measured by ALICE~\cite{ALICE:2019lga,ALICE:2020pvw}, compared to transport model calculations based on the TAMU transport and resonance recombination model employing Langevin transported charm-quark distribution functions. For the red (green) band  the $c$-quark spectra are 
taken at an average regeneration time, $\tau_f$=5.2~fm/$c$ with(out) SMCs, where the widths reflect a 15-25\% charm/onium shadowing range.
The dashed lines are for $\tau_f$=4.2~fm/$c$ with 15\% shadowing (brown (dark-green) dashed: with(out) SMCs). Inclusive results from
previous calculations~\cite{Du:2015wha,ALICE:2019lga} (grey bands) are shown for comparison; figures taken from Ref.~\cite{He:2021zej}.}
\label{fig_jpsi-raa-v2}
\end{figure}

In recent work~\cite{He:2021zej} several improvements of the semiclassical transport description~\cite{Du:2015wha} have been carried out.
First, charm-quark phase space distributions from a strongly coupled transport approach for HF diffusion that describes open-charm observables 
at the LHC~\cite{He:2019vgs,ALICE:2021rxa} (as discussed in Sec.~\ref{ssec_hf-obs}) were implemented into the regeneration process of charmonia.
Specifically, the four-momentum conserving resonance recombination model (RRM)~\cite{Ravagli:2007xx} has been employed whose total yields are 
constrained by the rate-equation results with the latest $c\bar{c}$ cross section data from $pp$ collisions~\cite{ALICE:2021dhb} (corresponding to 
Fig.~\ref{fig_jpsi-raa-502}). Both momentum and spatial dependencies of these distributions (through transported momentum spectra and 
so-called space-momentum correlations (SMCs) between the position and momentum of the recombining quarks) have been found to extend the reach 
of recombination processes up to $p_T\sim 8$\,GeV in semicentral Pb-Pb (5.02\,TeV) collisions. Second, an explicit event-by-event simulation of the 
suppression of primordially produced charmonia in a hydrodynamically expanding medium have been found to produce a significantly larger $v_2$ 
for this component than previous schematic estimates (mostly caused by the path-length dependent dissociation through the spatial density
profile of the hydrodynamic medium). In addition, the $p_T$ and azimuthal-angle dependence of the non-prompt component from feeddown of 
transported $B$-mesons was accounted for~\cite{He:2014cla}. Taken together, these improvements result in a much improved description 
of the experimental data for $v_2$ of inclusive $J/\psi$ , and also better capture the momentum dependence of the $R_{\rm AA}$ data, see 
Fig.~\ref{fig_jpsi-raa-v2}. These developments highlight the close connection between the transport of open- and hidden-charm 
particles that become empirically accessible through precision data.

\subsubsection{Bottomonia}
\label{sssec_bottomonia}

\begin{figure}[!t]
\begin{minipage}{0.49\linewidth}
\includegraphics[width=0.97\textwidth]{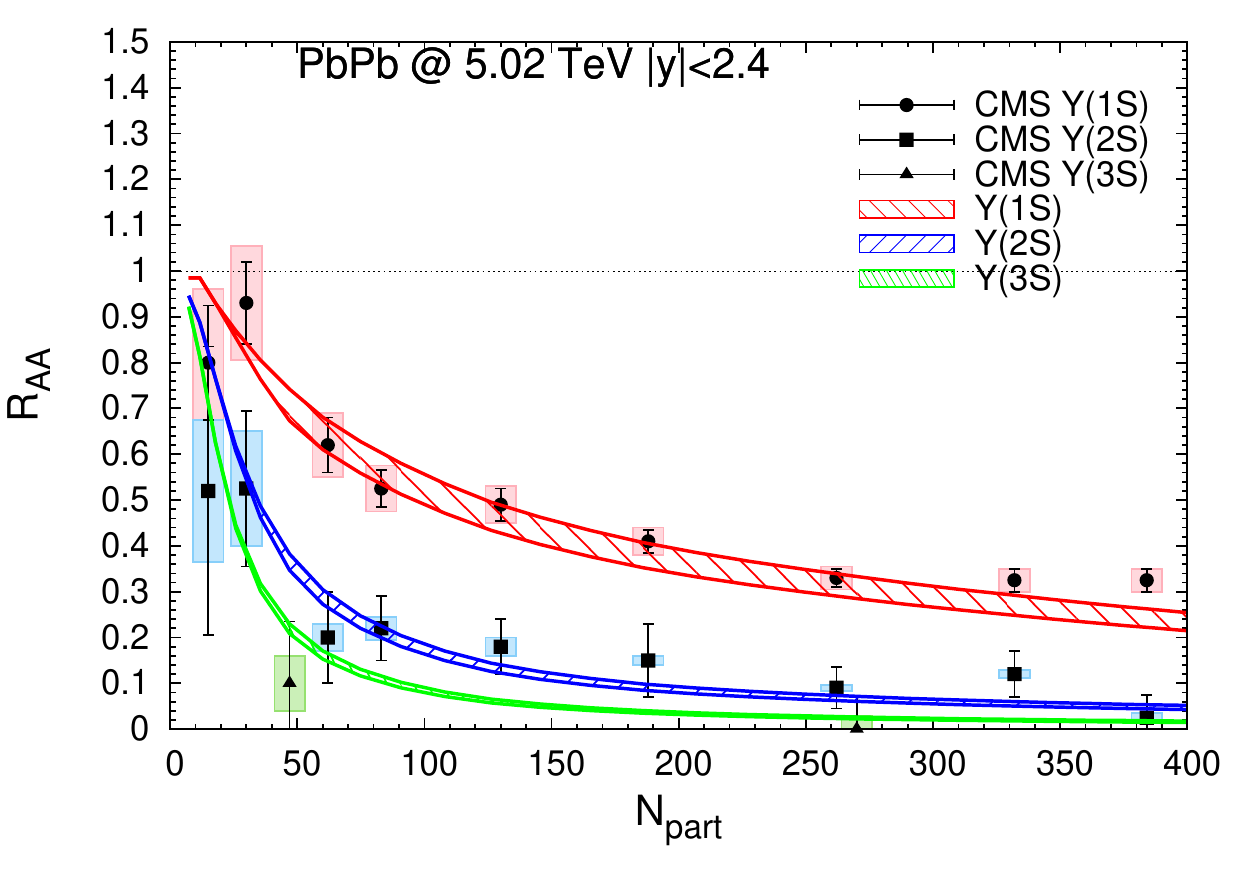}
\end{minipage}
\begin{minipage}{0.49\linewidth}
\hspace{-0.4cm}
\includegraphics[width=0.97\textwidth]{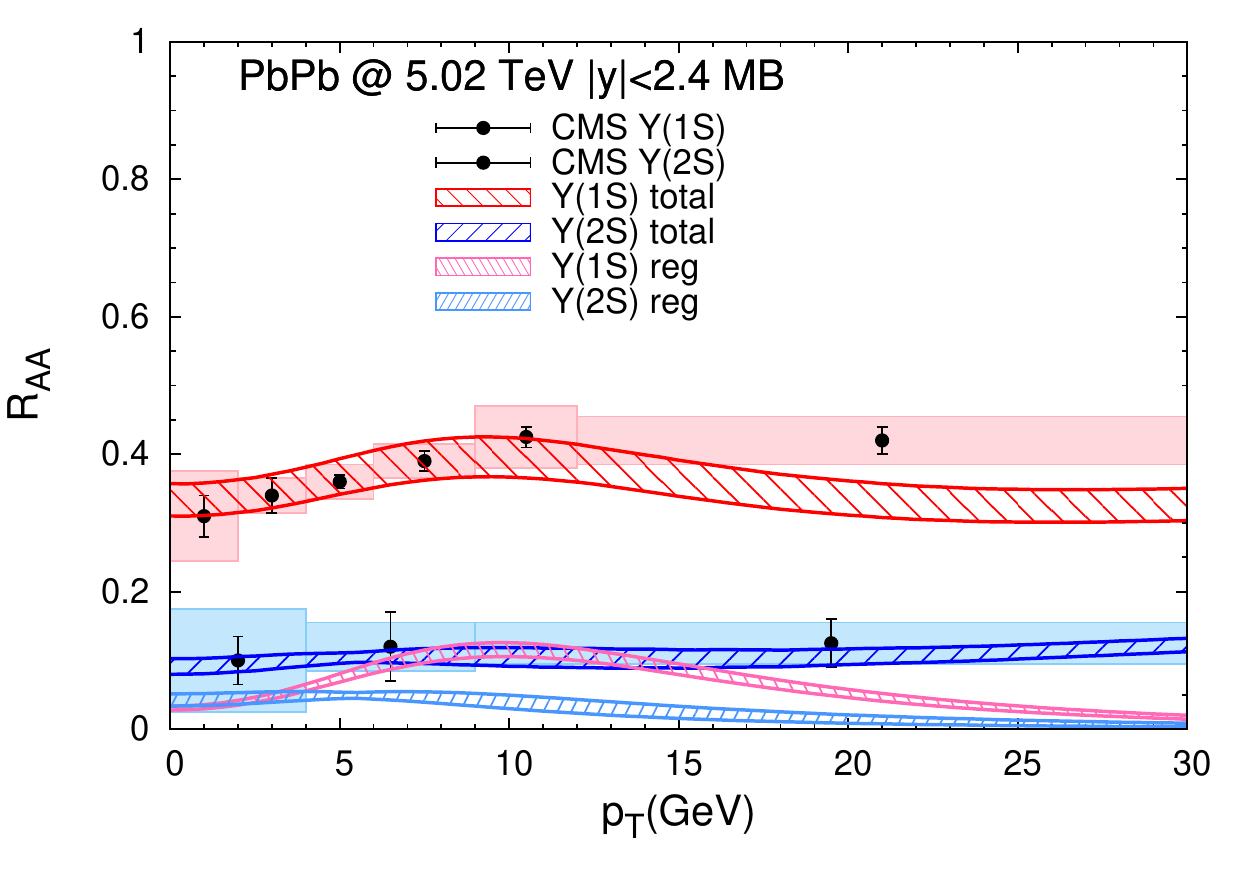}
\end{minipage}

\vspace{0.4cm}

\begin{minipage}{0.49\linewidth}
\hspace{0.2cm}
\includegraphics[width=0.9\textwidth]{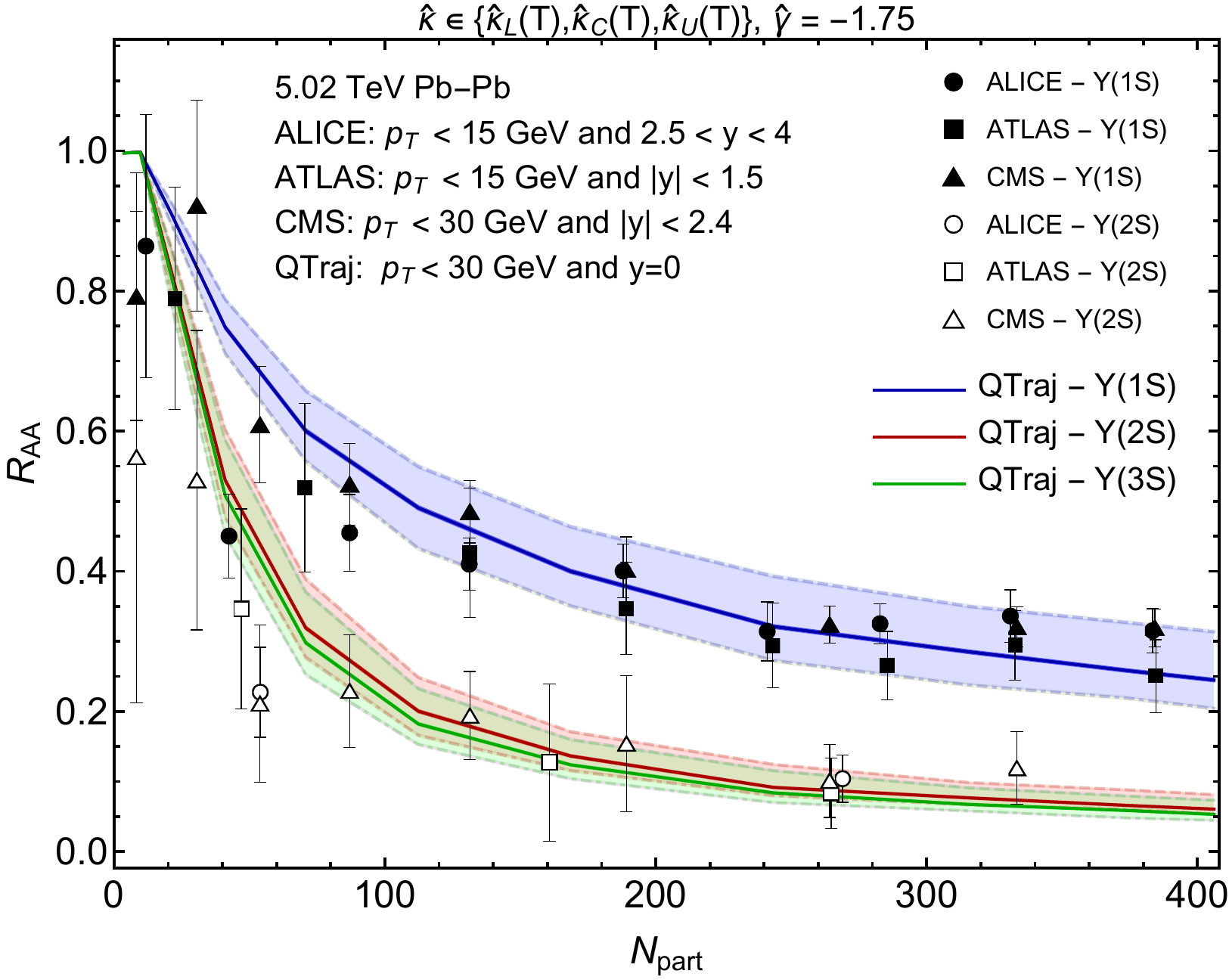}
\end{minipage}
\begin{minipage}{0.49\linewidth}
\includegraphics[width=0.9\textwidth]{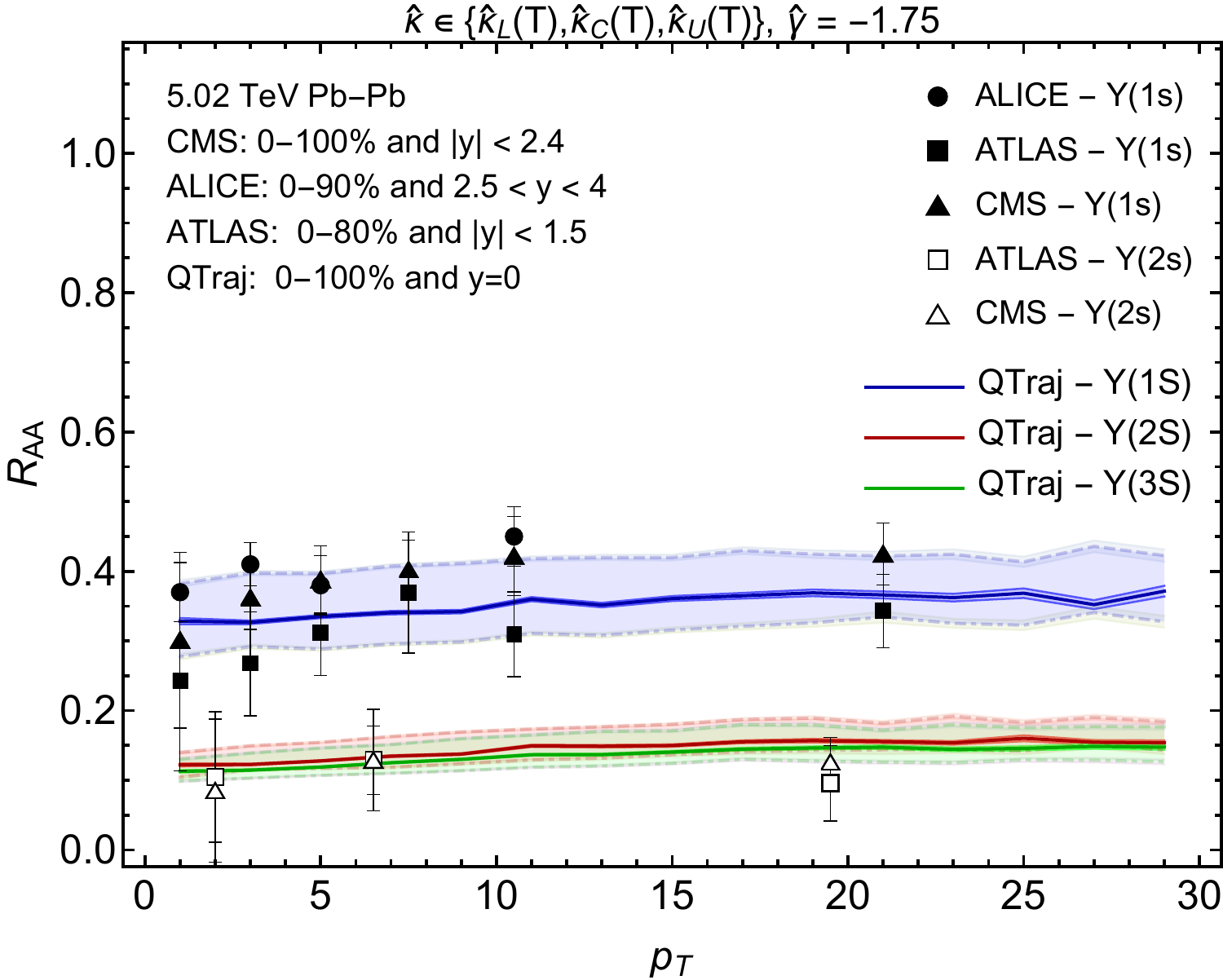}
\end{minipage}
\caption{Centrality (left column) and transverse-momentum (right column) dependence of the $R_{AA}$ for $\Upsilon(1S)$,  $\Upsilon(2S)$ and  
$\Upsilon(3S)$ production iin Pb-Pb(5.02\,TeV) collisions at the LHC as measured by CMS~\cite{CMS:2018zza}, ATLAS~\cite{Lee:2021vlb} and ALICE~\cite{ALICE:2020wwx}, compared to model calculations of the semiclassical TAMU transport approach~\cite{Du:2017qkv} (upper panels) 
and the quantum transport approach of TU Munich/Kent State~\cite{Brambilla:2021wkt}; figures taken from Refs.~\cite{Du:2017qkv} and  
\cite{Strickland:2021cox}.}
\label{fig_ups-raa}
\end{figure}
Contrary to charmonia, the regeneration contribution is expected to be much reduced for bottomonia, recall the right panel of Fig.~\ref{fig_excit-onia}.  
This opens a more direct and accurate window on the suppression mechanisms, in particular on the dissociation rates (closely related to the HQ interactions
with the medium) and their dependence on the binding energy, which in turn is closely related to the in-medium HQ potential.
Several investigations utilizing semiclassical transport approaches have established that the use of the free-energy proxy (taken from lQCD computations) 
for the in-medium HQ potential leads to an over-suppression of bottomonia compared to data from heavy-ion collisions, in particular for the $\Upsilon(1S)$ ~\cite{Emerick:2011xu,Strickland:2011aa,Zhou:2014hwa}. In Ref.~\cite{Du:2017qkv} in-medium binding energies from the $T$-matrix approach 
based on an internal-energy potential have been used, resulting in a fair description of LHC bottomonium data, cf.~the upper panels of 
Fig.~\ref{fig_ups-raa}; similar results have been obtained in Ref.~\cite{Krouppa:2016jcl} without the inclusion of regeneration contributions, and also 
more recently  with a ``lQCD vetted" potential~\cite{Krouppa:2017jlg} which is less attractive than the internal energy and thus produces a little more 
suppression which is still well compatible with the RHIC and LHC bottomonium data. A  state-of-the-art calculation employing quantum transport
is shown in the lower panels of Fig.~\ref{fig_ups-raa}~\cite{Brambilla:2021wkt}. It employs a vacuum Coulomb potential with in-medium widths that are related to the HQ momentum diffusion coefficient, $\kappa$, which has been taken from quenched lQCD computations~\cite{Brambilla:2020siz} 
(it is further related to the HQ diffusion coefficient as  $2\pi T\cD_s =4\pi T^3/\kappa$). For the $\Upsilon(1S)$, \eg, one has  
$\Gamma_{\Upsilon(1S)} = 3 a_0^2 \kappa$. 
With a Bohr radius of $a_0$=0.13\,fm and $\kappa/T^3$=1.3-3.6 at $T=1.5\,T_c\simeq 240$\,MeV, this translates into a width of $\simeq25-65$\,MeV, 
which is quite comparable to the values employed by the Kent-State group in their semiclassical approach, recall middle panel of Fig.~\ref{fig_gam-psi-ups}.  

\begin{figure}[!t]
\hspace{-0.3cm}
		\begin{minipage}[b]{0.32\linewidth}
			\includegraphics[width=1.15\textwidth]{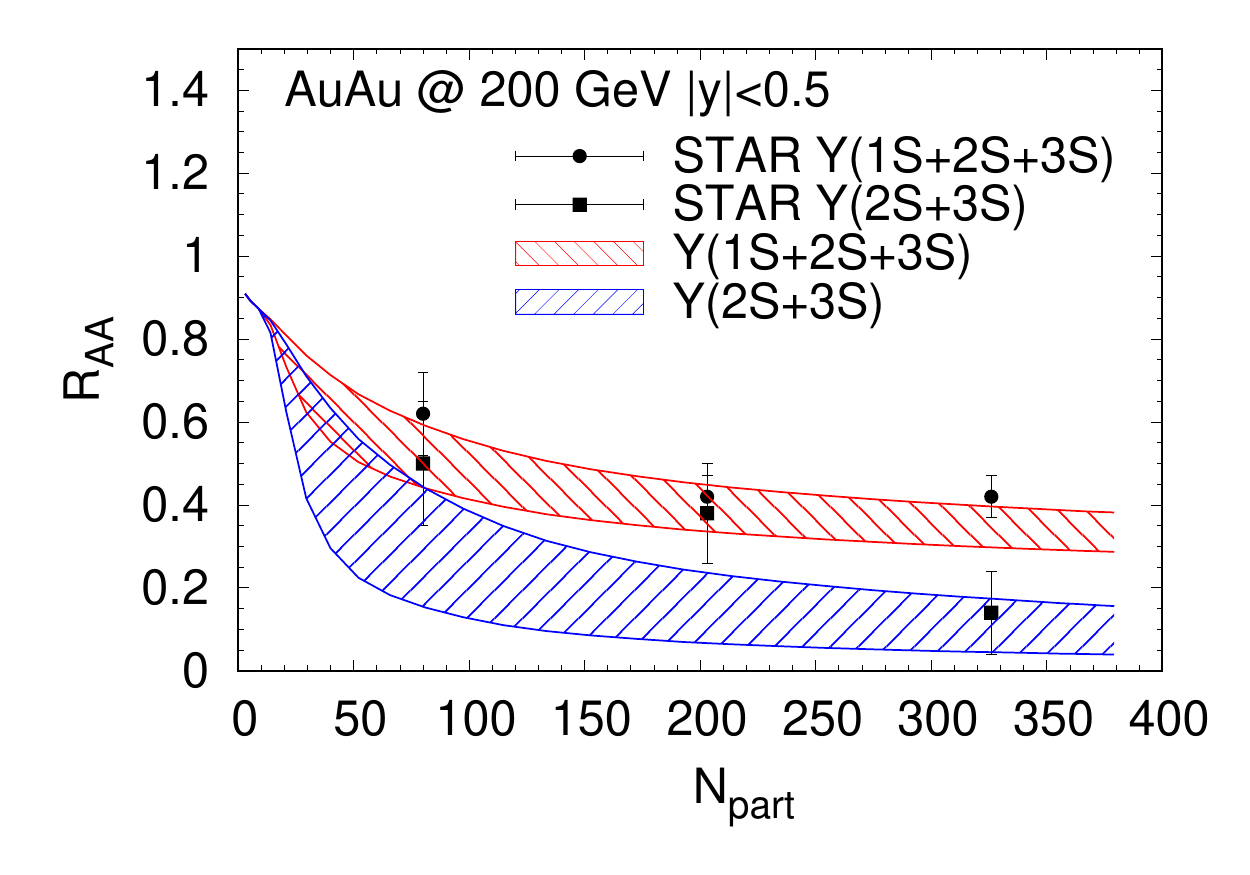}
		\end{minipage}
		\begin{minipage}[b]{0.32\linewidth}
			\includegraphics[width=1.15\textwidth]{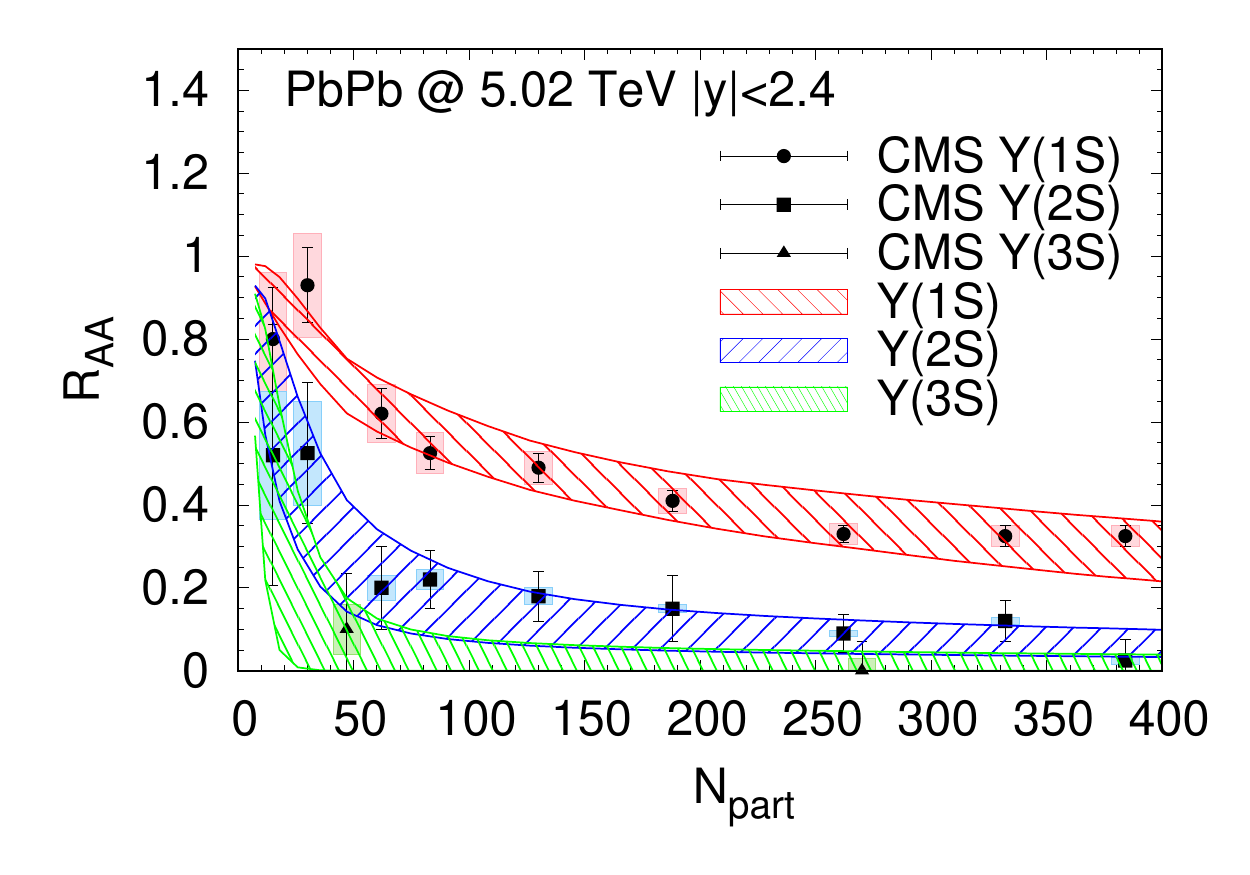}
		\end{minipage}
		\begin{minipage}[b]{0.32\linewidth}
			\includegraphics[width=1.15\textwidth]{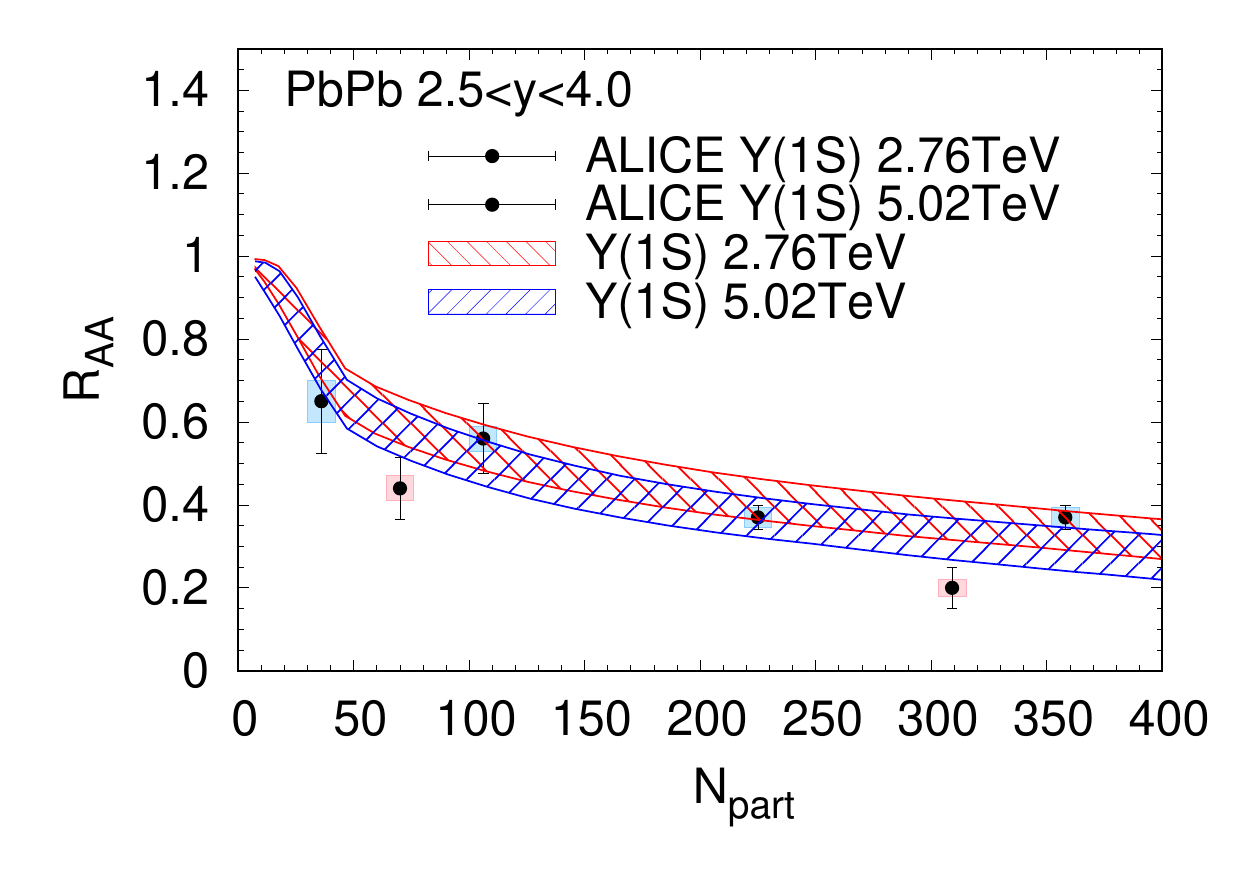}
		\end{minipage}
\caption{Fit results for bottomonium data from Au-Au collisions at RHIC and Pb-Pb collisions at the LHC utilizing an in-medium trial potential of Cornell type 
in combination with quasifree dissociation rates with a $K$-factor of 5 to simulate nonperturbative interaction strength for the HQ coupling to thermal partons.
The bands reflect 95\% confidence levels; figures taken from Ref.~\cite{Du:2019tjf}.}
\label{fig_ups-raa-fit}
\end{figure}

\begin{figure}[!t]
\begin{center}
\begin{minipage}{0.49\linewidth}
\includegraphics[width=1.0\textwidth]{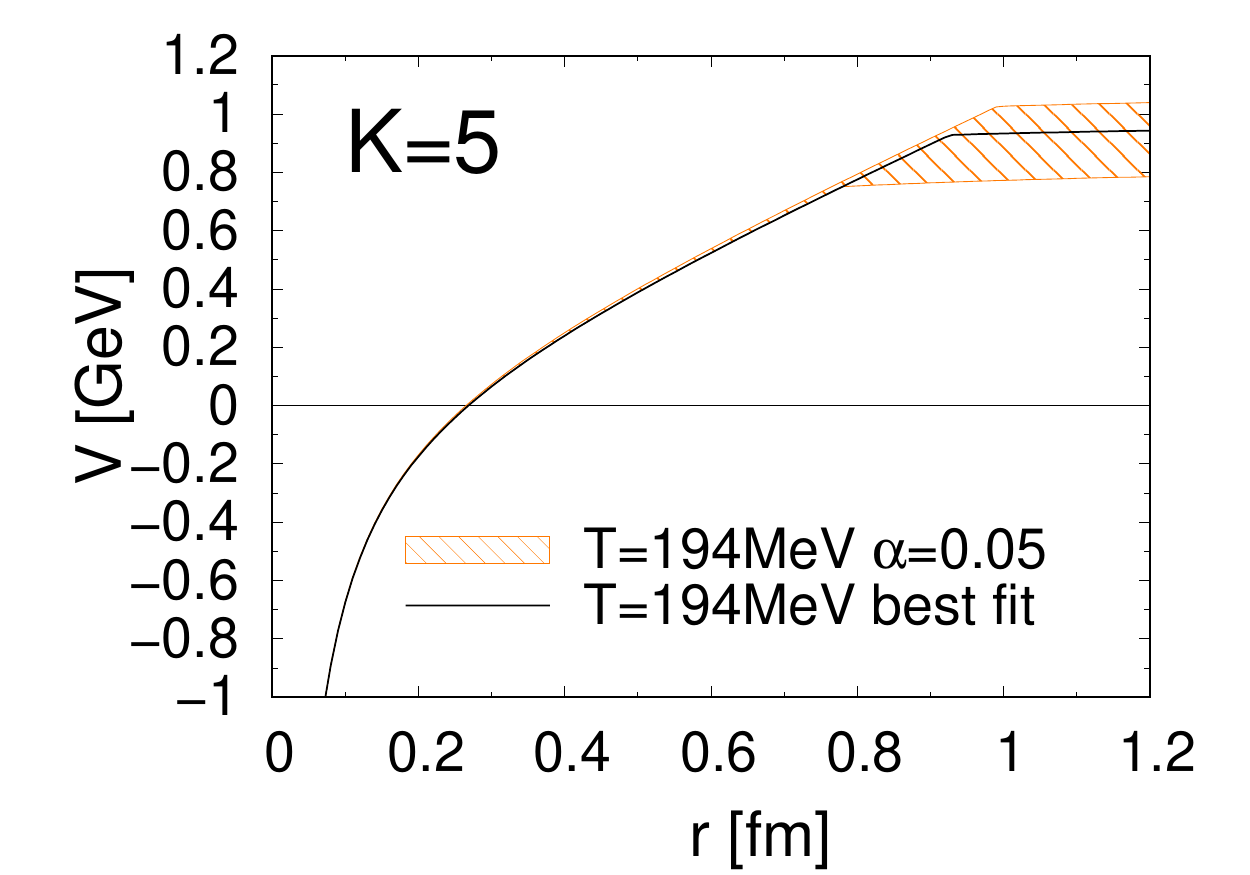}
\end{minipage}
\hspace{-0.8cm}
\begin{minipage}{0.49\linewidth}
\includegraphics[width=1.0\textwidth]{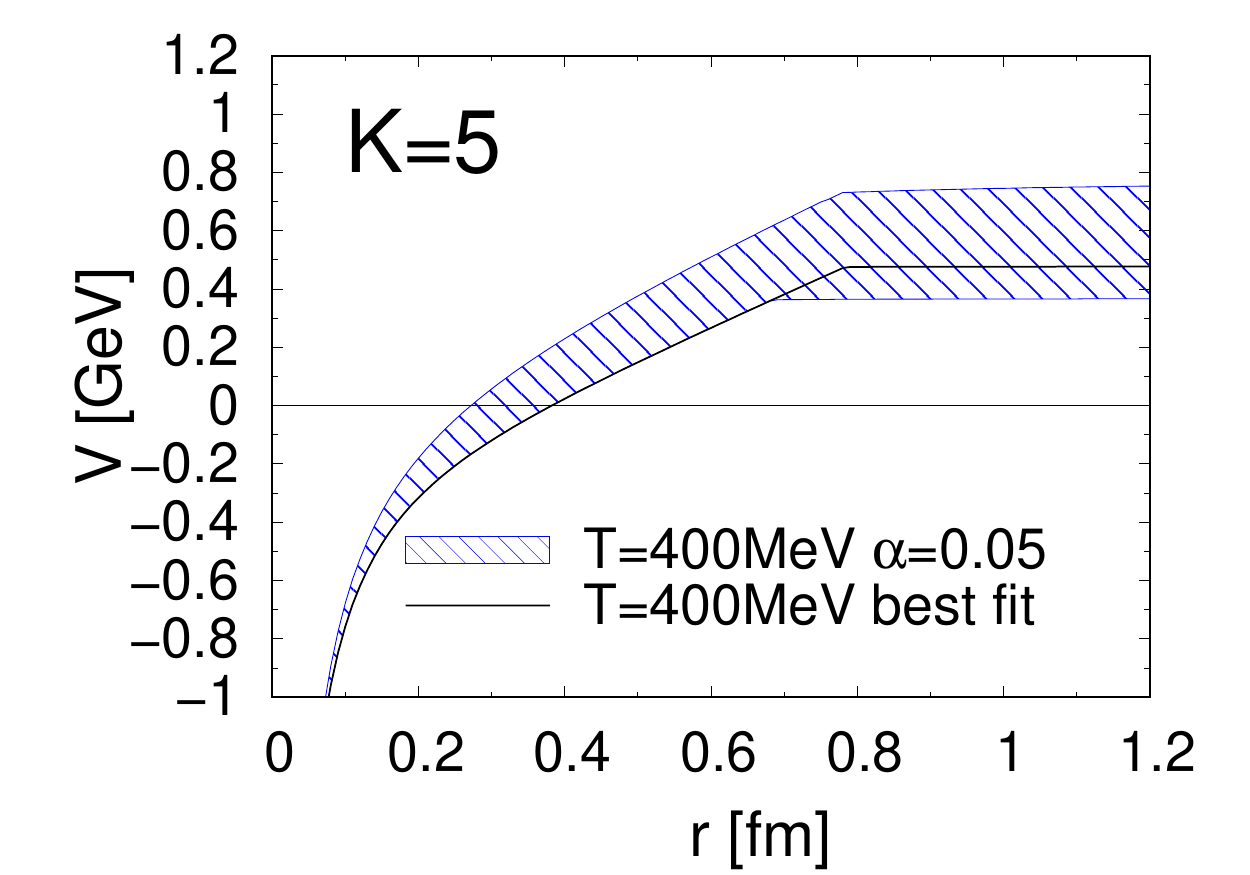}
\end{minipage}
\end{center}
\caption{Extracted in-medium HQ potentials determined through the fits to bottomonium data shown in Fig.~\ref{fig_ups-raa-fit} employing the 
TAMU transport approach with $K$-factor of 5 in the perturbative quasifree dissociation rates. The bands represent the 95\% confidence level for 
a given parameterization of the $r$- and $T$-dependence of $V(r,T)$, and the black lines show the best fit result, with a $\chi^2/dof$ of below one; 
figures taken from Ref. ~\cite{Du:2019tjf}.}
\label{fig_V-extract}
\end{figure}
In Ref.~\cite{Du:2019tjf} the TAMU transport approach~\cite{Du:2017qkv} has been employed to conduct a statistical extraction of the in-medium 
HQ potential from bottomonium data at RHIC and the LHC. Toward this end the in-medium $\Upsilon$ binding energies have been computed from 
the $T$-matrix approach utilizing trial input potentials of screened Cornell type, and then used to calculate the quasifree reaction rates
thereby allowing for a $K$-factor to simulate nonperturbative effects in the HQ coupling to the medium partons. 
For $K$=1, the $\Upsilon(1S)$ and $\Upsilon(2S)$ reaction rates recover the ones shown in the right two panels of 
Fig.~\ref{fig_gam-psi-ups}, and the pertinent potential is moderately screened at $T\simeq200$\,MeV and more strongly at higher temperatures.
It turns out that the quality of the fit to the experimental data changes very little when introducing $K$-factors larger than one in the quasifree rates, 
with a $\chi^2$ per degree of freedom of slightly below one up to at least $K$=10. However, for $K>1$ the underlying potentials require significantly 
stronger binding to keep the resulting reaction rates in a range which allows for a good description of the observed suppression.
As an example, we show in Fig.~\ref{fig_ups-raa-fit} the results for $K$=5, where the bands represent the 95\% confidence level that a given 
potential parameterization (as a function of $r$ and $T$) describes the data within a 2$\sigma$ level. 
The pertinent potentials are rather little screened for temperatures below $T\simeq250$\,MeV, and even for $T\simeq400$\,MeV
the residual confining force is still significant. When comparing these potentials to the $T$-matrix results shown in Fig.~\ref{fig_FvsV}, which 
are based on completely independent fits to  lQCD ``data", one finds a good agreement with  the strongly-coupled scenario (SCS). 
These stronger potentials are, in turn, required to obtain transport coefficients for HQ diffusion that are viable for open HF 
phenomenology~\cite{Rapp:2018qla} at RHIC and the LHC. This reiterates the additional constraining power of a combined analysis of open and 
hidden HF probes, and corroborates the importance of the remnant confining force in the QGP to generate its strong-coupling properties.


	\newpage

	\section{Summary and outlook}
       \label{sec_sum}
In this article we have reviewed recent research in the theory and phenomenology of Brownian motion of charm and bottom particles
in QCD matter as formed in high-energy heavy-ion collisions, and their role in studying the properties of the quark-gluon plasma and its 
transition back into hadrons. 
We have first discussed contemporary calculations of the basic entity for the diffusion process of low-momentum heavy quarks in the QGP, 
namely the elastic heavy-light quark scattering amplitude off thermal partons in the medium. In particular, we have laid out how 
nonperturbative interactions can be constructed by coupling quantum many-body theory to constraints from first-principle 
lattice QCD, taking advantage of the large HQ mass scale that enables the use of in-medium potential approaches. There is 
mounting evidence now that the in-medium HQ potential is subject to a rather weak screening at temperatures not too far above the 
pseudocritical one. This leaves significant remnants of the long-range confining force operative in the QGP which generates large scattering 
rates for the diffusing heavy quarks. It also dictates the need for resummations of the HQ interactions that can be carried out with the 
thermodynamic $T$-matrix approach. When approaching $\Tpc$ from above, the resummations generate broad partonic bound states 
which essentially correspond to the known spectroscopy of quarkonia and heavy-light mesons, as a conseqeunce of the vacuum benchmark 
of the potential approach.

While the strong in-medium potential enables the survival of quarkonium bound states in the QGP, in particular for the ground state, 
large HQ scattering rates in excess of 0.5 GeV are instrumental in melting the bound-state peaks. This is quite different from a weakly
coupled system, where a screening of the potential diminishes the binding energies, while small scattering rates play a minor role in the 
bound-state melting. We have also elaborated on the implications of a strong coupling environment on QGP structure: with scattering rates 
comparable to, or even larger than the thermal-parton masses, the light quarks and gluons can no longer be considered as quasiparticles 
and their contribution to the pressure appears to be taken over by emerging color-singlet bound states.   
The calculations of transport coefficients generally confirm that a large interaction strength in the heavy-light sector not only produces
a small diffusion coefficient for heavy quarks, but also implies values for bulk transport coefficients, such as the viscosity-over-entropy 
density ratio, that are close to conjectured lower bounds set by quantum mechanics. From this perspective, it is not surprising that the 
large off-shell effects, characterized by partonic spectral functions that are melted by their energy uncertainty, play a prominent role for 
the bulk structure. In addition, the broadening effects are essential for the HQ scattering to access the sub-threshold strength of 
the heavy-light resonances. 

We then provided a brief survey of the current HF diffusion phenomenology in heavy-ion colllisions. Recent advances in the precision of 
$D$-meson data for their nuclear modification factor and elliptic flow, in conjunction with model analyses, resulted in an appreciable step 
forward in the determination of the diffusion coefficient, narrowing it down to a range to 1.5-4.5 times the conjectured lower bound in 
the vicinity of $\Tpc$.  At the same time, progress in systematic measurements of the charm-hadrochemistry in AA collisions enabled much 
improved tests of hadronization models, thereby confirming the critical role of the HQ hadronization process through recombination with 
light quarks in interpreting observables. In other words, hadronization constitutes an essential interaction for the kinetics of 
HF particles in high-energy heavy-ion collisions. 
These findings suggest that a further improvement in the precision of charm-hadron data will be very significant, and that a similar data 
set for bottom hadrons would be much over-constraining the available model calculations.
We have also highlighted selected topics in the quarkonium sector, specifically those that bear a close connection to open heavy flavor. 
Recent work has shown that the non-equilibrated part of the charm-quark distributions at intermediate momenta can play a critical role in 
$J/\psi$ $p_T$ spectra, in particular to explain their large elliptic flow in this regime. Bottomonia are possibly the observable with the 
closest connection to the in-medium HQ potential; here, quantum approaches, when implemented nonperturbatively, are likely to play an 
important role going forward, especially for an internally consistent description of overlapping excited states. 

In  summary, we believe that HF particles will remain a premier asset for researching the QGP, to unravel the interactions and mechanisms 
that render it a strongly-coupled quantum liquid and result in its hadronization. With increasingly precise results from heavy-ion 
experiments in the charm sector, and independent stress tests in the bottom sector, microscopic models constrained by lattice-QCD data
will yield an unprecedented and robust understanding of the many-body physics of QCD matter.

	
	\newpage
	\section*{Acknowledgements}

	We thank our collaborators, in particular Xiaojian Du and  Shuai Y.F.Liu, for their contributions to the presented research, 
        and Biaogang Wu for providing Fig.~\ref{fig_jpsi-raa-v2}.   
	This work was supported by the US NSF under grant no. PHY-1913286 (RR), NSFC under grant no. 12075122 (MH), and DFG (Germany)     
         through the CRC-TR 211 (HvH).


	


	
	

\end{document}